\documentclass[review]{elsarticle}

\usepackage{lineno}
\usepackage{amssymb}
\usepackage{color}
\usepackage{amsmath}
\usepackage{tikz}
\usepackage{amsfonts}
\usepackage{mathrsfs}
\usepackage{amsthm}

\usepackage{bbding}
\usepackage{mathrsfs}
\usepackage{amsmath, amsfonts, amssymb, mathrsfs, txfonts}
\usepackage{graphicx, subfigure}
\usepackage{wasysym}
\usepackage{color}
\usepackage{bm}
\usepackage{enumerate}

\usepackage{pifont}
\usepackage{txfonts}
\usepackage{amsmath}
\usepackage{graphicx}
\usepackage{amsfonts}
\usepackage{amssymb}
\usepackage{mathrsfs,psfrag,eepic,epsfig}
\usepackage{epstopdf}

\modulolinenumbers[5]

\journal{Journal of \LaTeX\ Templates}

\makeatletter \@addtoreset{equation}{section}
\renewcommand{\theequation}{\arabic{section}.\arabic{equation}}

\renewcommand{\baselinestretch}{1.25}

\begin{document}

\begin{frontmatter}

\title{Riemann-Hilbert approach to the generalized variable -coefficient nonlinear Schr\"{o}dinger equation with non-vanishing boundary conditions \tnoteref{mytitlenote}}
\tnotetext[mytitlenote]{%Project supported by the Fundamental Research Fund for the Central Universities under the grant No. 2017XKQY101.\\
%\hspace*{3ex}$^{*}$
Corresponding author.\\
\hspace*{3ex}\emph{E-mail addresses}: sftian@cumt.edu.cn,
shoufu2006@126.com (S. F. Tian) }

\author{Zhi-Qiang Li, Shou-Fu Tian\tnoteref{mytitlenote} and Jin-Jie Yang}
\address{
School of Mathematics and Institute of Mathematical Physics, China University of Mining and Technology, Xuzhou 221116, People's Republic of China
%$^{2}$College of Mathematics and Systems Science, Shandong University of Science and Technology, Qingdao, 266590, China\\
}

\begin{abstract}
In this work, we consider the generalized variable-coefficient nonlinear Schr\"{o}dinger equation with non-vanishing boundary conditions at infinity  including the simple and double poles of the scattering coefficients. By introducing an appropriate Riemann surface and uniformization coordinate variable, we first convert the double-valued functions which occur in the process of direct scattering to single-value functions. Then, we  establish the direct scattering problem via analyzing   the analyticity, symmetries and asymptotic behaviors of Jost functions and  scattering matrix derived from Lax pairs of the equation. Based on these results, a generalized Riemann-Hilbert problem is successfully established for the equation. The discrete spectrum and residual conditions, trace foumulae and theta conditions are investigated systematically including the simple poles case and double poles case. Moreover, the inverse scattering problem is solved via the Riemann-Hilbert approach. Finally, under the condition of reflection-less potentials, the soliton and breather solutions are well derived. Via evaluating the impact of each parameters, some interesting phenomena of these solutions are analyzed graphically.
\end{abstract}


\end{frontmatter}

%\linenumbers 去掉论文中的行标号
%\tableofcontents
%\newpage

\section{Introduction}

Nonlinear Schr\"{o}dinger (NLS) equation
\begin{equation}
iu_{t}+u_{xx}+2|u|^{2}u=0,
\end{equation}
is a basic physical model. The NLS-type equations play an important role in various fields of nonlinear science such as deep water waves\cite{I-1}, plasma physics\cite{I-2,I-3}, nonlinear optical fibers\cite{I-6,I-7}, magneto-static spin waves\cite{I-8}. Because of the rich mathematical structure of the NLS-type equations, the research on such equations are more and more popular.
There are three well-known derivative
NLS equations, including   Kaup-Newell equation \cite{KN},   Chen-Lee-Liu
equation \cite{CLL} and   Gerdjikov-Ivanov equation \cite{GI,Tian-PAMS} which have been extensively studied. It is known that these three equations
may be transformed into each other by implicit gauge transformations, and the method of
gauge transformation can also be applied to some generalized cases \cite{Fan-dnls}.
In this respect, there are a lot of research for the extended NLS equations \cite{I-13}-\cite{I-18}.

As we know that exact solutions of these equations is one of important branch among these researches. However, in practise, there are many other nonlinear effects that need to be considered for the  solutions.
In this work, we investigate the generalized variable-coefficient nonlinear Schr\"{o}dinger (gvcNLS) equation\cite{I-19} with nonzero boundary conditions (NZBCs) at infinity which have extensive applications in physical fields. The gvcNLS equation takes the form
\begin{gather}
iq_{t}+\frac{1}{2}q_{xx}+q^{2}q^{*}-i\alpha(t)q_{xxx}-6i\alpha(t)q_{x}qq^{*}+4\gamma(t)q_{x}qq^{*}_{x}\notag +8\gamma(t)q_{xx}qq^{*}\\
+6\gamma(t)q_{x}^{2}q^{*}+6\gamma(t)q^{3}(q^{*})^{2}+2\gamma(t)q^{2}q^{*}_{xx}+\gamma(t)q_{xxxx}
=0,
\label{1.1}
\end{gather}
where $q$ is complex function of variables $x$, $t$, the coefficient $\alpha(t)$ and $\gamma(t)$ are both the real functions of $t$. Furthermore, some unique cases of the equation \eqref{1.1} can be obtained by fixing the coefficient $\alpha(t)$ and $\gamma(t)$. Eq. \eqref{1.1} have some reductions as follows.
\begin{itemize}
 \item When $\alpha(t)=\gamma(t)=0$, Eq. \eqref{1.1} degenerates into the NLS equation which can be used to describe the solitons and rogue waves\cite{I-20}.
 \item When $\alpha(t)=0$ and $\gamma(t)=$constant, Eq. \eqref{1.1} degenerates into the Lakshmanan-Porsezian-Daniel(LPD) equation which can be used to describe the nonlinear spin excitations \cite{Daniel, I-21}.
 \item When $\alpha(t)=$constant and $\gamma(t)=0$, Eq. \eqref{1.1} degenerates into the Hirota equation\cite{I-22}.
 \item When $\alpha(t)=$constant and $\gamma(t)=$constant, Eq. \eqref{1.1} degenerates into an extended NLS equation with higher-order odd and even terms with independent coefficients, and its soliton solutions have been obtained via applying the Darboux transformation(DT)\cite{I-23}.
 \end{itemize}

To the best of our knowledge, the research using the inverse scattering transform (IST) for the gvcNLS equation with NZBCs has not been reported yet.
The IST is first presented to solve exactly the famous Korteweg-de Vries equation \cite{I-29}. Then, Zakharov and Shabat\cite{I-3} show that the method can be applied to physically significant nonlinear evolution equation, namely, the nonlinear Schr\"{o}dinger equation. After that, it is more and more popular to use the IST to study the NLS equations. Although the IST is extended to study the NLS equations with NZBCs \cite{I-30}-\cite{I-48}, there are very few work   to study the case of variable coefficients.

In this work, we study the gvcNLS equation \eqref{1.1} with the following NZBCs
\begin{gather}\label{1.2}
q(x,t)\sim q_{\pm}e^{i(q_{0}^{2}+6\gamma q_{0}^{4})t}, ~~~~~~ x\rightarrow\pm\infty,
\end{gather}
where $|q_{\pm}|=q_{0}>0$, $q_{\pm}$ are independent of $x$, $t$, and $\gamma$ is a constant.
Here, two goals will be achieved. On one hand, based on the Lax pairs of the equation \eqref{1.1}, some results are established, involving the analyticity, symmetries and asymptotic behaviors of the scattering coefficients, the establishment of a generalized Riemann-Hilbert problem and the construction of the discrete spectrum and residual conditions. On the other hand, based on the above results, the solutions of the gvcNLS equation with NZBCs will be derived for the simple and double poles. In view of the complexity of the gvcNLS equation and its Lax pairs, some computational skills are used here, and a lot of calculations are made during the analysis to achieve that two goals. Meanwhile, to obtain the solutions of the gvcNLS equation with NZBCs for the double poles, we need to make a substitution.

The outline of this work is as follows. In section 2, we introduce an appropriate Riemann surface and uniformization variable to deal with the double-valued functions.  In section 3 and section 4, we analyze the analytical properties of the Jost functions and scattering matrix. In section 5 and section 6, we analyze the symmetries and asymptotic properties of the Jost functions and scattering matrix. In section 7, a generalized Riemann-Hilbert problem is constructed for the equation. In section 8, in the case of simple poles, we discuss the discrete spectrum and residue condition, establish reconstructed formula, and obtain the trace formulate as well as theta condition. Then, under the reflectionless condition, the $N$-soliton solution is well derived. Finally,  some interesting phenomena of the soliton solutions  are analyzed graphically. In section 9, in the case of double poles, the similar content is also presented. Some conclusions and discussions are given in the last section.

\section{Riemann surface and uniformization coordinate}
In this section, we need to overcome the difficulty that the double-valued functions will arise during the analysis. Firstly, to make the analysis easier, we introduce an appropriate transformation to change the boundary conditions\eqref{1.2} into constant boundary.

The Lax pair of the equation \eqref{1.1} reads
\begin{gather}
\Psi_{x}=X\Psi,\qquad \Psi_{t}=T\Psi, \notag \\
\Psi=(\Psi_{1},\Psi_{2})^{T}, \label{2.1}
\end{gather}
where $\Psi_{i}, (i=1,2)$ are eigenfunctions, and
\begin{gather}
X=ik\sigma_{3}+Q, \notag \\
T=\left(                 %左括号
  \begin{array}{cc}   %该矩阵一共2列，每一列都居中放置
    T_{11} & T_{12}  \\  %第一行元素
    T_{21} & -T_{11} \\  %第二行元
  \end{array}
\right),\label{2.2}
\end{gather}
with
\begin{equation}
Q=\left(                 %左括号
  \begin{array}{cc}   %该矩阵一共2列，每一列都居中放置
    0 & iq^{*}  \\  %第一行元素
    iq & 0 \\  %第二行元
  \end{array}
\right), \qquad
\sigma_{3}=\left(                 %左括号
  \begin{array}{cc}   %该矩阵一共2列，每一列都居中放置
    1 & 0  \\  %第一行元素
    0 & -1 \\  %第二行元素
  \end{array}
\right), \notag \\
\end{equation}\\
\begin{gather}
T_{11}=ik^{2}-4ik^{3}\alpha(t)-8ik^{4}\gamma(t)-\frac{1}{2}iqq^{*}+2ik\alpha(t)qq^{*}+4ik^{2}\gamma(t)qq^{*} \notag \\
-3i\gamma(t)q^{2}(q^{*})^{2}-\alpha(t)q_{x}q^{*}-2k\gamma(t)q_{x}q^{*}+\alpha(t)qq^{*}_{x}+2k\gamma(t)qq^{*}_{x} \notag \\
+i\gamma(t)q_{x}q^{*}_{x}-i\gamma(t)q_{xx}q^{*}-i\gamma(t)qq^{*}_{xx} ,\notag\\
T_{12}=ikq^{*}-4ik^{2}\alpha(t)q^{*}-8ik^{3}\gamma(t)q^{*}+2i\alpha(t)q(q^{*})^{2}+4ik\gamma(t)q(q^{*})^{2} \notag\\
+\frac{1}{2}q^{*}_{x}-2k\alpha(t)q^{*}_{x}-4k^{2}\gamma(t)q^{*}_{x}+6\gamma(t)qq^{*}q^{*}_{x}+i\alpha(t)q^{*}_{xx} \notag\\
+2ik\gamma(t)q^{*}_{xx}+\gamma(t)q^{*}_{xxx},\notag\\
T_{21}=ikq-4ik^{2}\alpha(t)q-8ik^{3}\gamma(t)q+2i\alpha(t)q^{2}q^{*}+4ik\gamma(t)q^{2}q^{*} \notag\\
-\frac{1}{2}q_{x}+2k\alpha(t)q_{x}+4k^{2}\gamma(t)q_{x}-6\gamma(t)qq^{*}q_{x}+i\alpha(t)q_{xx} \notag\\
+2ik\gamma(t)q_{xx}-\gamma(t)q_{xxx}.
\end{gather}
In equation \eqref{2.2}, the complex parameter $k$ is independent of $x$ and $t$, and the compatibility condition $X_{t}-T_{x}+[X,T]=0$ leads to equation \eqref{1.1}.

We take a transformation
\begin{align}
\begin{split}
&q= qe^{i(q_{0}^{2}+6\gamma q_{0}^{4})t},\\
&\Psi= \phi e^{i(q_{0}^{2}+6\gamma q_{0}^{4})t\sigma_{3}}, \notag
\end{split}
\end{align}
then   equation \eqref{1.1} changes to
\begin{gather}
iq_{t}+\frac{1}{2}q_{xx}-(q_{0}^{2}+3\gamma q_{0}^{4})q+q^{2}q^{*}-i\alpha(t)q_{xxx}-6i\alpha(t)q_{x}qq^{*}+4\gamma q_{x}qq^{*}_{x}\notag \\
+8\gamma q_{xx}qq^{*}+6\gamma q_{x}^{2}q^{*}+6\gamma q^{3}(q^{*})^{2}+2\gamma q^{2}q^{*}_{xx}
=0,
\label{2.3}
\end{gather}
and the corresponding boundary becomes
\begin{gather}
\lim_{x\rightarrow \pm \infty}q(x,t)=q_{\pm}, \qquad \mid q_{\pm}\mid=q_{0}\neq 0. \label{asb}
\end{gather}
The Lax pair \eqref{2.2} is changed to
\begin{gather}
\phi_{x}=X\phi,\qquad \phi_{t}=T\phi, \notag \\
\phi=(\phi_{1},\phi_{2})^{T}, \label{2.4}
\end{gather}
where
\begin{gather}
X=-ik\sigma_{3}+Q, \quad Q=\left(                 %左括号
  \begin{array}{cc}   %该矩阵一共2列，每一列都居中放置
    0 & iq^{*}(x,t)  \\  %第一行元素
    iq(x,t) & 0 \\  %第二行元
  \end{array}
\right)\notag \\
T=\left(                 %左括号
  \begin{array}{cc}   %该矩阵一共2列，每一列都居中放置
    T_{11}-i(\frac{1}{2}q_{0}^{2}+3\gamma q_{0}^{4})\sigma_{3} & T_{12}  \\  %第一行元素
    T_{21} & -T_{11}+i(\frac{1}{2}q_{0}^{2}+3\gamma q_{0}^{4})\sigma_{3} \\  %第二行元
  \end{array}
\right),\notag
\end{gather}
and the coefficient $\alpha(t)$ is the real functions of $t$ and $\gamma$ is a constant.
Based on the asymptotic boundary \eqref{asb}, letting $x\rightarrow\pm\infty$, the limit spectral problem can be obtained as
\begin{gather}
\psi_{x}=X_{\pm}\psi,\qquad \psi_{t}=T_{\pm}\psi, \label{2.5}
\end{gather}
where
\begin{gather}
X_{\pm}=-ik\sigma_{3}+Q_{\pm},  Q_{\pm}=\left(                 %左括号
  \begin{array}{cc}   %该矩阵一共2列，每一列都居中放置
    0 & iq^{*}_{\pm}  \\  %第一行元素
    iq_{\pm} & 0 \\  %第二行元
  \end{array}
\right),\notag \\
T_{\pm}=(-8\gamma k^{3}-4\alpha(t)k^{2}+(4\gamma q_{0}^{2}+1)k+2\alpha(t)q_{0}^{2})X_{\pm}.\notag
\end{gather}
After a simple calculation, we get the eigenvalues $\pm i\lambda=\pm i\sqrt{k^{2}+q_{0}^{2}}$ of the matrix $X_{\pm}$. Now, we need to deal with the difficulty that the eigenvalues are doubly branched. The branch points of the eigenvalues are $k=\pm iq_{0}$. Therefore, we introduce a Riemann surface which is completed by gluing together two copies of extended complex $k$-plane $S_{1}$ and $S_{2}$ along the cut $iq_{0}[-1,1]$ between the branch points $k=\pm iq_{0}$. To get a single-valued function, we set
\begin{align}
k+iq_{0}=r_{1}e^{i\theta_{1}},\quad k-iq_{0}=r_{2}e^{i\theta_{2}},\quad -\frac{\pi}{2}<\theta_{1},\theta_{2}<\frac{3\pi}{2}.
\end{align}
Then, the single-valued analytical function on the Riemann surface can be obtained as
\begin{align}
\lambda(k)=&\left\{\begin{aligned}
&(r_{1}r_{2})^\frac{1}{2}e^\frac{{\theta_{1}+\theta_{2}}}{2}, \quad &on\quad S_{1},\\
-&(r_{1}r_{2})^\frac{1}{2}e^\frac{{\theta_{1}+\theta_{2}}}{2}, \quad &on\quad S_{2}.
\end{aligned} \right. \label{2.6}
\end{align}
From the transformation \eqref{2.6}, the following properties can be obtained easily as
\begin{itemize}
  \item  $Im k>0$ of sheet $S_{1}$ and $Im k<0$ of sheet $S_{2}$ are mapped into $Im \lambda >0$,
  \item  $Im k<0$  of sheet $S_{1}$ and $Im k>0$ of sheet $S_{2}$ are mapped into $Im \lambda <0$.
\end{itemize}
To ensure that the eigenvalue $\lambda$ is single-valued, we define a uniformization variable $z$  as
\begin{align}
z=k+\lambda,
\end{align}
by using the results presented in \cite{I-50}.
Then, we obtain the following two single-value functions
\begin{align}\label{2.7}
\lambda(z)=\frac{1}{2}\left(z+\frac{q_{0}^{2}}{z}\right),\quad k(z)=\frac{1}{2}\left(z-\frac{q_{0}^{2}}{z}\right).
\end{align}
Interestingly, there are two asymptotic relationships between two planes $k$ and $z$.
When $k\in S_{1}$, we have
\begin{align*}
z=k+\sqrt{k^{2}+q_{0}^{2}}=k+k\left(1+\frac{q_{0}^{2}}{k^{2}}+\cdots\right)^{1/2}\sim 2k+o(k^{-1}), \quad k\rightarrow\infty.\notag
\end{align*}
That means $z\rightarrow\infty$ as $k\rightarrow\infty$ $(k\in S_{1})$. Meanwhile, when $k\in S_{2}$,
$z\rightarrow 0(k\rightarrow\infty)$ $(k\in S_{2})$ can be obtained, similarly.
Now, we consider the correspondence between $\lambda-$plane and $z-$plane.
Evaluating the imaginary part of the Joukowsky transformation
\begin{align}\label{2.8}
\lambda=\frac{z(|z|^{2}-q_{0}^{2})+2q_{0}^{2}Rez}{2|z|^{2}},
\end{align}
i.e., $Im\lambda=\frac{|z|^{2}-q_{0}^{2}}{2|z|^{2}}Im z$, we can easily obtain the following properties
\begin{itemize}
  \item  $Im \lambda>0$ is mapped into $\left\{z\in \mathbb{C}:\left(|z|^{2}-q_{0}^{2}\right)Im z>0\right\}\triangleq D_{+}$,
  \item  $Im \lambda<0$ is mapped into $\left\{z\in \mathbb{C}:\left(|z|^{2}-q_{0}^{2}\right)Im z<0\right\}\triangleq D_{-}$.
\end{itemize}
We can summarize the above analysis into the following illustrations in figure 1.
\\

\centerline{\begin{tikzpicture}
\path [fill=green] (-4.5,2.5) -- (-0.5,2.5) to
(-0.5,4.5) -- (-4.5,4.5);
\draw[-][thick](-4.5,2.5)--(-2.5,2.5);
\draw[fill] (-2.5,2.5) circle [radius=0.035];
\draw[->][thick](-2.5,2.5)--(-0.5,2.5)node[above]{$Rek$};
\draw[<-][thick](-2.5,4.5)node[right]{$Imk$}--(-2.5,3.5)node[right]{$iq_{0}$};
\draw[fill] (-2.5,3.5) circle [radius=0.035];
\draw[-][thick](-2.5,3.5)--(-2.5,2.5);
\draw[-][thick](-2.5,2.5)--(-2.5,1.5)node[right]{$-iq_{0}$};
\draw[fill] (-2.5,1.5) circle [radius=0.035];
\draw[-][thick](-2.5,1.5)--(-2.5,0.5);
\draw[fill] (-4.5,4) node[right]{$S_{1}$};
\draw[fill] (-2.5,2.2) node[right]{$0$};
\draw[fill] (-1.8,3.8) circle [radius=0.035] node[right]{$Im k>0$};
\draw[fill] (-1.8,1.7) circle [radius=0.035] node[right]{$Im k<0$};
\path [fill=green] (0.5,0.5) -- (4.5,0.5) to
(4.5,2.5) -- (0.5,2.5);
\draw[-][thick](0.5,2.5)--(2.5,2.5);
\draw[fill] (2.5,2.5) circle [radius=0.035];
\draw[->][thick](2.5,2.5)--(4.5,2.5)node[above]{$Rek$};
\draw[<-][thick](2.5,4.5)node[right]{$Imk$}--(2.5,3.5)node[right]{$iq_{0}$};
\draw[fill] (2.5,3.5) circle [radius=0.035];
\draw[-][thick](2.5,3.5)--(2.5,2.5);
\draw[-][thick](2.5,2.5)--(2.5,1.5)node[right]{$-iq_{0}$};
\draw[fill] (2.5,1.5) circle [radius=0.035];
\draw[-][thick](2.5,1.5)--(2.5,0.5);
\draw[fill] (0.5,4) node[right]{$S_{2}$};
\draw[fill] (-2.5,2.2) node[right]{$0$};
\draw[fill] (3.1,3.8) circle [radius=0.035] node[right]{$Im k>0$};
\draw[fill] (3.1,1.7) circle [radius=0.035] node[right]{$Im k<0$};
\path [fill=green] (-4.5,-2.5) -- (-0.5,-2.5) to
(-0.5,-0.5) -- (-4.5,-0.5);
\draw[-][thick](-4.5,-2.5)--(-2.5,-2.5);
\draw[fill] (-2.5,-2.5) circle [radius=0.035];
\draw[->][thick](-2.5,-2.5)--(-0.5,-2.5)node[above]{$Rek$};
\draw[<-][thick](-2.5,-0.5)node[right]{$Imk$}--(-2.5,-1.5)node[right]{$iq_{0}$};
\draw[fill] (-2.5,-1.5) circle [radius=0.035];
\draw[-][thick](-2.5,-1.5)--(-2.5,-2.5);
\draw[-][thick](-2.5,-2.5)--(-2.5,-3.5)node[right]{$-iq_{0}$};
\draw[fill] (-2.5,-3.5) circle [radius=0.035];
\draw[-][thick](-2.5,-3.5)--(-2.5,-4.5);
\draw[fill] (-4.5,-1) node[right]{$\lambda-$plane};
\draw[fill] (-2.5,-2.7) node[right]{$0$};
\draw[fill] (-1.9,-1.7) circle [radius=0.035] node[right]{$Im \lambda>0$};
\draw[fill] (-1.9,-3.8) circle [radius=0.035] node[right]{$Im \lambda<0$};
\path [fill=green] (0.5,-2.5) -- (4.5,-2.5) to
(4.5,-0.5) -- (0.5,-0.5);
\filldraw[white, line width=0.5](3.5,-2.5) arc (0:180:1);
\filldraw[green, line width=0.5](1.5,-2.5) arc (-180:0:1);
\draw[->][thick](0.5,-2.5)--(1,-2.5);
\draw[-][thick](1,-2.5)--(2,-2.5);
\draw[<-][thick](2,-2.5)--(2.5,-2.5);
\draw[fill] (2.5,-2.5) circle [radius=0.035];
\draw[-][thick](2.5,-2.5)--(3,-2.5);
\draw[<->][thick](3,-2.5)--(4,-2.5);
\draw[-][thick](4,-2.5)--(4.5,-2.5)node[above]{$Rez$};
\draw[-][thick](2.5,-0.5)node[right]{$Imz$}--(2.5,-2.5);
\draw[-][thick](2.5,-2.5)--(2.5,-4.5);
\draw[fill] (2.5,-2.8) node[right]{$0$};
\draw[fill] (2.5,-1.5) circle [radius=0.035];
\draw[fill] (2.5,-3.5) circle [radius=0.035];
\draw[fill] (2.5,-1.2) node[right]{$iq_{0}$};
\draw[fill] (2.5,-3.8) node[right]{$-iq_{0}$};
\draw[-][thick](3.5,-2.5) arc(0:360:1);
\draw[-<][thick](3.5,-2.5) arc(0:30:1);
\draw[-<][thick](3.5,-2.5) arc(0:150:1);
\draw[->][thick](3.5,-2.5) arc(0:210:1);
\draw[->][thick](3.5,-2.5) arc(0:330:1);
\draw[->][thick](-1.5,1.5)--(-1.5,-0.3);
\draw[->][thick](1.5,1.5)--(-1,-0.4);
\draw[->][thick](-0.8,- 3.2)--(0.8,-3.2);
\end{tikzpicture}}
\noindent { \small \textbf{Figure 1.} (Color online) The mapping relationship between two-sheeted Riemann surface, $\lambda-$plane and $z-$plane.}

We know that all the values of $k (\in S_{1}, S_{2})$ which is included in the continuous spectrum $\Sigma_{k}$ satisfy $\lambda(k)\in\mathbb{R}$. That means $\Sigma_{k}=\mathbb{R}\cup i[-q_{0},q_{0}]$. Then, via introducing the uniformization variable, the $\Sigma_{k}$ is changed into the $\Sigma_{z}=\mathbb{R}\cup C_{0}$. The subscript $z$ implies that the set is in complex $z$-plane and $C_{0}$ is a circle with $0$ as the center and $q_{0}$ as the radius. To simplify analysis without causing confusion, we omit the subscript i.e., $\Sigma_{z}\rightarrow\Sigma$.

\section{Jost functions}
In this section, we present the following main results for the Jost functions.

\noindent \textbf {Theorem 1}
\emph{The functions $\mu_{-,1}, \mu_{+,2}$ are analytic in $D_{-}$ and $\mu_{-,2}, \mu_{+,1}$ are analytic in $D_{+}$, and they can be recorded as $\mu^{-}_{-,1}, \mu^{-}_{+,2}, \mu^{+}_{-,2}, \mu^{+}_{+,1}$, respectively.
The functions $\mu_{\pm,j} (j=1,2)$ is the $j$-th column of $\mu_{\pm}$. For conveniently, $\mu_{+}$ and $\mu_{-}$ can be rewritten as}
\begin{align}
\mu_{+}=(\mu^{+}_{+,1}, \mu^{-}_{+,2}), \quad \mu_{-}=(\mu^{-}_{-,1}, \mu^{+}_{-,2}), \notag
\end{align}
\emph{where the $\mu_{\pm}$ will be defined in the following analysis.}\\

According to the above analysis, we know that the matrix $X_{\pm}$ has two eigenvalues $\pm i\lambda$.
Because of the linear relationship between $X_{\pm}$ and $T_{\pm}$, the matrix $T_{\pm}$ also has two eigenvalues. That means $X_{\pm}$ and $T_{\pm}$ can be transformed to diagonal matrices with the same characteristic  matrix i.e.,
\begin{align}\label{2.9}
\begin{split}
  X_{\pm}(x,t;z)&=Y_{\pm}(z)(i\lambda \sigma_{3})Y_{\pm}^{-1}(z),\\
  T_{\pm}(x,t;z)&=Y_{\pm}(z)[i\lambda(-8\gamma k^{3}-4\alpha(t)k^{2}+(4\gamma q_{0}^{2}+1)k+2\alpha(t)q_{0}^{2})\sigma_{3}]Y_{\pm}^{-1}(z),
  \end{split}
\end{align}
where
\begin{align}\label{2.10}
Y_{\pm}(z)=\left(
  \begin{array}{cc}
     1 & -\frac{q^{*}_{\pm}}{z} \\
     \frac{q_{\pm}}{z} & 1 \\
  \end{array}
\right)=\mathbb{I}+(i/z)\sigma_{3}Q_{\pm}.
\end{align}
Substituting   \eqref{2.10} into  \eqref{2.5}, we can derive that
\begin{align}
\psi_{\pm}(x,t;z)=Y_{\pm}(z)e^{i\theta(x,t;z)\sigma_{3}},
\end{align}
where $\theta(x,t;z)=\lambda(z)[x+(-8\gamma k^{3}-4\alpha(t)k^{2}+(4\gamma q_{0}^{2}+1)k+2\alpha(t)q_{0}^{2})t]$. Then, the Jost solutions of the Lax pair \eqref{2.4} can be defined as
\begin{align}\label{djs}
\phi_{\pm}(x,t;z)\thicksim\psi_{\pm}(x,t;z)=Y_{\pm}(z)e^{i\theta(x,t;z)\sigma_{3}},
\quad x\rightarrow \pm\infty.
\end{align}
Then, we define that
\begin{align}\label{2.11}
\mu_{\pm}(x,t;z)=\phi_{\pm}(x,t;z)e^{-i\theta(x,t;z)\sigma_{3}}  \sim  Y_{\pm}(z), \quad x\rightarrow\pm\infty.
\end{align}
Consequently, we derive that
\begin{equation} \label{2.12}
\begin{split}
&(Y_{\pm}^{-1}(z)\mu_{\pm}(z))_{x}+i\lambda [Y_{\pm}^{-1}(z)\mu_{\pm}(z),\sigma_{3}]=Y_{\pm}^{-1}(z)\Delta Q_{\pm}(z)\mu_{\pm}(z), \\
&(Y_{\pm}^{-1}(z)\mu_{\pm}(z))_{t}+i\lambda(-8\gamma k^{3}-4\alpha(t)k^{2}+(4\gamma q_{0}^{2}+1)k+2\alpha(t)q_{0}^{2})[Y_{\pm}^{-1}(z)\mu_{\pm}(z),\sigma_{3}] \\
&=Y_{\pm}^{-1}(z)\Delta T_{\pm}(z)\mu_{\pm}(z),
 \end{split}
 \end{equation}
which equivalent to the Lax pair \eqref{2.4}. The  $\Delta Q_{\pm}(z)$ and $\Delta T_{\pm}(z)$ which appear in \eqref{2.12} mean $\Delta Q_{\pm}(z)=Q-Q_{\pm}$ and $\Delta T_{\pm}(z)=T-T_{\pm}$. Furthermore, the equation \eqref{2.12} can be written in full derivative form. Based on this, two Volterra integral equations can be derived as
\begin{align}
\begin{matrix}
\mu_{-}(x,t;z)=Y_{-}+\int_{-\infty}^{x}Y_{-}e^{i\lambda(x-y)\hat{\sigma}_{3}}[Y_{-}^{-1}\Delta Q_{-}(y,t)\mu_{-}(y,t;z)]\, dy,\\
\mu_{+}(x,t;z)=Y_{+}-\int_{x}^{\infty}Y_{+}e^{i\lambda(x-y)\hat{\sigma}_{3}}[Y_{+}^{-1}\Delta Q_{+}(y,t)\mu_{+}(y,t;z)]\, dy,
\end{matrix}
\end{align}
where the $e^{\hat{\sigma}_{3}}A=e^{\sigma_{3}}Ae^{-\sigma_{3}}$. From the Volterra integral equations, one can derive that
\begin{align}
Y^{-1}_{-}\mu_{-,1}(y,t;z)=\left(                 % 左括号
  \begin{array}{c}   %该矩阵一共1列，每一列都居中放置
    1  \\  %第一行元素
    0  \\  %第二行元素
  \end{array}
\right)+\int_{-\infty}^{x}G_{0}(x-y,z)\Delta Q_{-}(y)\mu_{-,1}(y,t;z)\, dy,
\end{align}
where
\begin{align}
G_{0}(x-y,z)=\frac{1}{\tau}\left(                 %左括号
  \begin{array}{cc}   %该矩阵一共2列，每一列都居中放置
    1 & \frac{q^{*}_{-}}{z}  \\  %第一行元素
    -\frac{q_{-}}{z}e^{-2i\lambda(x-y)} & e^{-2i\lambda(x-y)} \\ %第二行元素
  \end{array}
\right)
\end{align} with
\begin{align}\label{tau}
\tau=\det(Y_{\pm})=1+\frac{q^{2}_{0}}{z}.
\end{align}
Via considering the analytical properties of $G_{0}$, the corresponding analytical properties of the first column of $\mu_{-}$ can be obtained. Similarly, the analytical properties of $\mu_{-,2}, \mu_{+,1}, \mu_{+,2}$ can be obtained. Then, the we have the \emph{Theorem 1}.

\section{Scattering matrix}
In this section, we will get the following main results for scattering matrix.

\noindent \textbf {Theorem 2}
\emph{The function $s_{11}$ is analytic in $D_{+}$ and $s_{22}$ is analytic in $D_{-}$. However, the functions $s_{12}$ and $s_{21}$ are nowhere analytic. The $s_{11}$, $s_{22}$, $s_{12}$ and $s_{21}$  will be defined in the following analysis.} \\

Necessarily, a lemma needs to be introduced.

\noindent\textbf{Lemma 1.}
%\begin{lemma}
\emph{If $A(x)$ and $Y(x)$ are $n$-order matrix matrices, satisfying $Y_{x}=AY$, then $(\det Y)_{x}=tr(A)\det Y$ and $\det Y(x)=[\det Y(x_{0})]e^{\int_{x_{0}}^{x}tr[A(y)]\,dy}$.}%\label{lemma1}
%\end{lemma}

The Lemma 1 can be proved easily and one can refer to the reference\cite{Lizq}. Now, considering the expression of $X$ and $T$ in \eqref{2.4} and applying the Lemma 1, we have $(\det \phi)_{x}=(\det \phi)_{t}=0$. According to  \eqref{djs} and \eqref{tau}, we obtain that $\det(\phi_{\pm}(x,t;z))=\det (Y_{\pm}(z))=\tau(z)$. Since $\phi_{\pm}$ are two fundamental matrix solutions of scattering problem, a linear relationship between the $\phi_{+}$ and $\phi_{-}$ can be expressed as
\begin{align}\label{2.13}
 \phi_{+}(x,t;z)=\phi_{-}(x,t;z)S(z),
\end{align}
where $S(z)=(s_{ij})_{2\times2}$ is a matrix, and is independent of the variable $x$ and $t$. The reflection coefficients are defined as
\begin{align}\label{RC}
\rho(z)=s_{21}(z)/s_{11}(z),\quad \tilde{\rho}(z)=s_{12}(z)/s_{22}(z),\quad \forall z\in\Sigma.
\end{align}
Furthermore, from \eqref{2.13}, it is easy to calculate that
\addtocounter{equation}{1}
\begin{align}
&s_{11}(z)=\frac{Wr\left(\phi_{+,1},\phi_{-,2}\right)}{\tau},\quad
s_{22}(z)=\frac{Wr\left(\phi_{-,1},\phi_{+,2}\right)}{\tau}, \tag{\theequation a} \label{2.14}\\
&s_{12}(z)=\frac{Wr\left(\phi_{+,2},\phi_{-,2}\right)}{\tau},\quad
s_{21}(z)=\frac{Wr\left(\phi_{-,1},\phi_{+,1}\right)}{\tau}, \tag{\theequation b}
\end{align}
where the subscript of $\phi_{\pm,j}$ mean the $j$-column of $\phi_{\pm}$. Now, we consider the analytical properties of $s_{ij}(i,j=1,2)$ i.e., the \emph{Theorem 2}. From the definitions \eqref{2.11} and \eqref{2.13}, we have
\begin{align}\label{2.15}
\mu_{+}=\mu_{-}e^{i\theta(z)\hat{\sigma}_{3}}S(z), \quad (\det \mu_{\pm})_{x}=(\det \phi_{\pm})_{x}=0, \notag \\
\det \mu_{\pm}=\det(\lim_{x\rightarrow\pm\infty}\mu_{\pm}=\det Y_{\pm})_{x}=\tau\neq0,
\end{align}
which implies that $\mu_{\pm}$ is reversible. Therefore, $S(z)$ can be represented by $\mu_{\pm}$. Then, according to the \emph{Theorem 1}, the analytical properties of $S(z)$ can be obtained i.e., \emph{Theorem 2.}

\section{Symmetries}
In this section, we will get the following main results for symmetries.

\noindent \textbf {Theorem 3}
\emph{The symmetries of $\mu_{\pm}$ are}
\begin{align}
\mu_{\pm}(x,t;z)&=-\sigma\mu_{\pm}^{*}(x,t;z^{*})\sigma, \label{S-1}\\
\mu_{\pm}(x,t;z)&=\frac{i}{z}\mu_{\pm}\left(x,t;-\frac{q_{0}^{2}}{z}\right)\sigma_{3}Q_{\pm},\label{S-2}
\end{align}
\emph{where $\sigma=\left(
      \begin{array}{cc}
        0 & 1  \\
        -1 & 0  \\
      \end{array}
    \right)$.}
\emph{Furthermore, by considering the \eqref{S-1} and \eqref{S-2}, the individual columns of $\mu_{\pm}$ can be written as }
\begin{align}
\mu_{\pm,1}(x,t;z)&=\sigma\mu_{\pm,2}^{*}(x,t;z^{*}),\quad \qquad
\mu_{\pm,2}(x,t;z)=-\sigma\mu_{\pm,1}^{*}(x,t;z^{*}),\label{S-3}\\
\mu_{\pm,1}(x,t;z)&=\left(\frac{q_{\pm}}{z}\right)
\mu_{\pm,2}\left(x,t;-\frac{q_{0}^{2}}{z}\right),\quad
\mu_{\pm,2}(x,t;z)=\left(-\frac{q^{*}_{\pm}}{z}\right)
\mu_{\pm,1}\left(x,t;-\frac{q_{0}^{2}}{z}\right).\label{S-4}
\end{align}

\noindent \textbf {Theorem 4}
\emph{The scattering matrix $S(z)$ possesses the symmetries which are expressed as }
\begin{align}
S(z)&=-\sigma S^{*}(z^{*})\sigma, \label{S-5}\\
S(z)&=(\sigma_{3}Q_{-})^{-1}S\left(-\frac{q_{0}^{2}}{z}\right)\sigma_{3}Q_{+}.\label{S-6}
\end{align}
\emph{Furthermore, based on the \eqref{S-5} and the \eqref{S-6}, the relationships among the elements of $S(z)$ can be derived as}
\begin{gather}
s_{22}(z)=s^{*}_{11}(z^{*}), ~ s_{12}(z)=-s^{*}_{21}(z^{*}), \label{S-7}\\
s_{11}(z)=-\frac{q_{+}}{q_{-}}s_{22}\left(-\frac{q_{0}^{2}}{z}\right),\label{S-8}\\
s_{12}(z)=\frac{q^{*}_{+}}{q_{-}}s_{21}\left(-\frac{q_{0}^{2}}{z}\right).\label{S-9}
\end{gather}\\

To obtain the symmetries of $\mu_{\pm}(x,t;z)$ and scattering matrix, we have to consider the map $k\rightarrow k^{*}$ and the sheets of the Riemann surface i.e., $(a)$ $z\rightarrow z^{*}$ ($\in$ $z$-plane) indicates $(k,\lambda)\rightarrow(k^{*},\lambda^{*})$ ($\in$ $k$-plane), $(b)$ $z\rightarrow -q_{0}^{2}/z$ ($\in$ $z$-plane) indicates $(k,\lambda)\rightarrow(k,-\lambda)$ ($\in$ $k$-plane). Through the analysis of the above two aspects i.e., $(a)$ and $(b)$, two types of symmetry will be derived. Now, we prove the \emph{Theorem 3} and \emph{Theorem 4}.

The proof of the \emph{Theorem 3} is given as follows.
\begin{proof}
Let $\omega(x,t;z)=\sigma\mu_{\pm}^{*}(x,t;z^{*})\sigma$. It is easy to calculate that
\begin{gather}
\sigma\sigma=-I, \quad \sigma\sigma_{3}\sigma=\sigma_{3}, \notag \\
\sigma Y_{\pm}^{*}(z^{*})\sigma=-Y_{\pm}^{-1}(z),\quad \sigma\Delta Q_{\pm}^{*}\sigma=-\Delta Q_{\pm}. \notag
\end{gather}
Then, it is easy to check that $\omega$ is the solution of   equation \eqref{2.12}. According to   \eqref{2.11}, $\omega$ satisfies that
\begin{align}
\omega_{\pm}(x,t;z)\sigma=-(Y_{\pm}(z)+o(1)),\quad x\rightarrow\pm\infty.
\end{align}
Meanwhile, since the solution of the scattering problem is unique under the condition of given boundary, the $-\omega_{\pm}=\mu_{\pm}$ is derived, i.e., the \eqref{S-1} is proved. The equation \eqref{S-2} can be proved in a similar way.
\end{proof}

The proof of the \emph{Theorem 4} is given as follows.
\begin{proof}
From \emph{Theorem 3} and \eqref{2.15}, it is easy to derive that
\begin{align}
-\sigma S^{*}(z^{*})\sigma&=\sigma e^{i\theta(z)\sigma_{3}}\sigma\sigma(\mu_{-}^{-1})^{*}(x,t;z^{*})\sigma\sigma
\mu^{*}_{+}(x,t;z^{*})\sigma\sigma e^{-i\theta(z)\sigma_{3}}\sigma, \notag\\
&=e^{-i\theta(z)\hat{\sigma}_{3}}\mu_{-}^{-1}(x,t;z)\mu_{+}(x,t;z)=S(z).\notag
\end{align}
Therefore, the \eqref{S-5} is proved. The equation \eqref{S-6} can be proved in a similar way.
\end{proof}
In addition, According to the relationships which have been shown in \emph{theorem 3} and \emph{theorem 4}, it is easy to derive the relationship between the reflection coefficients as
\begin{align}
\rho(z)=-\tilde{\rho}^{*}(z^{*})=-\frac{q_{-}}{q^{*}_{-}}\tilde{\rho}(-q_{0}^{2}/z).\notag
\end{align}

\section{Asymptotic behavior of $\mu_{\pm}$ and scattering matrix}
In this section, we will get the following main results for asymptotic behavior of $\mu_{\pm}$ and scattering matrix.

\noindent \textbf {Theorem 5}
\emph{The asymptotic properties of the $\mu_{\pm}$ are }
\begin{align}\label{A-1}
\mu_{\pm}(x,t;z)=
\left\{
\begin{aligned}
&I+\frac{i}{z}\sigma_{3}Q+O(z^{-2}),\quad\quad &z\rightarrow\infty,\\
&\frac{i}{z}\sigma_{3}Q_{\pm}+O(1),\quad &z\rightarrow0.
\end{aligned}
\right.
\end{align}

\noindent \textbf {Theorem 6}
\emph{The asymptotic properties of the scattering matrix are }
\begin{align}\label{A-2}
S(z)=
\left\{
\begin{aligned}
&I+O(1/z),\quad\quad\quad\quad &z\rightarrow\infty,\\
&diag(q_{+}/q_{-},q_{-}/q_{+})+O(z),\quad &z\rightarrow0.
\end{aligned}
\right.
\end{align}

Because of the significance of the asymptotic behaviors of the eigenfunction and the scattering matrix during the construction of the Riemann Hilbert problem which will be analysed in later section. Recall that when $k\in S_{1}$, $z\rightarrow\infty(k\rightarrow\infty)$, and when $k\in S_{2}$
$z\rightarrow 0(k\rightarrow\infty)$. Now, we prove the {Theorem 5} and {Theorem 6}.

The proof of the \emph{Theorem 5} is given as follows.
\begin{proof}
Firstly, we consider the case that $z\rightarrow\infty$. Let
\begin{align}\label{A-3}
Y^{-1}_{\pm}\mu_{\pm}&=\chi_{\pm}^{(0)}+\frac{\chi_{\pm}^{(1)}}{z}
+o\left(\frac{1}{z^{2}}\right)\notag \\
&=Y^{-1}_{\pm}\left(\mu_{\pm}^{(0)}+\frac{\mu_{\pm}^{(1)}}{z}+o\left(\frac{1}{z^{2}}\right)\right),\quad z\rightarrow\infty.
\end{align}
According to the \eqref{2.10}, we can derive that
\begin{align}\label{A-4}
Y^{-1}_{\pm}(z)=\frac{1}{1+q_{0}^{2}/z^{2}}\left(
  \begin{array}{cc}
     1 & \frac{q^{*}_{\pm}}{z} \\
     -\frac{q_{\pm}}{z} & 1 \\
  \end{array}
\right)=\frac{1}{1+q_{0}^{2}/z^{2}}(\mathbb{I}-(i/z)\sigma_{3}Q_{\pm}).
\end{align}
Combining \eqref{A-3}and \eqref{A-4}, we can derive the relationship between $\chi_{\pm}^{(i)}$ and $\mu_{\pm}^{(i)}$ $(i=0,1,2)$. Furthermore, substituting \eqref{A-3} into the Lax pair \eqref{2.12}, comparing the same power coefficients of $z$, and applying the \eqref{2.11} and the relationship between $\chi_{\pm}^{(i)}$ and $\mu_{\pm}^{(i)}$, we derive that
\begin{gather}
\mu_{\pm}^{(0)}=\mathbb{I}, \quad \mu_{\pm}^{(1)}=i\sigma_{3}Q.
\end{gather}
Therefore, the asymptotic behavior of $\mu_{\pm}$ is derived, i.e., $\mu_{\pm}(x,t;z)=I+\frac{i}{z}\sigma_{3}Q+O(z^{-2})$ as $z\rightarrow\infty$. The other case that $z\rightarrow0$ can be proved in a similar way.
\end{proof}

The proof of the \emph{Theorem 6} is given as follows.
\begin{proof}
According to the \eqref{2.15} and \emph{Theorem 5}, the conclusion of \emph{Theorem 6} is obvious.
\end{proof}

\section{Generalized Riemann-Hilbert problem}
In this subsection, the main results are given as follows.

\noindent \textbf {Theorem 7}
\emph{The generalized Riemann-Hilbert problem}
\begin{itemize}
  \item  $M(x,t;z)$ is meromorphic in $C\setminus\Sigma$;
  \emph{where the $M(x,t;z)$ is defined in \eqref{Matrix}.}
  \item  $M^{+}(x,t;z)=M^{-}(x,t;z)(\mathbb{I}-G(x,t;z))$,~~~$z\in\Sigma$;
  \emph{where} \begin{align*}
G(x,t;z)=e^{i\theta(z)\hat{\sigma}_{3}}\left(
\begin{array}{ccc}
  0 & -\tilde{\rho}(z) \\
  \rho(z) & \rho(z)\tilde{\rho}(z)
\end{array} \right). \notag
\end{align*}
  \item  $M(x,t;z)$ satisfies residue conditions at zero points $\{z| s_{11}(z)=s_{22}(z)=0\}$;
  \item  $M^{\pm}(x,t;z)\thicksim\mathbb{I}+O(1/z)$, ~~$z\rightarrow\infty$;
  \item  $M^{\pm}(x,t;z)\thicksim\frac{i}{z}\sigma_{3}Q_{-}+O(1)$,~~$z\rightarrow0$,
\end{itemize}
\emph{where the $M(x,t;z)$ is defined in \eqref{Matrix}.}\\

In the above section, we have analyzed the analytical properties, symmetries and asymptotic behavior of the $\mu_{\pm}$ and scattering matrix. Now, we construct a generalized Riemann-Hilbert problem (RHP). According to \eqref{2.15}, we can derive that
\begin{align}\label{G-1}
\begin{split}
\mu_{+,1}(z)&=s_{11}(z)\mu_{-,1}(z)+s_{21}(z)e^{2i\theta(z)}\mu_{-,2}(z), \\
\mu_{+,2}(z)&=s_{12}(z)e^{-2i\theta(z)}\mu_{-,1}(z)+s_{22}(z)\mu_{-,2}(z).
\end{split}
\end{align}
Considering the analytical properties of $\mu_{\pm}$ and scattering matrix, we define a sectionally  meromorphic matrix
\begin{align}\label{Matrix}
M(x,t;z)=\left\{\begin{aligned}
&M^{+}(x,t;z)=\left(\frac{\mu_{+,1}(x,t;z)}{s_{11}(z)},\mu_{-,2}(x,t;z)\right), \quad z\in D_{+},\\
&M^{-}(x,t;z)=\left(\mu_{-,1}(x,t;z),\frac{\mu_{+,2}(x,t;z)}{s_{22}(z)}\right), \quad z\in D_{-}.
\end{aligned}\right.
\end{align}
Now, we prove the \emph{Theorem 7}.
\begin{proof}
Firstly, based on the analytical properties of $\mu_{\pm}$ and scattering matrix, the analytical property can be acquired easily. Then, based on   \eqref{G-1} and the expression of reflection coefficients $\rho(z)$ and $\tilde{\rho}(z)$ in \eqref{RC}, the jump condition $G(x,t;z)$ can be derived. Finally, according to the asymptotic behavior of the $\mu_{\pm}$ and scattering matrix $S(z)$ that have been shown in \emph{Theorem 5} and \emph{Theorem 6}, respectively, it is not difficult to derive the asymptotic behavior of $M^{\pm}$ that have been shown in \emph{Theorem 7}.
\end{proof}

\section{The gvcNLS equation with NZBCs: simple poles}
\subsection{Discrete spectrum and residue condition}
In this section, we will get the following main results.

\begin{itemize}
  \item \emph{The set of the discrete spectrum are that}
  \begin{align}
\mathbb{Z}=\left\{z_{n}, -\frac{q_{0}^{2}}{z_{n}^{*}},
  z_{n}^{*}, -\frac{q_{0}^{2}}{z_{n}}\right\},\quad s_{11}(z_{n})=0, \quad n=1,2,\ldots,N. \notag
\end{align}
  \item \emph{The residue conditions are that}
  \addtocounter{equation}{1}
  \begin{align}
  \mathop{Res}_{z=z_{n}}\left[\frac{\mu_{+,1}(z)}{s_{11}(z)}\right]&=C_{n}[z_{n}]
e^{-2i\theta(z_{n})}\mu_{-,2}(z_{n}), \tag{\theequation a} \label{D-1}\\
  \mathop{Res}_{z=z_{n}^{*}}\left[\frac{\mu_{+,2}(z)}{s_{22}(z)}\right]&=\tilde{C}_{n}[z_{n}]
e^{2i\theta(z_{n}^{*})}\mu_{-,1}(z_{n}^{*}), \tag{\theequation b} \label{D-2}\\
  \mathop{Res}_{z=-\frac{q_{0}^{2}}{z^{*}_{n}}}\left[\frac{\mu_{+,1}(z)}{s_{11}(z)}\right]&=
C_{N+n}e^{-2i\theta(-\frac{q_{0}^{2}}{z^{*}_{n}})}\mu_{-,2}\left(-\frac{q_{0}^{2}}{z^{*}_{n}}\right), \tag{\theequation c} \label{D-3}\\
  \mathop{Res}_{z=-\frac{q_{0}^{2}}{z_{n}}}\left[\frac{\mu_{+,2}(z)}{s_{22}(z)}\right]&=
\tilde{C}_{N+n}e^{2i\theta(-\frac{q_{0}^{2}}{z_{n}})}\mu_{-,1}\left(-\frac{q_{0}^{2}}{z_{n}}\right) \tag{\theequation d}, \label{D-4}
  \end{align}
\emph{where the $C_{n}[z_{n}]$, $\tilde{C}_{n}[z_{n}]$, $C_{N+n}$ and $\tilde{C}_{N+n}$ are defined in \eqref{D-8} and \eqref{D-9} , respectively.}
\end{itemize}

We know that the discrete spectrum of the scattering problem is a set that is composed of the values $z\in\mathbb{C}\setminus\Sigma$ such that the eigenfunctions exist in $L^{2}(\mathbb{R})$. These values satisfy that $s_{11}(z)=0$ $(z\in D_{+})$ and $s_{22}(z)=0$ $(z\in D_{-})$. We suppose that   $z_{n}(\in D_{+}\cap\{z\in\mathbb{C}: Imz>0\}, n=1,2,\ldots,N)$ are the simple poles of $s_{11}(z)$ i.e., $s_{11}(z_{n})=0$ but $s'_{11}(z_{n})\neq0$, $n=1, 2,\ldots, N$. According to the \emph{Theorem 4}, we can obtain that
\begin{align}
s_{11}(z_{n})=s_{22}(z_{n}^{*})=s_{22}(-q_{0}^{2}/z_{n})=s_{11}(-q_{0}^{2}/z_{n}^{*})=0. \notag
\end{align}
Therefore, the first result is obvious in this section.

Next, we consider the residue condition which will play an important role in later analysis. Because of $s_{11}(z_{n})=s_{22}(z_{n}^{*})=0$ and considering the expression of $s_{11}$ and $s_{22}$ in \eqref{2.14}, we have
\begin{align}\label{D-5}
\begin{split}
\phi_{+,1}(z_{n})=b_{n}(z_{n})\phi_{-,2}(z_{n}), \\
\phi_{+,2}(z_{n}^{*})=d_{n}(z_{n}^{*})\phi_{-,1}(z_{n}^{*}),
\end{split}
\end{align}
where $b_{n}$ and $d_{n}$ are constants. Then, applying the equation \eqref{2.11}, one can obtain \begin{align}\label{D-6}
\begin{split}
\mu_{+,1}(z_{n})=b_{n}(z_{n})e^{-2i\theta(z_{n})}\mu_{-,2}(z_{n}),\\
\mu_{+,2}(z_{n}^{*})=d_{n}(z_{n}^{*})e^{2i\theta(z_{n}^{*})}\mu_{-,1}(z_{n}^{*}).
\end{split}
\end{align}
Therefore, we get that
\begin{align}\label{D-7}
\begin{split}
\mathop{Res}_{z=z_{n}}\left[\frac{\mu_{+,1}(z)}{s_{11}(z)}\right]=
\frac{\mu_{+,1}(z_{n})}{s'_{11}(z_{n})}=\frac{b_{n}(z_{n})}{s'_{11}(z_{n})}
e^{-2i\theta(z_{n})}\mu_{-,2}(z_{n}),\\
\mathop{Res}_{z=z_{n}^{*}}\left[\frac{\mu_{+,2}(z)}{s_{22}(z)}\right]=
\frac{\mu_{+,2}(z_{n}^{*})}{s'_{22}(z_{n}^{*})}=
\frac{d_{n}(z_{n}^{*})}{s'_{22}(z_{n}^{*})}
e^{2i\theta(z_{n}^{*})}\mu_{-,1}(z_{n}^{*}).
\end{split}
\end{align}
For convenient, with a transformation
\begin{align}\label{D-8}
 C_{n}[z_{n}]=\frac{b_{n}(z_{n})}{s'_{11}(z_{n})},\quad
\tilde{C}_{n}[z_{n}]=\frac{d_{n}(z_{n}^{*})}{s'_{22}(z_{n}^{*})}.
\end{align}
The first two in the second result i.e., equations \eqref{D-1} and \eqref{D-2}, are obtained in this section. Following the similar way, we can get equations \eqref{D-3} and \eqref{D-4}, where
\begin{align}\label{D-9}
C_{N+n}=-\frac{q_{-}d_{n}(z_{n}^{*})}{q^{*}_{+}s'_{11}(-q_{0}^{2}/z^{*}_{n})},
\tilde{C}_{N+n}=-\frac{q^{*}_{-}b_{n}(z_{n})}{q_{+}s'_{22}(-q_{0}^{2}/z_{n})}.
\end{align}
It is worth noting that there is a fixed relationship between $b_{n}$ and $d_{n}$. Applying \emph{Theorem 3} to one of  equations in \eqref{D-6} and comparing it with another one, it is not difficult to get that
\begin{align}\label{D-10}
b_{n}(z_{n})=-d^{*}_{n}(z_{n}^{*}).
\end{align}
Consequently, $C_{n}[z_{n}]$ and $\tilde{C}_{n}[z_{n}]$ have the relationship
\begin{align}\label{D-11}
-C^{*}_{n}[z_{n}]=\tilde{C}_{n}[z_{n}^{*}].
\end{align}

\subsection{Reconstruct the formula for potential}
In this subsection, we will get the following main results.
\begin{itemize}
\item  \emph{The reconstruction formula for the potential are that}
\begin{align}\label{R-1}
 q(x,t)=&q_{-}+\sum_{n=1}^{2N}C_{n}[\xi_{n}]e^{-2i\theta(x,t;\xi_{n})}\mu_{-,22}(x,t;\xi_{n})
\notag\\&-\frac{1}{2\pi i}\int_{\Sigma}(M^{+}(x,t;s)G(x,t;s))_{21}\,ds.
\end{align}
\end{itemize}

To make the following analysis more convenient, we introduce a transformation
\begin{align}\label{R-2}
\xi_{n}=z_{n}, \quad \xi_{N+n}=-\frac{q_{0}^{2}}{z^{*}_{n}}, \quad \xi^{*}_{n}=z^{*}_{n},\quad \xi^{*}_{N+n}=-\frac{q_{0}^{2}}{z_{n}}.
\end{align}
That means $s_{11}(\xi_{n})=0, (n=1,2,\cdots, 2N)$ and $s_{22}(\xi^{*}_{n})=0, (n=1,2,\cdots, 2N)$. Therefore, according to the residue conditions which have been shown in \eqref{D-1}, \eqref{D-2}, \eqref{D-3} and \eqref{D-4}, and the expression of $M(x,t;z)$, we can derive that
\begin{align}\label{R-3}
\mathop{Res}_{z=\xi_{n}}M^{+}=(C_{n}[\xi_{n}]e^{-2i\theta(\xi_{n})}\mu_{-,2}(\xi_{n}),0),\quad
n=1,2,\cdots,2N,\notag \\
\mathop{Res}_{z=\xi^{*}_{n}}M^{-}=(0,\tilde{C}_{n}[\xi^{*}_{n}]e^{2i\theta(\xi^{*}_{n})}\mu_{-,1}(\xi^{*}_{n})), \quad
n=1,2,\cdots,2N.
\end{align}
To get a regular RHP, we subtract out the asymptotic behavior and the pole contributions and have
\begin{align}\label{R-4}
\begin{split}
&M^{-}(x,t;z)-\mathbb{I}-\frac{i}{z}\sigma_{3}Q_{-}-\sum_{n=1}^{2N}\frac
{\mathop{Res}_{z=\xi^{*}_{n}}M^{-}(z)}{z-\xi^{*}_{n}}-\sum_{n=1}^{2N}\frac
{\mathop{Res}_{z=\xi_{n}}M^{+}(z)}{z-\xi_{n}}\\=&
M^{+}(x,t;z)-\mathbb{I}-\frac{i}{z}\sigma_{3}Q_{-}-\sum_{n=1}^{2N}\frac
{\mathop{Res}_{z=\xi^{*}_{n}}M^{-}(z)}{z-\xi^{*}_{n}}-\sum_{n=1}^{2N}\frac
{\mathop{Res}_{z=\xi_{n}}M^{+}(z)}{z-\xi_{n}}-M^{+}(z)G(z).
\end{split}
\end{align}
It is obvious that the left side of \eqref{R-4} is analytic in $D_{-}$ and the right side of \eqref{R-4}, except the item $M^{+}(z)G(z)$, is analytic in $D_{-}$. At the same time, combining \emph{Theorem 6} and \emph{Theorem 7}, it is apparent that the asymptotic behavior of both sides of the equation \eqref{R-4} are $O(1/z)(z\rightarrow\infty)$ and $O(1)(z\rightarrow0)$. Now, we introduce the projection operators $P_{\pm}$ over $\Sigma$ by
\begin{align}\label{Cauchy}
P_{\pm}[f](z)=\frac{1}{2\pi i}\int_{\Sigma}\frac{f(\zeta)}{\zeta-(z\pm i0)}\,d\zeta,
\end{align}
where the symbol $\int_{\Sigma}$ implies the integral along the oriented contour shown in Fig. 1 and the $z\pm i0$ mean the limit which is taken from the left/right of $z(z\in\Sigma)$, respectively. Via applying the projection operators, the solution of the RHP can be obtained as
\begin{align}\label{R-5}
\begin{split}
M(x,t;z)=&\mathbb{I}+\frac{i}{z}\sigma_{3}Q_{-}+\sum_{n=1}^{2N}\frac
{\mathop{Res}_{z=\xi^{*}_{n}}M^{-}(z)}{z-\xi^{*}_{n}}+\sum_{n=1}^{2N}\frac
{\mathop{Res}_{z=\xi_{n}}M^{+}(z)}{z-\xi_{n}}\\
&+\frac{1}{2i\pi}\int_{\Sigma}\frac{M(x,t;s)^{+}G(x,t;s)}{s-z}\,ds,\quad
z\in\mathbb{C}\setminus\Sigma.
\end{split}
\end{align}
To get a closed linear algebraic integral system for the solution of the RHP, we evaluate the second column of the \eqref{R-5} at $z=\xi_{n}$ in $D_{+}$ and the first column of the \eqref{R-5} at $z=\xi^{*}_{n}$ in $D_{-}$ through the concrete expression of $\mathop{Res}_{z=\xi_{n}}M^{+}$ and $\mathop{Res}_{z=\xi^{*}_{n}}M^{-}$ i.e., \eqref{R-3}. Then, for $n=1,2,\cdots,2N$, we can obtain that
\begin{align}\label{R-6}
\begin{split}
\mu_{-,2}(\xi_{n})=\left(\begin{array}{cc}
                       -\frac{q^{*}_{-}}{\xi_{n}} \\
                        1
                     \end{array}\right)+\sum_{k=1}^{2N}
\frac{\tilde{C}_{k}[\xi^{*}_{k}]e^{2i\theta(\xi^{*}_{k})}}{\xi_{n}-\xi^{*}_{k}}
\mu_{-,1}(\xi^{*}_{k})
+\frac{1}{2\pi i}\int_{\Sigma}\frac{(M^{+}G)_{2}(\xi)}{s-\xi_{n}}\,ds, \\
\mu_{-,1}(\xi^{*}_{n})=\left(\begin{array}{cc}
                            1 \\
                        \frac{q_{-}}{\xi^{*}_{n}}
                     \end{array}\right)+\sum_{j=1}^{2N}
\frac{C_{j}[\xi_{j}]e^{-2i\theta(\xi_{j})}}{\xi^{*}_{n}-\xi_{j}}
\mu_{-,2}(\xi_{j})
+\frac{1}{2\pi i}\int_{\Sigma}\frac{(M^{+}G)_{1}(\xi)}{s-\xi^{*}_{n}}\,ds,
\end{split}
\end{align}
where $(M^{+}G)_{j}$ denotes the $j-th$ column of $(M^{+}G)$.
Finally, we reconstruct the potential from the solution of RHP. Through considering the asymptotic behavior of \eqref{R-5}, we can obtain that
\begin{align}\label{R-7}
\begin{split}
M(x,t;z)=&\mathbb{I}+\frac{1}{z}\left\{i\sigma_{3}Q_{-}+\sum_{n=1}^{2N}
\mathop{Res}_{z=\xi^{*}_{n}}M^{-}(z)
+\sum_{n=1}^{2N}\mathop{Res}_{z=\xi_{n}}M^{+}(z)\right. \\
&\left.-\frac{1}{2\pi i}\int_{\Sigma}M^{+}(x,t;s)G(x,t;s)\,ds\right\}+O(z^{-2}),\quad
z\rightarrow\infty.
\end{split}
\end{align}
%\begin{align}\label{R-7}
%M(x,t;z)=&\mathbb{I}+\frac{1}{z}\left\{i\sigma_{3}Q_{-}+\sum_{n=1}^{2N}
%\mathop{Res}_{z=\xi^{*}_{n}}M^{-}(z)
%+\sum_{n=1}^{2N}\mathop{Res}_{z=\xi_{n}}M^{+}(z) \notag \\
%&\left.-\frac{1}{2\pi i}\int_{\Sigma}M^{+}(x,t;s)G(x,t;s)\,ds\right\}+O(z^{-2}),\quad
%z\rightarrow\infty.
%\end{align}
Then, by taking $M=M^{-}$ and combining the $(2,1)$-element of \eqref{R-7} and the \emph{Theorem 5}, the reconstruction formula for the potential can be acquired, i.e., the result shown at the beginning of this section.

\subsection{Trace formulate and theta condition}
In this subsection, we will get the following main results.
\begin{itemize}
  \item \emph{The trace formulate are that}
  \begin{align}
s_{11}(z)&=exp\left(\frac{1}{2\pi i}\int_{\Sigma}\frac{\log[1-\rho(s)\tilde{\rho}(s)]}{s-z}
\,ds\right)\prod_{n=1}^{N}\frac{(z-z_{n})(z+q_{0}^{2}/z_{n}^{*})}
{(z-z_{n}^{*})(z+q_{0}^{2}/z_{n})},\label{TT-4}\\
s_{22}(z)&=exp\left(-\frac{1}{2\pi i}\int_{\Sigma}\frac{\log[1-\rho(s)\tilde{\rho}(s)]}{s-z}
\,ds\right)\prod_{n=1}^{N}\frac{(z-z_{n}^{*})(z+q_{0}^{2}/z_{n})}
{(z-z_{n})(z+q_{0}^{2}/z_{n}^{*})}.\label{TT-5}
\end{align}
  \item \emph{The theta condition are that}
  \begin{align}
\arg\frac{q_{+}}{q_{-}}=\frac{1}{2\pi}\int_{\Sigma}
\frac{\log[1-\rho(s)\tilde{\rho}(s)]}{s}\,ds+4\sum_{n=1}^{N}\arg z_{n}.
\end{align}
\end{itemize}

According to the analytic properties of $s_{11}$ and $s_{22}$ that have been shown in \emph{Theorem 2}, and the discrete spectrum, i.e., \begin{align}
\mathbb{Z}=\left\{z_{n}, -\frac{q_{0}^{2}}{z_{n}^{*}},
  z_{n}^{*}, -\frac{q_{0}^{2}}{z_{n}}\right\}, \quad n=1,2,...,N, \notag
\end{align}
which have been analysed in the \emph{subsection 8.1}, we can construct the following function
\begin{align}\label{TT-1}
\zeta^{+}_{1}(z)=s_{11}(z)\prod_{n=1}^{N}\frac{(z-z_{n}^{*})(z+q_{0}^{2}/z_{n})}
{(z-z_{n})(z+q_{0}^{2}/z_{n}^{*})},\notag\\
\zeta^{-}_{1}(z)=s_{22}(z)\prod_{n=1}^{N}\frac{(z-z_{n})(z+q_{0}^{2}/z_{n}^{*})}
{(z-z_{n}^{*})(z+q_{0}^{2}/z_{n})}.
\end{align}
Therefore, the $\zeta^{+}_{1}$ and $\zeta^{-}_{1}$ are analytic in $D_{\pm}$, respectively, and have no zeros. Meanwhile, based on the asymptotic behavior of $S(z)$ in \emph{Theorem 6}, it is obvious that $\zeta^{\pm}_{1}(z)\rightarrow1$ as $z\rightarrow\infty$. Then, considering that $\det S(z)=1$ and the expression of the reflection coefficients, we obtain that
\begin{align}\label{TT-2}
\zeta^{+}_{1}(z)\zeta_{1}^{-}(z)=\frac{1}{1-\rho(z)\tilde{\rho}(z)},\quad z\in\Sigma.
\end{align}
Furthermore, by taking the logarithm of   \eqref{TT-2}, and applying the Plemelj's formulae and projection operators, we obtain that
\begin{align}\label{TT-3}
\log\zeta^{\pm}_{1}(z)=\pm\frac{1}{2\pi i}\int_{\Sigma}
\frac{\log[1-\rho(s)\tilde{\rho}(s)]}{s-z}\,ds,\quad z\in D_{\pm}.
\end{align}
Then, through substituting   \eqref{TT-3} into \eqref{TT-1}, we can obtain the trace formula, i.e., the first result in this section.

Then we consider the theta condition. According to the \emph{Theorem 6}, we know that $s_{11}(z)\rightarrow q_{+}/q_{-}$ as $z\rightarrow0$. Meanwhile, since
\begin{align}
\prod_{n=1}^{N}\frac{(z-z_{n}^{*})(z+q_{0}^{2}/z_{n})}
{(z-z_{n})(z+q_{0}^{2}/z_{n}^{*})}\rightarrow 1, \quad as \quad z\rightarrow0, \notag
\end{align}
we can obtain the theta condition, i.e., the second result that has been shown at the beginning of this section.

\subsection{Solving the Generalized Riemann-Hilbert problem}
\subsubsection{Soliton solutions}
In this subsection, we will get a type of $N$-soliton solutions.
\begin{itemize}
  \item \emph{The specific formula of the $N$-soliton solution is}
  \begin{align}\label{SS-1}
q(x,t)=q_{-}-\frac{\det M^{\sharp}}{\det M},
\end{align}
\emph{where the $M$ and $M^{\sharp}$ are defined in \eqref{SS-5} and \eqref{SS-6}, respectively.}
\end{itemize}

It is interesting to study the case that the reflection coefficients $\rho(z)$ and $\tilde{\rho}(z)$ disappear. Consequently, the jump from $M^{+}$ to $M^{-}$  vanishes, i.e., $G(x,t;z)=0$. In this case, through   \eqref{R-6}, it is easy to acquire that
\begin{align}\label{SS-2}
u_{-,21}(x,t;\xi^{*}_{j})
=\frac{q_{-}}{\xi^{*}_{j}}+\sum_{k=1}^{2N}
\frac{C_{k}[\xi_{k}]e^{-2i\theta(\xi_{k})}}{\xi^{*}_{j}-\xi_{k}}
u_{-,22}(x,t;\xi_{k}), \notag\\
u_{-,22}(x,t;\xi_{n})=1+\sum_{j=1}^{2N}
\frac{\tilde{C}_{j}[\xi^{*}_{j}]e^{2i\theta(\xi^{*}_{j})}}{\xi_{n}-\xi^{*}_{j}}
u_{-,21}(x,t;\xi^{*}_{j}).
\end{align}
For convenience, we introduce a transformation
\begin{align}\label{SS-3}
c_{j}(x,t;z)=\frac{C_{j}[\xi_{j}]e^{-2i\theta(\xi_{j})}}{z-\xi_{j}},\quad j=1,2,\cdots,2N.
\end{align}
Substituting the first formula in \eqref{SS-2} into the second formula and applying  \eqref{SS-3}, we can obtain the formula as
\begin{align}\label{SS-4}
u_{-,22}(x,t;\xi_{n})=1-q_{-}\sum_{j=1}^{2N}\frac{c^{*}_{j}(\xi^{*}_{n})}{\xi^{*}_{j}}-
\sum_{j=1}^{2N}\sum_{k=1}^{2N}c^{*}_{j}(\xi^{*}_{n})c_{k}^{*}(\xi^{*}_{j})\mu_{-,22}(x,t;\xi_{k}).
\end{align}
To make the form of the $N$-soliton solutions more elegant, we introduce that
\begin{gather}
X_{n}=\mu_{-,22}(x,t;\xi_{n}),\quad X=(X_{1},\cdots,X_{2N})^{T}, \notag \\
U_{n}=1-q_{-}\sum_{j=1}^{2N}\frac{c^{*}_{j}(\xi^{*}_{n})}{\xi^{*}_{j}}, \quad U=(U_{1},\cdots,U_{2N})^{T},\notag\\
P=(P_{n,k})_{2N\times2N},\quad P_{n,k}=\sum_{j=1}^{2N}c^{*}_{j}(\xi_{n}^{*})c_{k}(\xi^{*}_{j}),\notag\\
n,k=1,2,\cdots,2N, \notag
\end{gather}
where the superscript $T$ implies transposition.
Then, through defining
\begin{gather}\label{SS-5}
M=\mathbb{I}+P,
\end{gather}
we can derive that $MX=U$. Furthermore, we define that
\begin{gather}\label{SS-6}
\begin{split}
M^{\sharp}=\left(\begin{array}{cc}
                      0 & \Upsilon \\
                      U & M
                    \end{array}\right), \Upsilon=(\Upsilon_{1},\cdots,\Upsilon_{2N}),\\
 \Upsilon_{n}=C_{n}[\xi_{n}]e^{-2i\theta(x,t;\xi_{n})} (n=1,2,\cdots,2N).
\end{split}
\end{gather}
Then, it is not difficult to derive the expression of the $N$-soliton solutions which has been shown at the beginning of this subsection.

\subsubsection{The phenomenon of the soliton solutions}
In this subsection, we are going to study the dynamic behavior of the soliton solutions. In the above section, we have obtained the specific expression of the $N$-soliton solutions, i.e., the equation \eqref{SS-1}. Then, by selecting appropriate parameters, some figures are illustrated. Firstly, we consider the case that the coefficient $\alpha(t)$ and $\gamma$ are zeros. Consequently, the gvcNLS equation is reduced to the classical nonlinear Schr\"{o}dinger equation. Then, when $N=1$, via using the appropriate parameters, we get the following images.\\

{\rotatebox{0}{\includegraphics[width=3.6cm,height=3.0cm,angle=0]{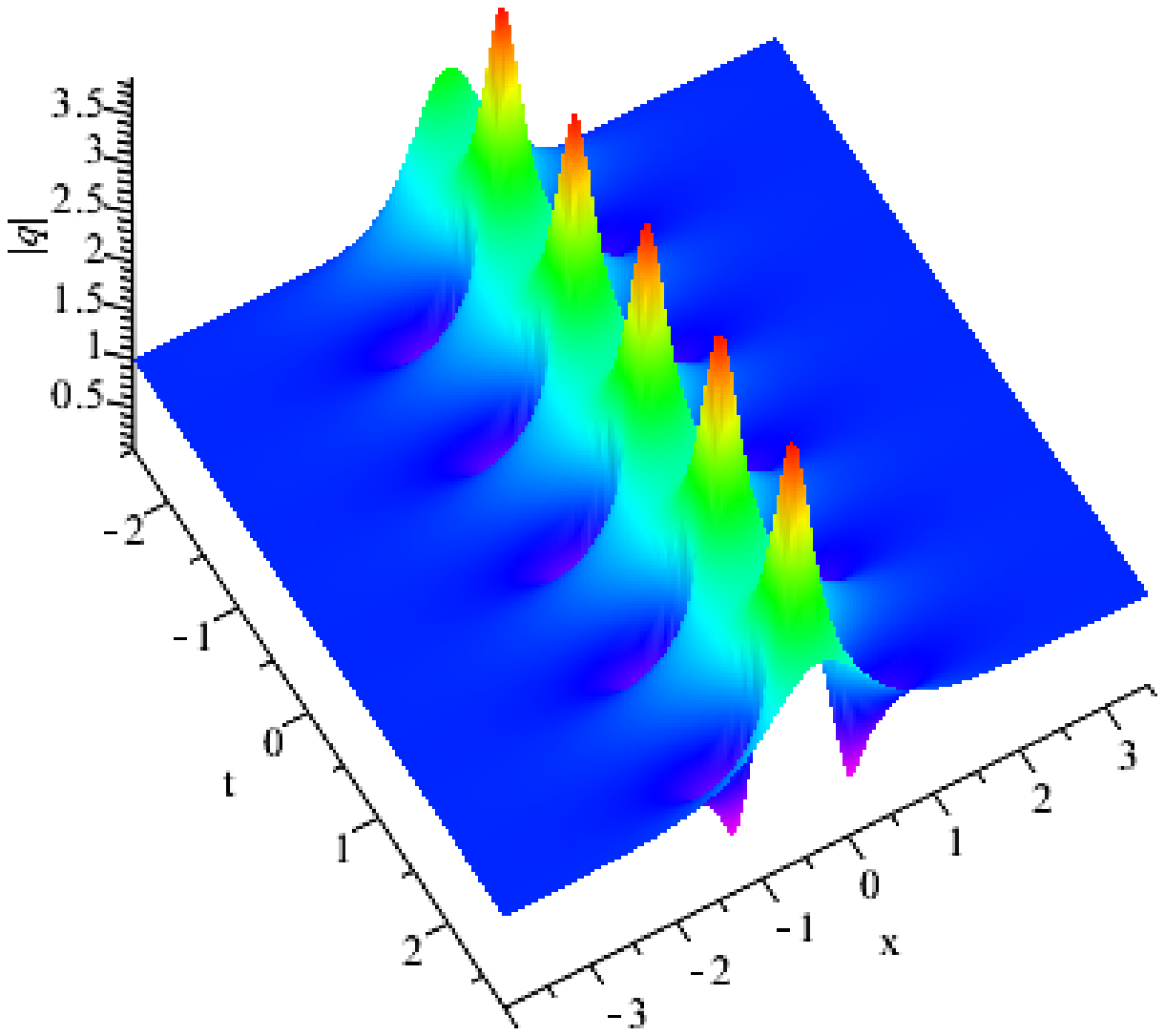}}}
~~~~
{\rotatebox{0}{\includegraphics[width=3.6cm,height=3.0cm,angle=0]{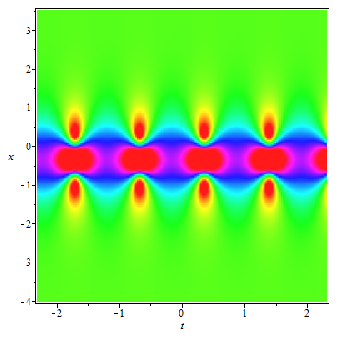}}}
~~~~
{\rotatebox{0}{\includegraphics[width=3.6cm,height=3.0cm,angle=0]{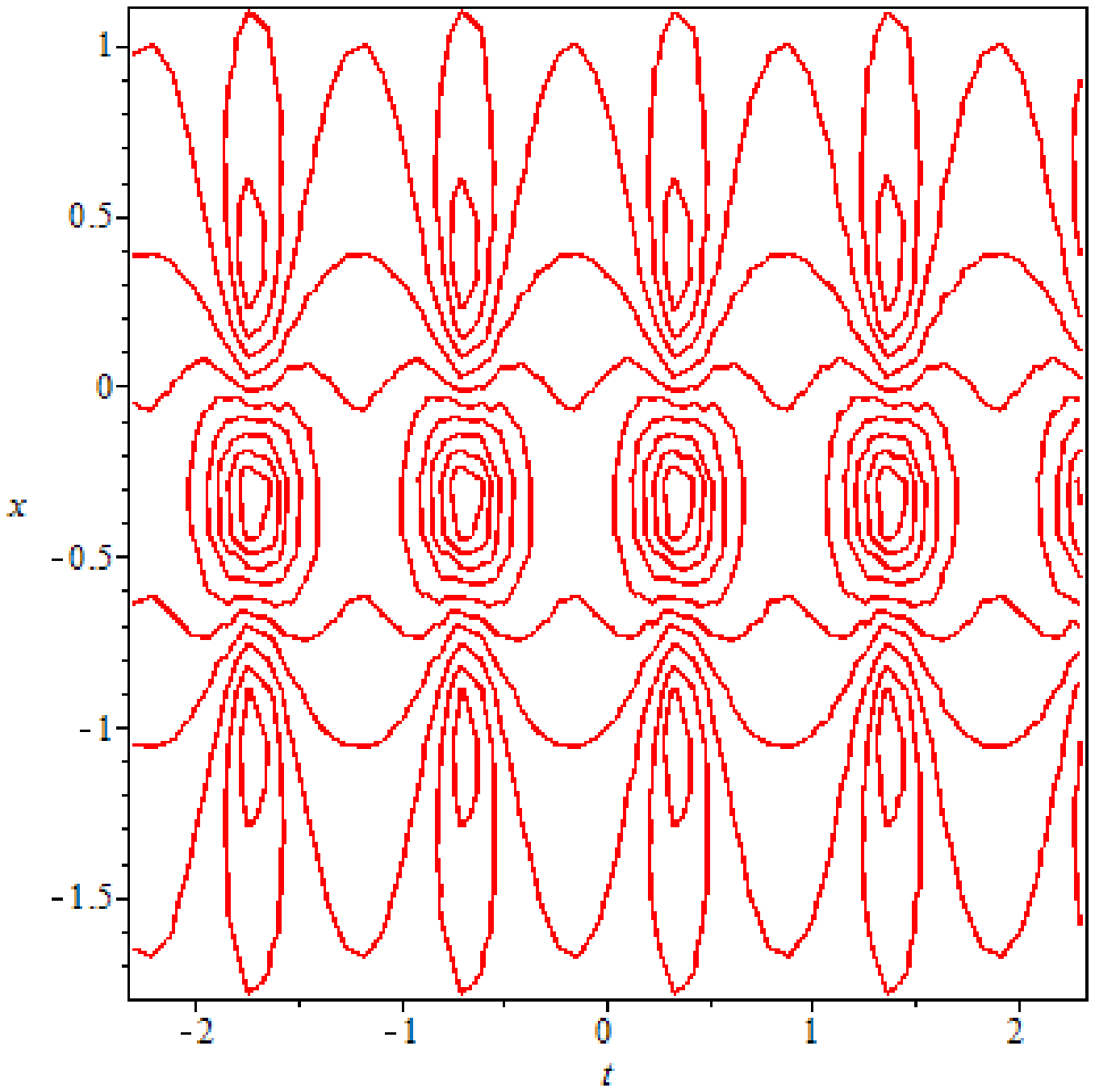}}}

$\qquad~~~~~~~~~(\textbf{a})\qquad \ \qquad\qquad\qquad\qquad~~~(\textbf{b})
\qquad\qquad\qquad\qquad\qquad~(\textbf{c})$\\

{\rotatebox{0}{\includegraphics[width=3.6cm,height=3.0cm,angle=0]{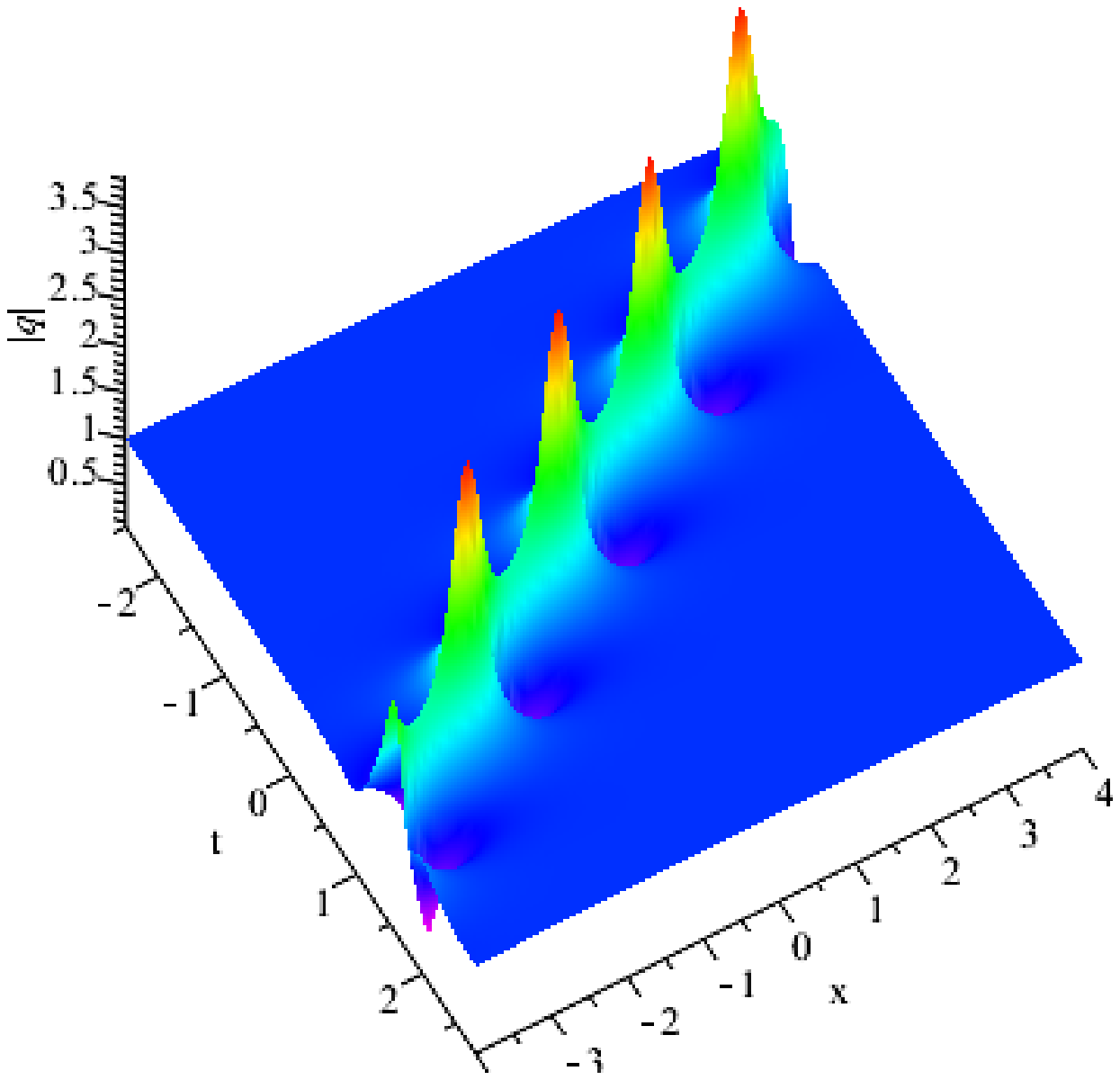}}}
~~~~
{\rotatebox{0}{\includegraphics[width=3.6cm,height=3.0cm,angle=0]{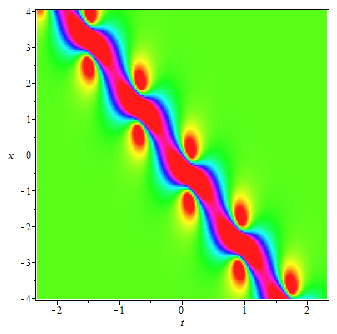}}}
~~~~
{\rotatebox{0}{\includegraphics[width=3.6cm,height=3.0cm,angle=0]{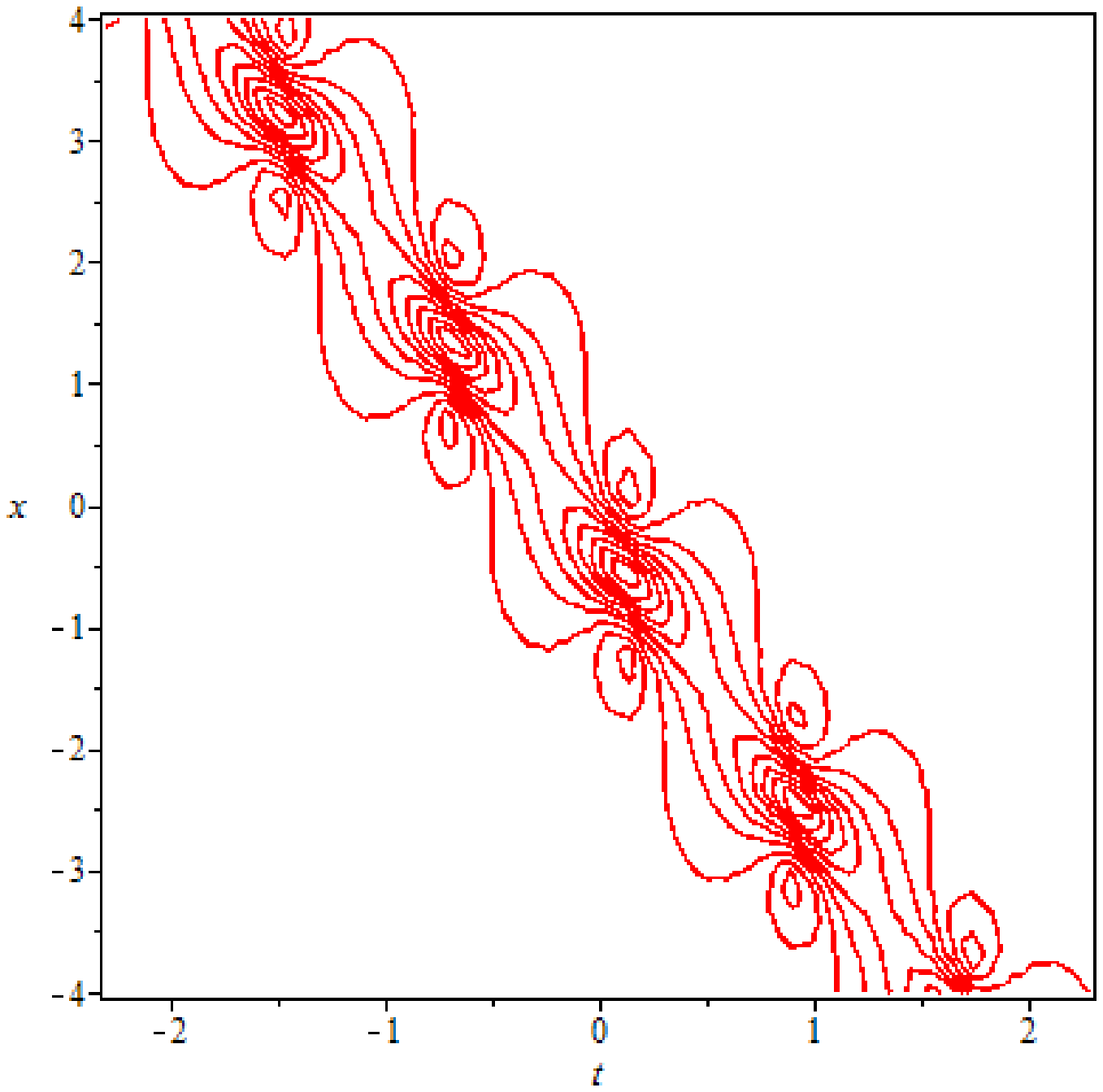}}}

$\qquad~~~~~~~~~(\textbf{d})\qquad \ \qquad\qquad\qquad\qquad~~~(\textbf{e})
\qquad\qquad\qquad\qquad\qquad~(\textbf{f})$\\
\noindent { \small \textbf{Figure 2.} (Color online) Plots of the breather solution of the equation  with the parameters $\alpha(t)=\gamma=0$, $q_{-}=1$ and $b_{1}=e^{2+i}$.
$\textbf{(a)}$: the breather solution with $\xi_{1}=2.5i$,
$\textbf{(b)}$: the density plot corresponding to $(a)$,
$\textbf{(c)}$: the contour line of the breather solution corresponding to $(a)$,
$\textbf{(d)}$: the soliton solution with $\xi_{1}=1+2.5i$,
$\textbf{(e)}$: the density plot corresponding to $(d)$,
$\textbf{(f)}$: the contour line of the breather solution corresponding to $(d)$.} \\

The above illustrations in Fig. 2 show two cases. The first case is that the propagation of the solution is parallel to the time axis when the discrete spectrum $\xi$ is pure imaginary, i.e., Fig. 2$(a)$, and the solution of this case is called the stationary breather solution. The other case is that the propagation of the solution is parallel to neither the $x$-axis nor the time axis when the discrete spectrum $\xi$ is a complex number where both the real and imaginary parts are not zero, i.e., Fig. 2$(d)$, and the solution of this case is called the non-stationary breather solution. Therefore, it is obvious that the value of  the discrete spectrum $\xi$ determines whether the breather solution is stationary or non-stationary.  Then, we change the boundary value $q_{-}$ and obtain the following graphs.\\

{\rotatebox{0}{\includegraphics[width=3.6cm,height=3.0cm,angle=0]{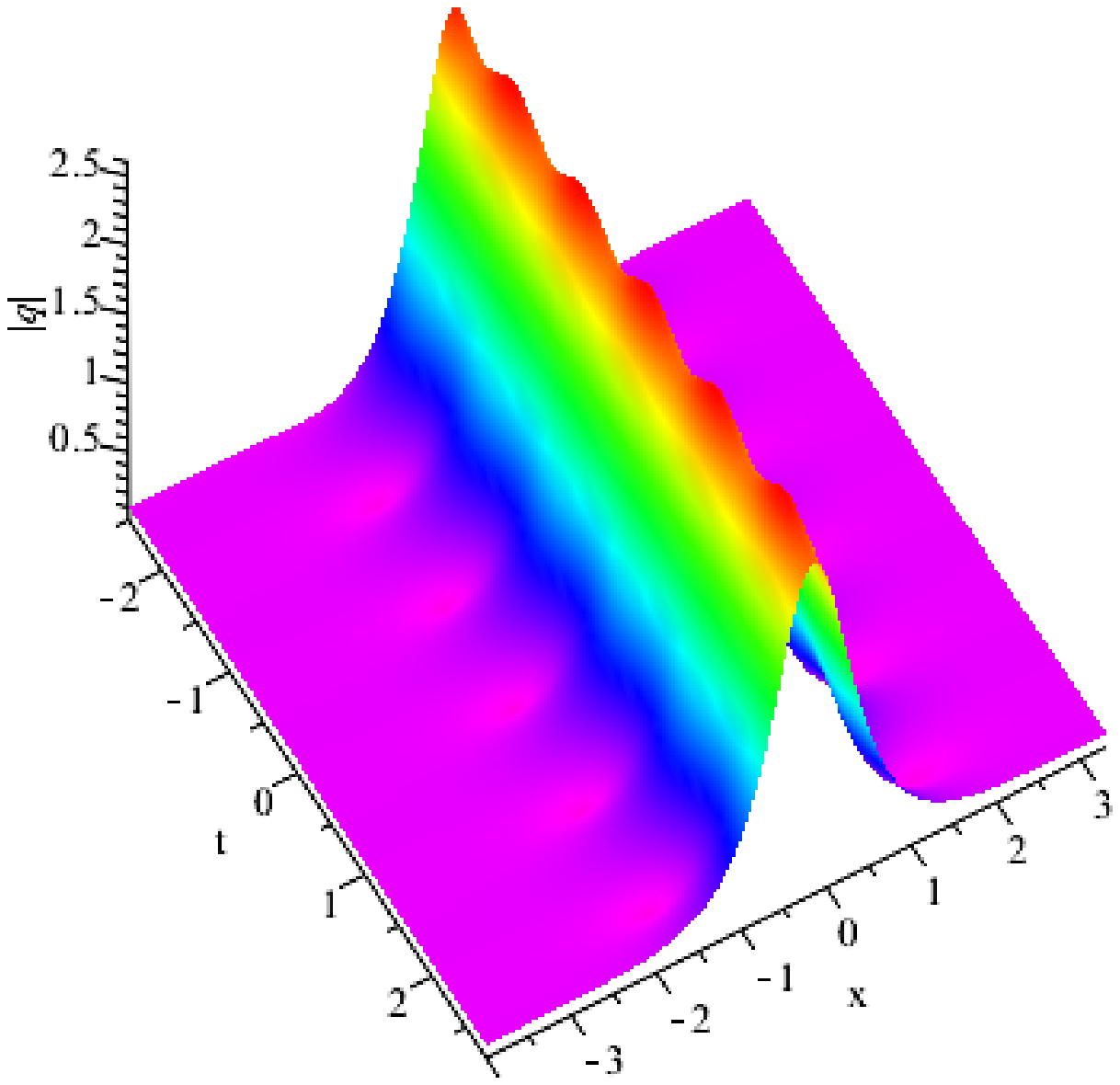}}}
~~~~
{\rotatebox{0}{\includegraphics[width=3.6cm,height=3.0cm,angle=0]{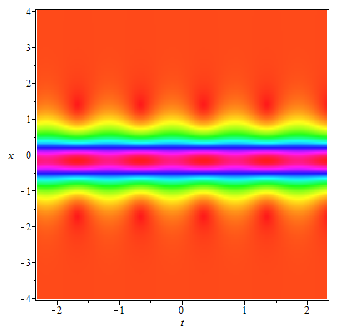}}}
~~~~
{\rotatebox{0}{\includegraphics[width=3.6cm,height=3.0cm,angle=0]{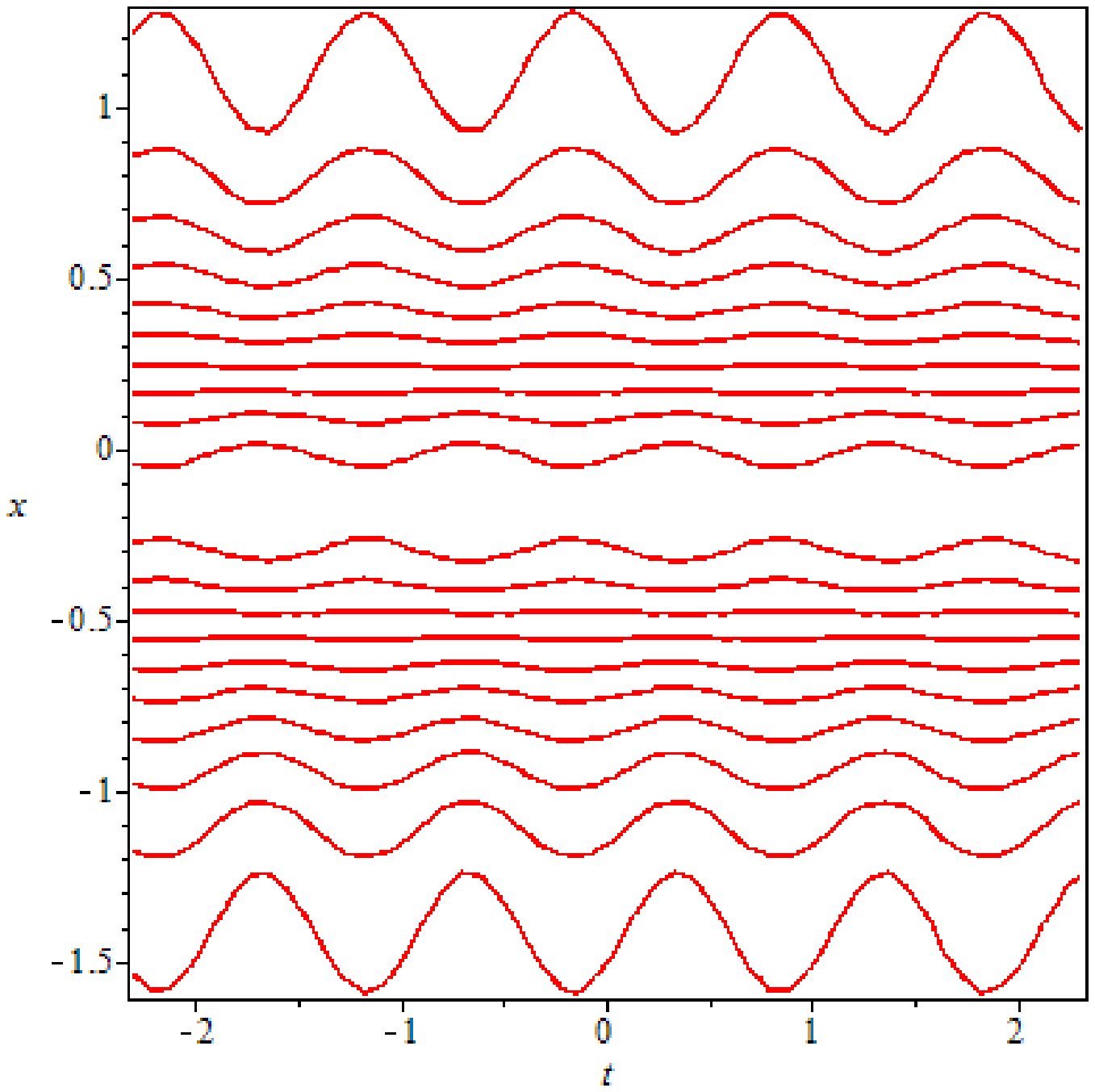}}}

$\qquad~~~~~~~~~(\textbf{a})\qquad \ \qquad\qquad\qquad\qquad~~~(\textbf{b})
\qquad\qquad\qquad\qquad\qquad~(\textbf{c})$\\

{\rotatebox{0}{\includegraphics[width=3.6cm,height=3.0cm,angle=0]{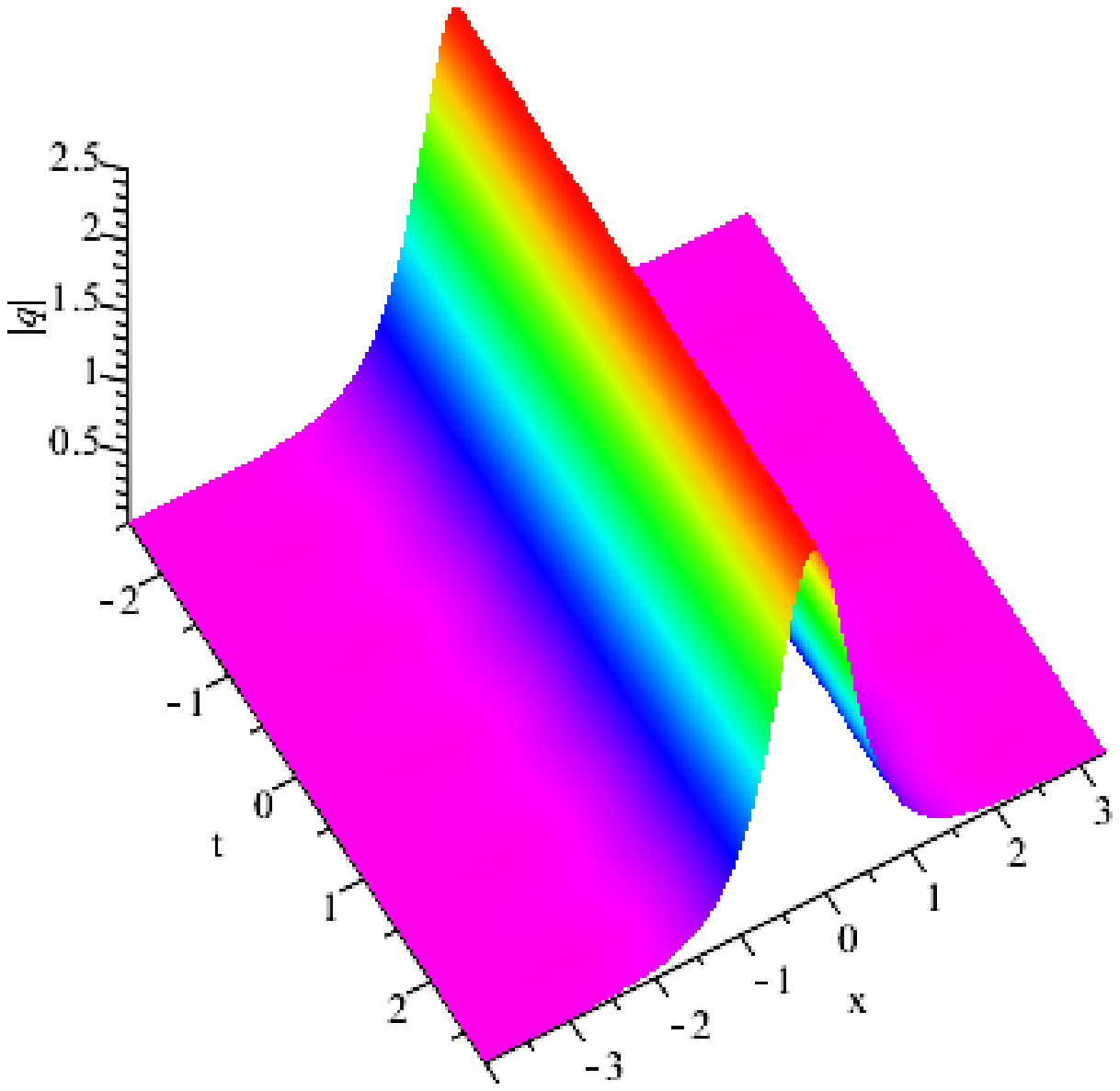}}}
~~~~
{\rotatebox{0}{\includegraphics[width=3.6cm,height=3.0cm,angle=0]{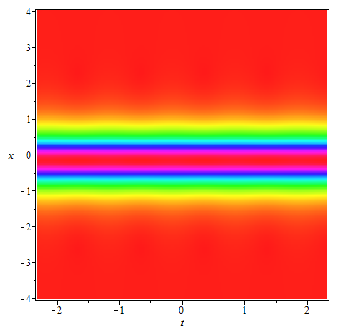}}}
~~~~
{\rotatebox{0}{\includegraphics[width=3.6cm,height=3.0cm,angle=0]{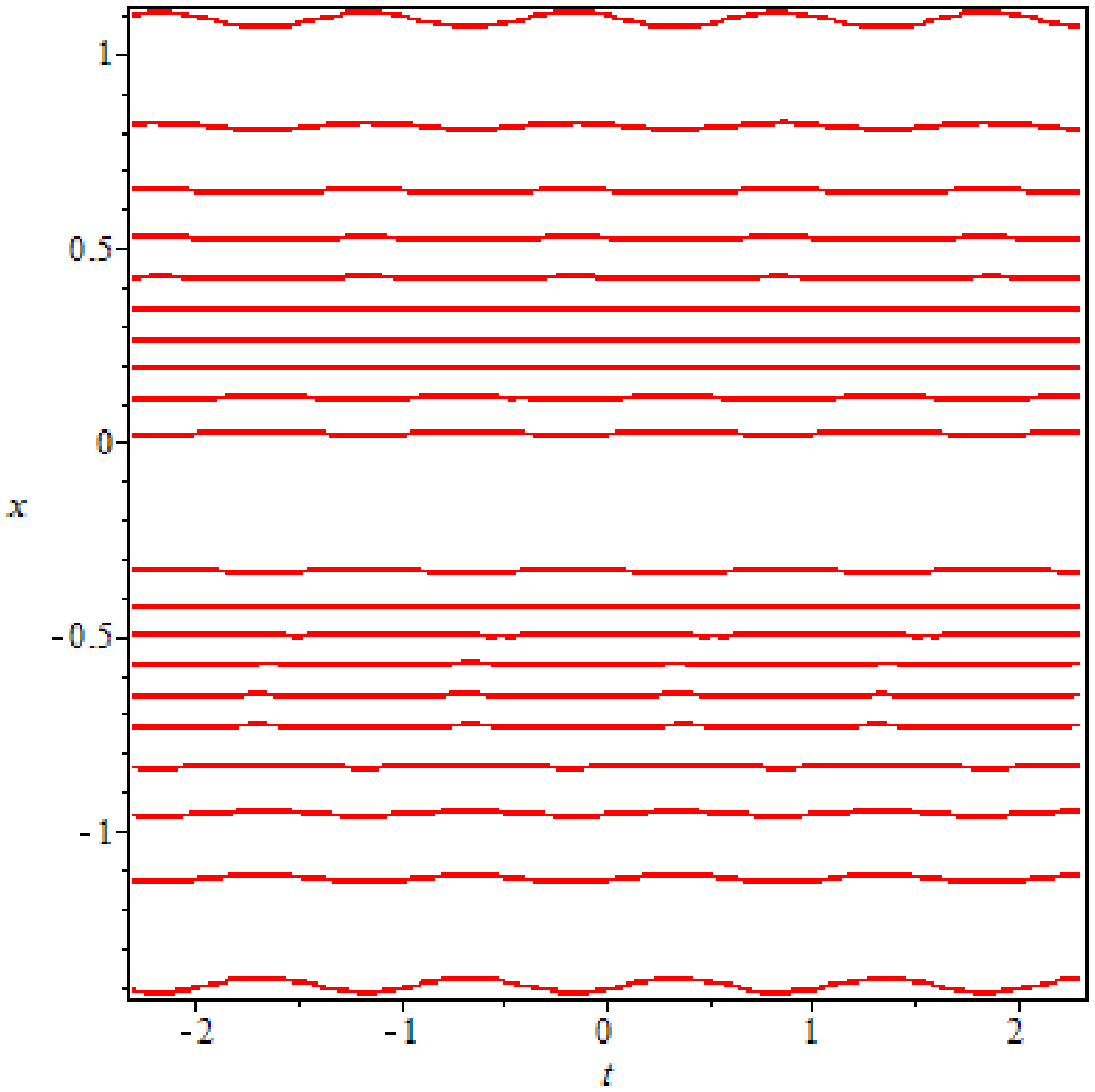}}}

$\qquad~~~~~~~~~(\textbf{d})\qquad \ \qquad\qquad\qquad\qquad~~~(\textbf{e})
\qquad\qquad\qquad\qquad\qquad~(\textbf{f})$\\
\noindent { \small \textbf{Figure 3.} (Color online) Plots of the soliton solution of the equation  with the parameters $\xi_{1}=2.5i$ and $b_{1}=e^{2+i}$.
$\textbf{(a)}$: the soliton solution with $q_{-}=0.1$,
$\textbf{(b)}$: the density plot corresponding to $(a)$,
$\textbf{(c)}$: the contour line corresponding to $(a)$,
$\textbf{(d)}$: the soliton solution with $q_{-}=0.01$,
$\textbf{(e)}$: the density plot corresponding to $(d)$,
$\textbf{(f)}$: the contour line corresponding to $(d)$.}\\

The images in Fig. 3 reveal  an interesting trend  that  the stationary breather solution is getting closer to a bell soliton solution when the boundary value $q_{-}$ becomes smaller and smaller. However, it is just shaped-like a bell soliton solution.

Furthermore, we consider the case that $\alpha(t)=0$ and $\gamma\neq0$. Then, via selecting appropriate parameters and changing $\gamma$, we can get the following images.\\

{\rotatebox{0}{\includegraphics[width=3.6cm,height=3.0cm,angle=0]{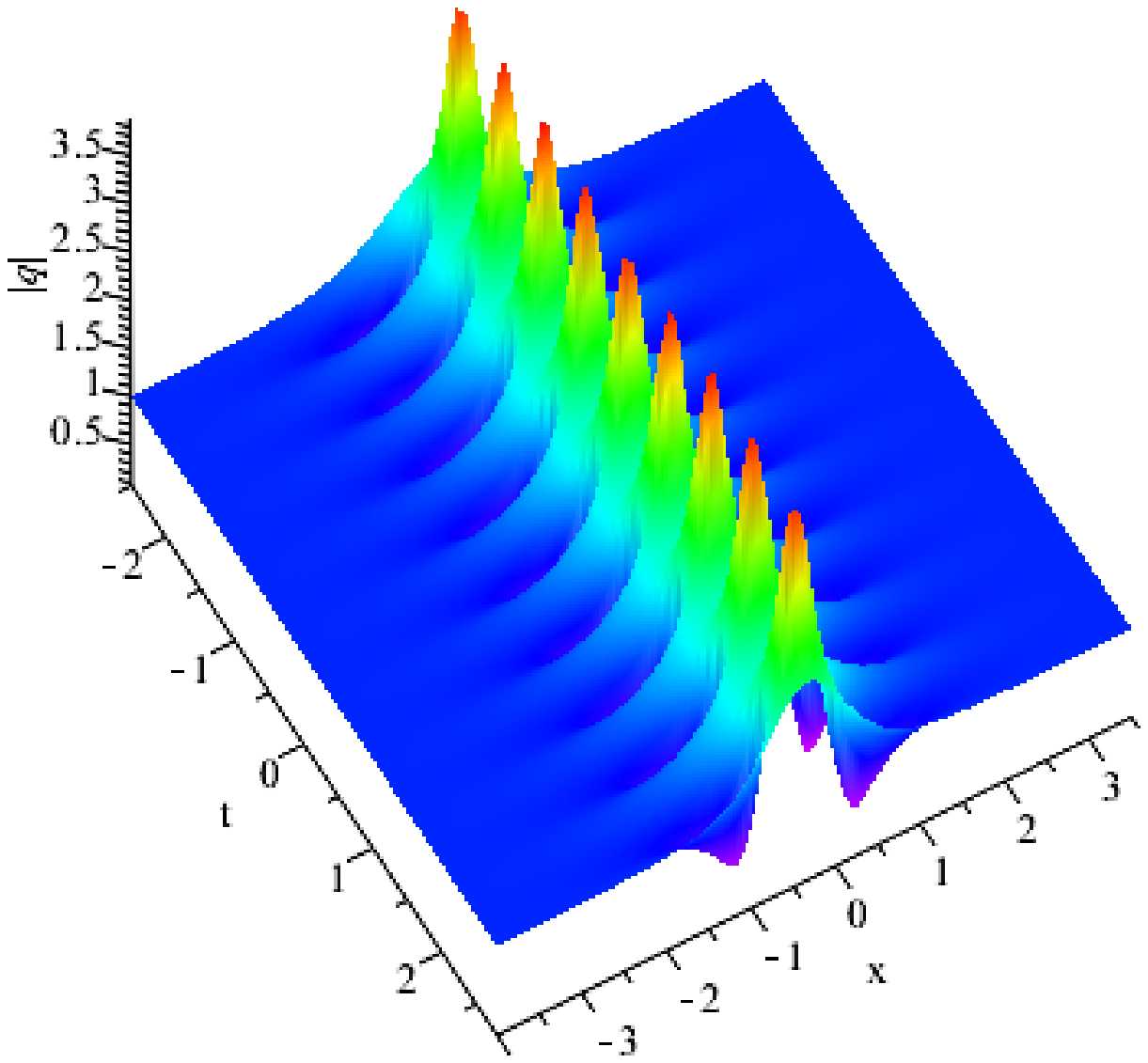}}}
~~~~
{\rotatebox{0}{\includegraphics[width=3.6cm,height=3.0cm,angle=0]{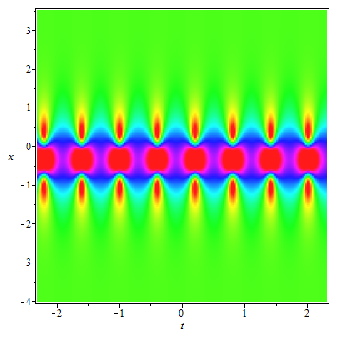}}}
~~~~
{\rotatebox{0}{\includegraphics[width=3.6cm,height=3.0cm,angle=0]{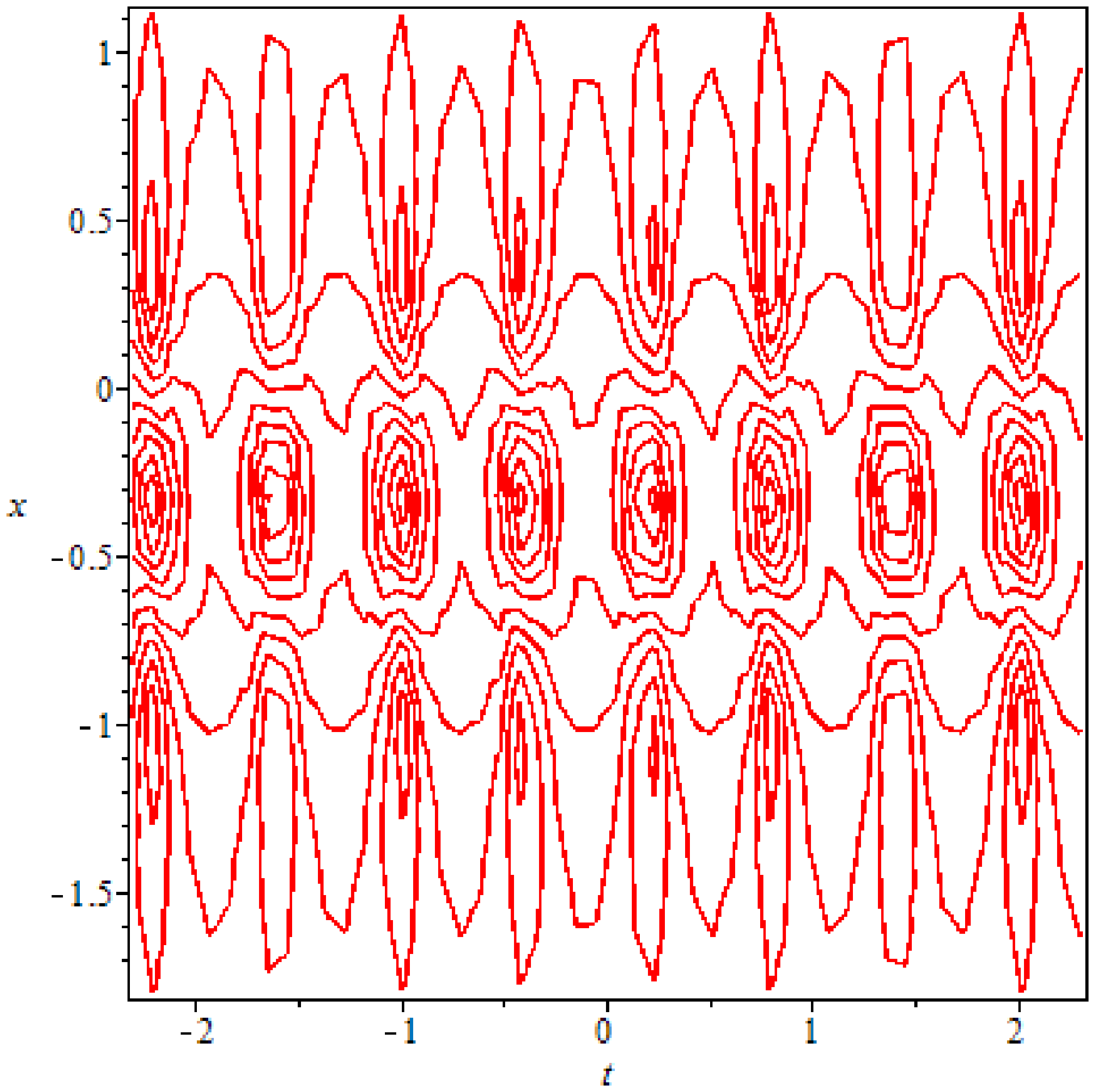}}}

$\qquad~~~~~~~~~(\textbf{a})\qquad \ \qquad\qquad\qquad\qquad~~~(\textbf{b})
\qquad\qquad\qquad\qquad\qquad~(\textbf{c})$\\

{\rotatebox{0}{\includegraphics[width=3.6cm,height=3.0cm,angle=0]{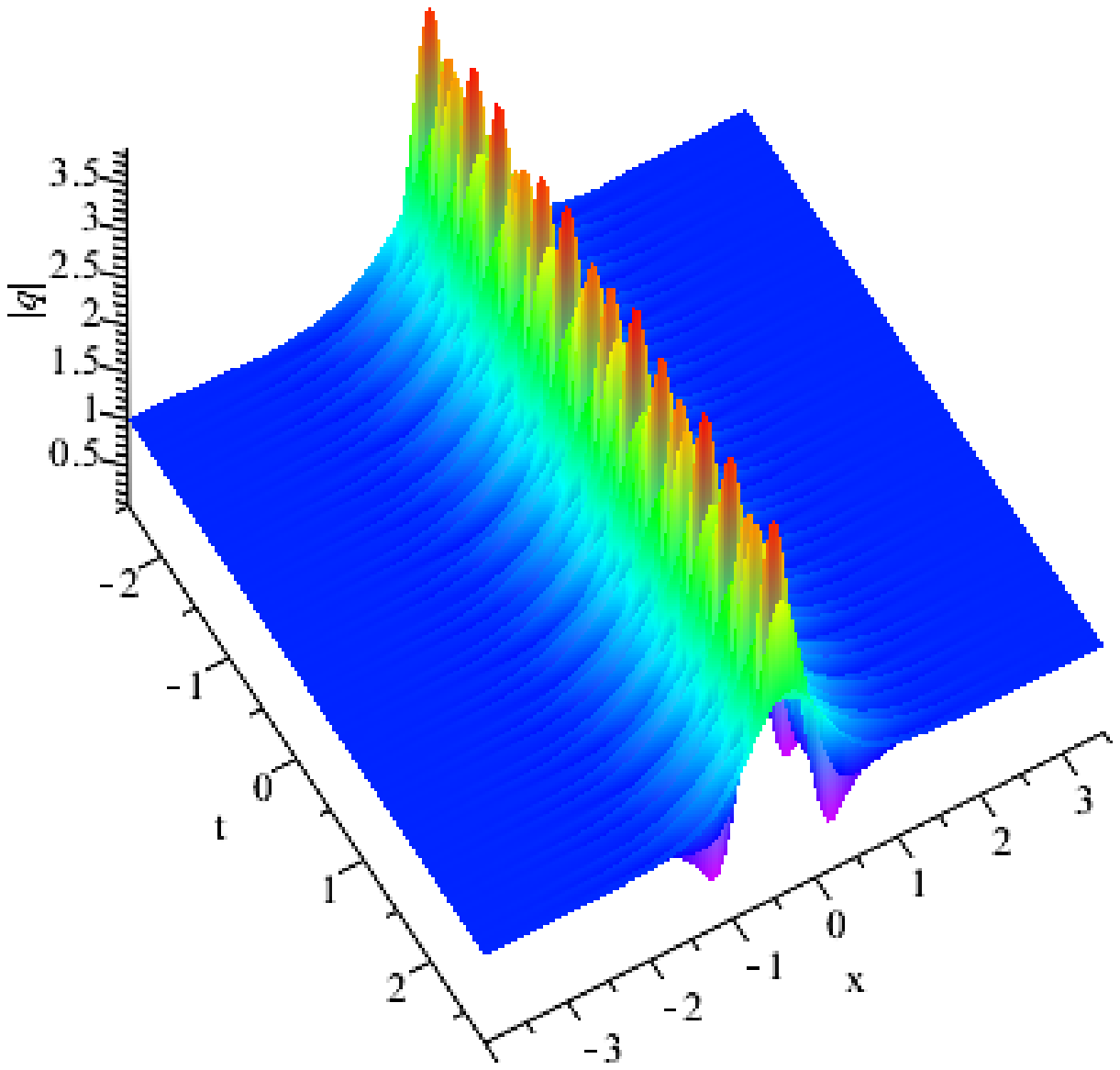}}}
~~~~
{\rotatebox{0}{\includegraphics[width=3.6cm,height=3.0cm,angle=0]{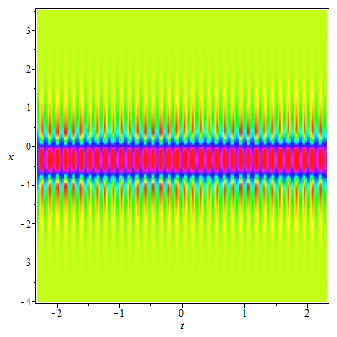}}}
~~~~
{\rotatebox{0}{\includegraphics[width=3.6cm,height=3.0cm,angle=0]{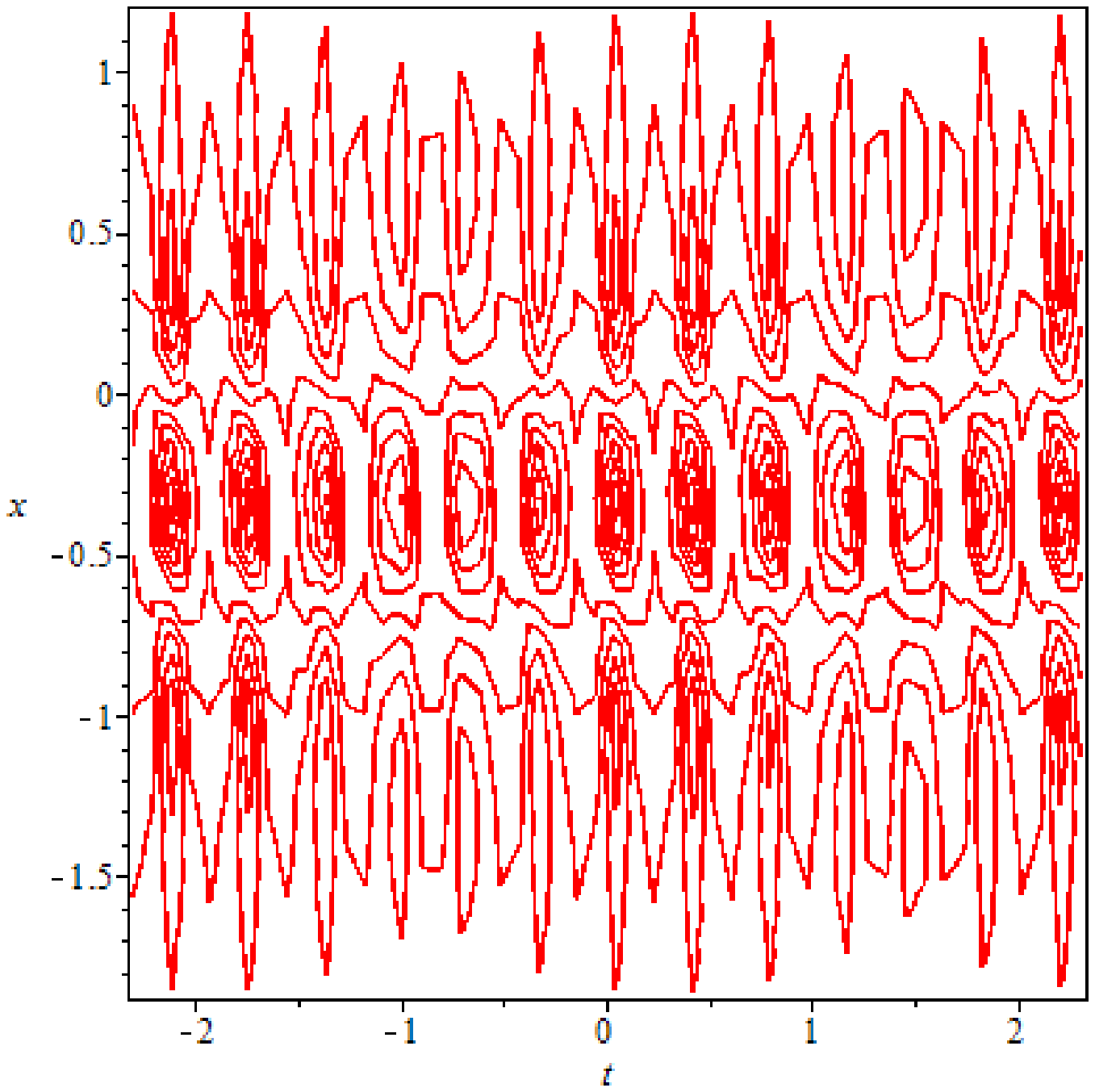}}}

$\qquad~~~~~~~~~(\textbf{d})\qquad \ \qquad\qquad\qquad\qquad~~~(\textbf{e})
\qquad\qquad\qquad\qquad\qquad~(\textbf{f})$\\

%{\rotatebox{0}{\includegraphics[width=3.6cm,height=3.0cm,angle=0]{1-7-1.eps}}}
%~~~~
%{\rotatebox{0}{\includegraphics[width=3.6cm,height=3.0cm,angle=0]{1-7-2.eps}}}
%~~~~
%{\rotatebox{0}{\includegraphics[width=3.6cm,height=3.0cm,angle=0]{1-7-3.eps}}}

%$\qquad~~~~~~~~~(\textbf{g})\qquad \ \qquad\qquad\qquad\qquad~~~(\textbf{h})
%\qquad\qquad\qquad\qquad\qquad~(\textbf{i})$\\
\noindent { \small \textbf{Figure 4.} (Color online) Plots of the soliton solution of the equation  with the parameters $q_{-}=1$, $\alpha(t)=0$, $\xi_{1}=2.5i$ and $b_{1}=e^{2+i}$.
$\textbf{(a)}$: the soliton solution with $\gamma=0.01$,
$\textbf{(b)}$: the density plot corresponding to $(a)$,
$\textbf{(c)}$: the contour line of the soliton solution corresponding to $(a)$,
$\textbf{(d)}$: the soliton solution with $\gamma=0.1$,
$\textbf{(e)}$: the density plot corresponding to $(d)$,
$\textbf{(f)}$: the contour line of the soliton solution corresponding to $(d)$.}\\

The graphs in Fig. 4 reveal  that  the soliton solutions are arranged more closely when the parameter $\gamma$ becomes larger. It is worth noting that the existence of $\gamma$ will change the shape of the solution but will not change the structure of the solution.

Furthermore, we consider the case that $\alpha(t)\neq0$ and $\gamma=0$. Then, via selecting appropriate parameters and changing $\alpha(t)$, we can get the following images.\\

{\rotatebox{0}{\includegraphics[width=3.6cm,height=3.0cm,angle=0]{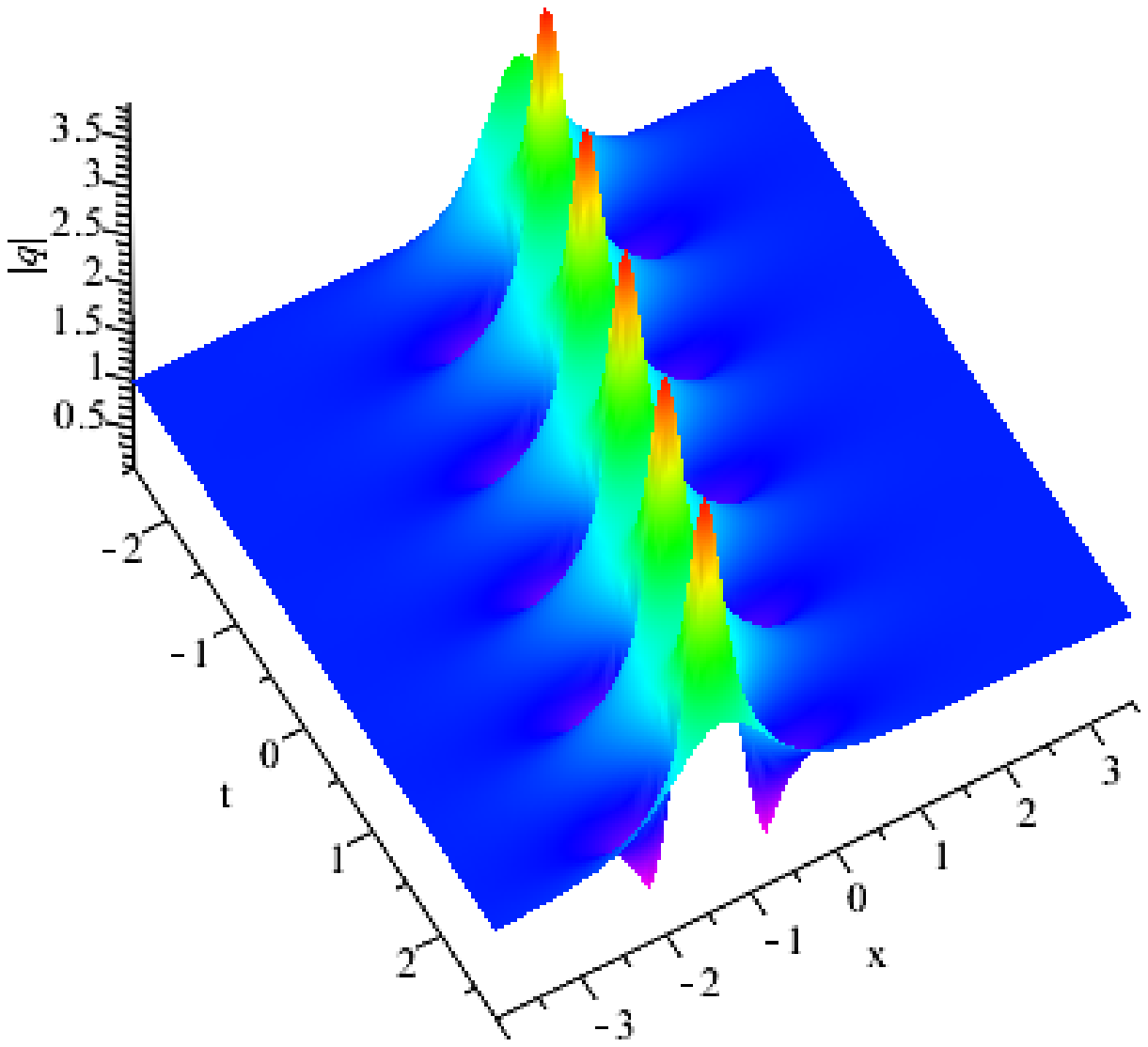}}}
~~~~
{\rotatebox{0}{\includegraphics[width=3.6cm,height=3.0cm,angle=0]{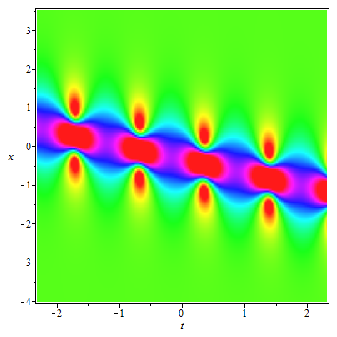}}}
~~~~
{\rotatebox{0}{\includegraphics[width=3.6cm,height=3.0cm,angle=0]{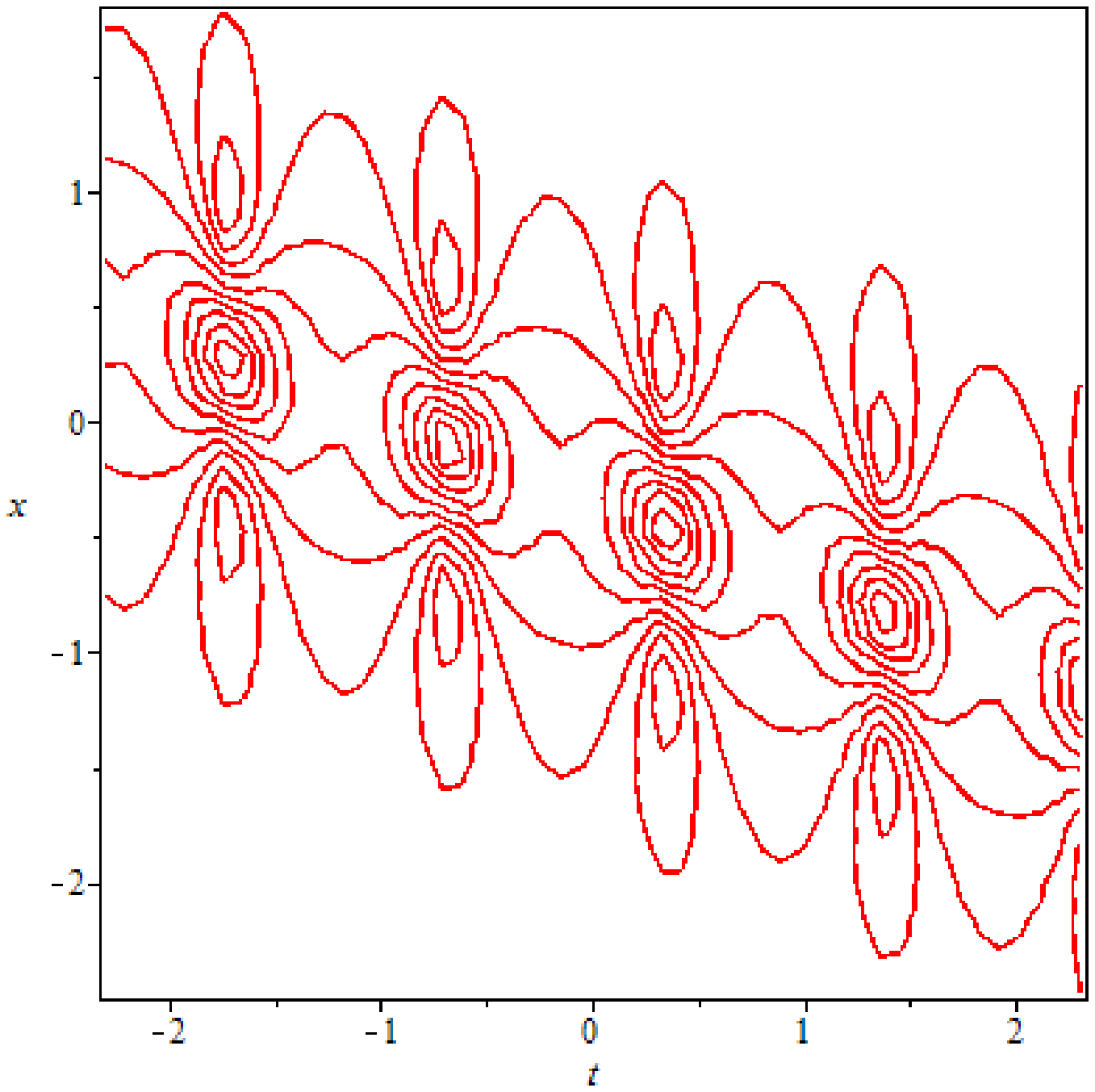}}}

$\qquad~~~~~~~~~(\textbf{a})\qquad \ \qquad\qquad\qquad\qquad~~~(\textbf{b})
\qquad\qquad\qquad\qquad\qquad~(\textbf{c})$\\

{\rotatebox{0}{\includegraphics[width=3.6cm,height=3.0cm,angle=0]{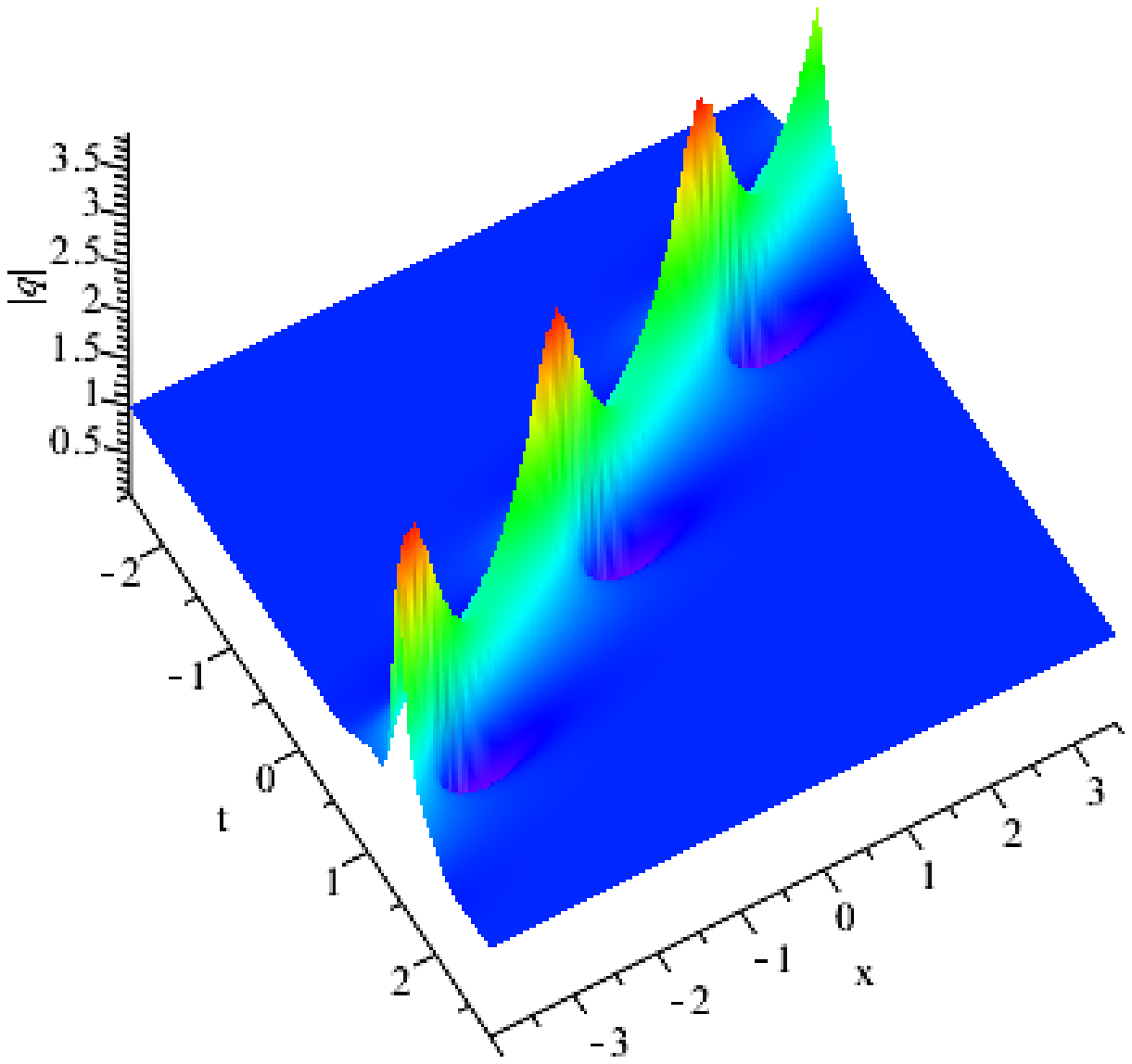}}}
~~~~
{\rotatebox{0}{\includegraphics[width=3.6cm,height=3.0cm,angle=0]{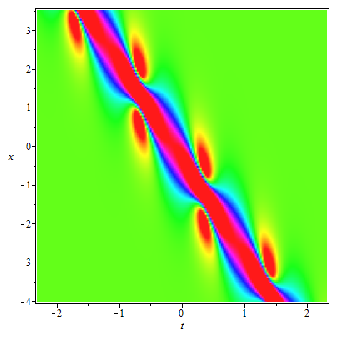}}}
~~~~
{\rotatebox{0}{\includegraphics[width=3.6cm,height=3.0cm,angle=0]{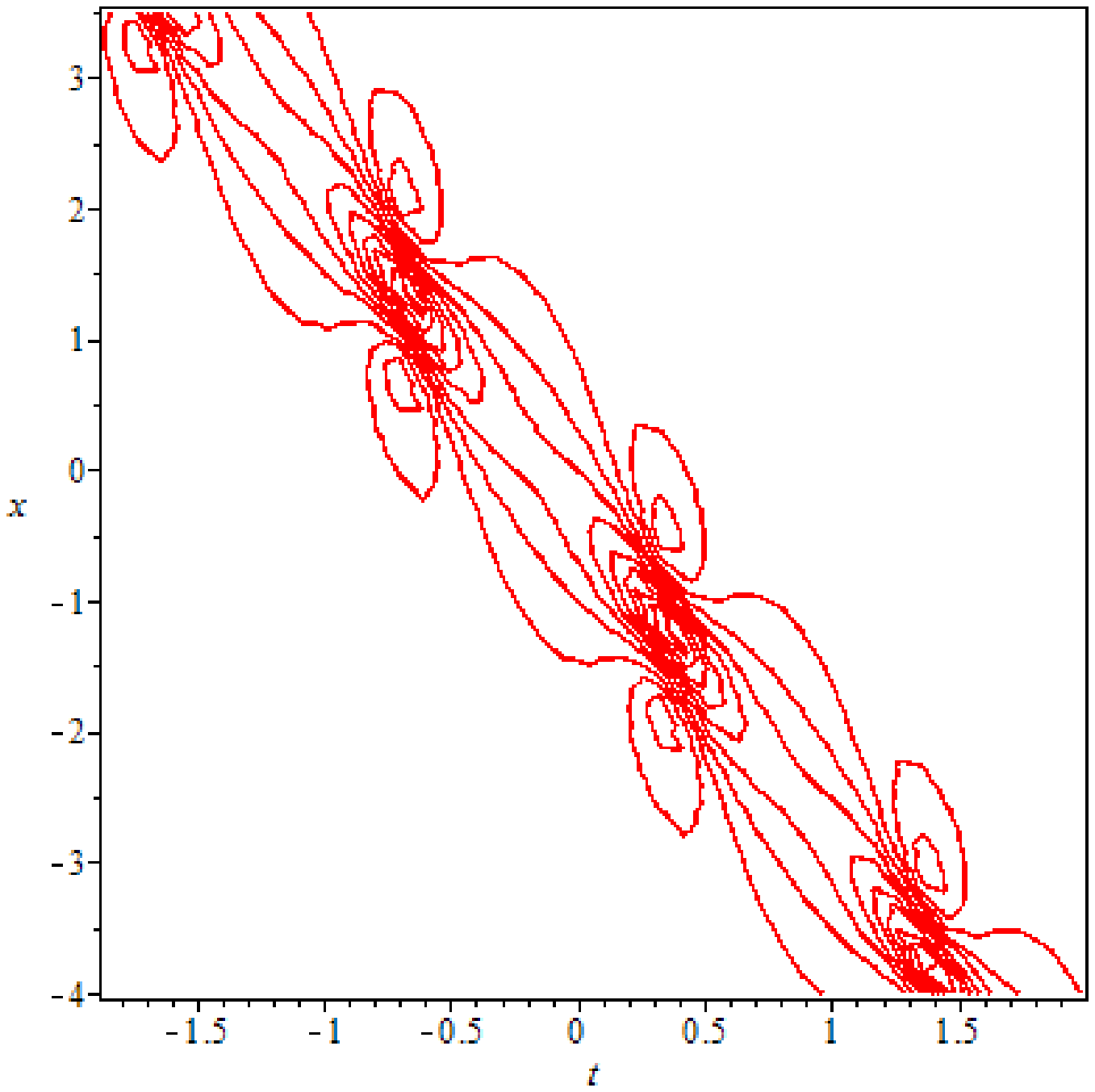}}}

$\qquad~~~~~~~~~(\textbf{d})\qquad \ \qquad\qquad\qquad\qquad~~~(\textbf{e})
\qquad\qquad\qquad\qquad\qquad~(\textbf{f})$\\

{\rotatebox{0}{\includegraphics[width=3.6cm,height=3.0cm,angle=0]{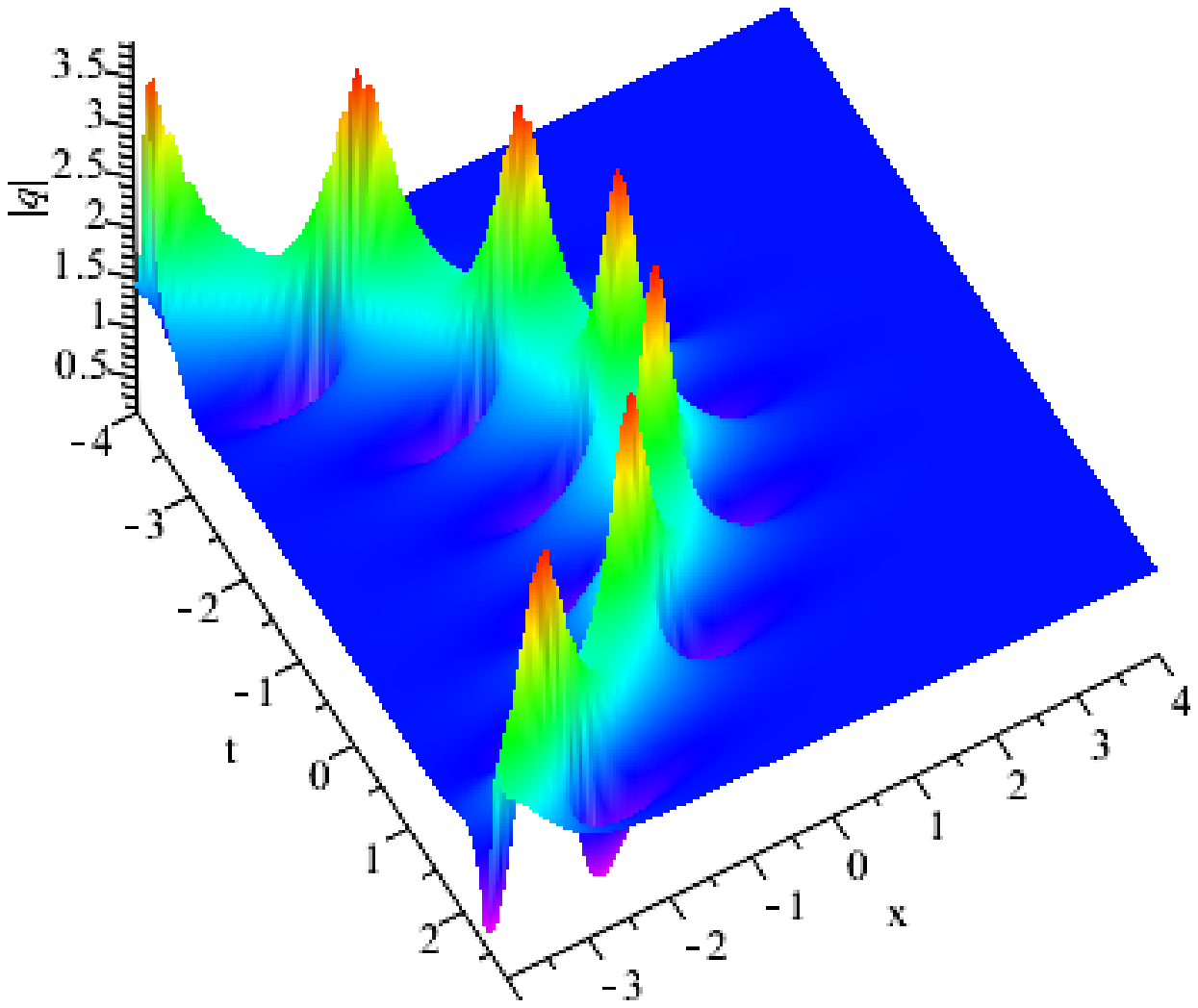}}}
~~~~
{\rotatebox{0}{\includegraphics[width=3.6cm,height=3.0cm,angle=0]{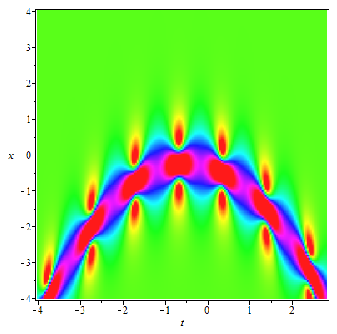}}}
~~~~
{\rotatebox{0}{\includegraphics[width=3.6cm,height=3.0cm,angle=0]{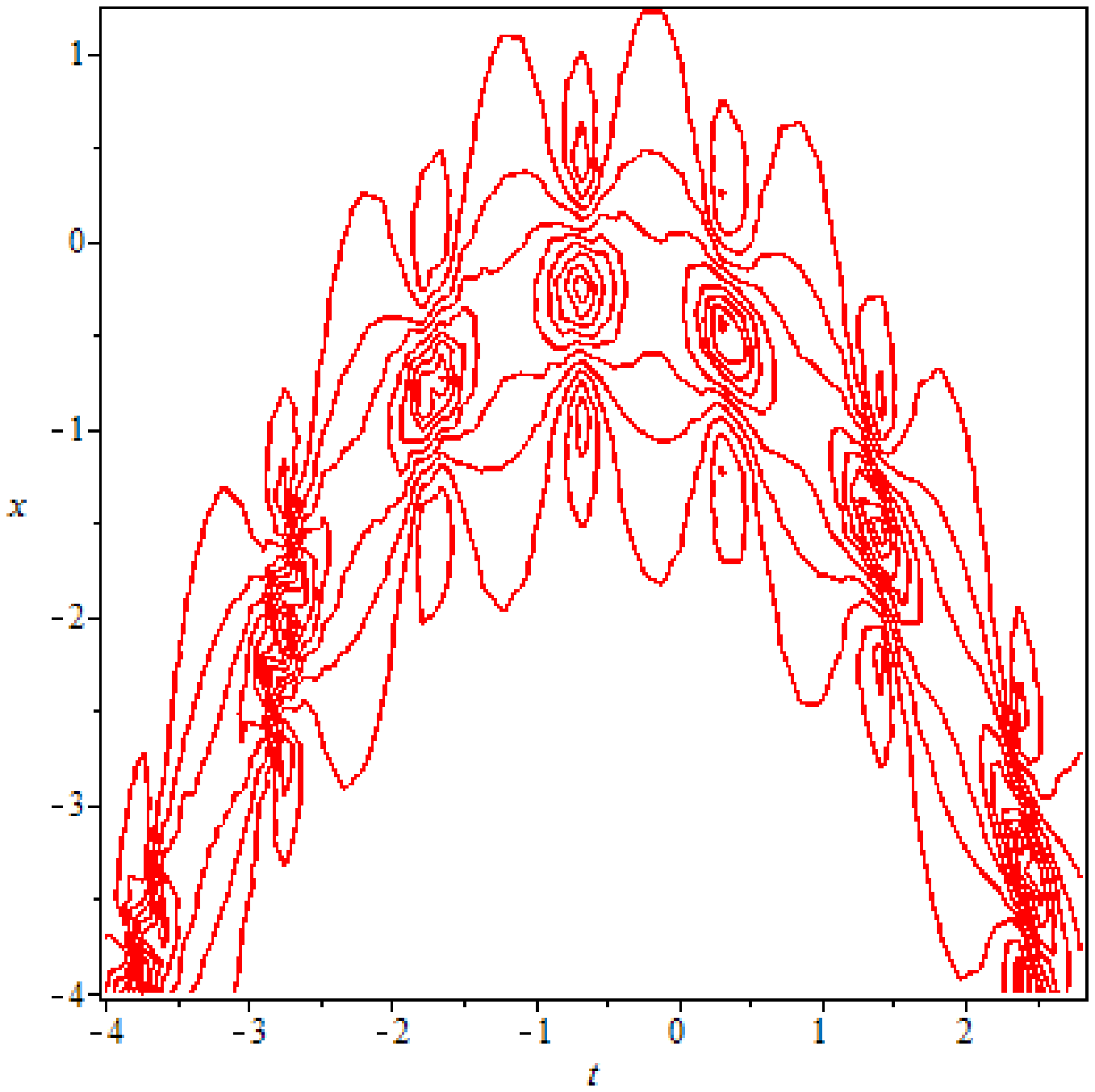}}}

$\qquad~~~~~~~~~(\textbf{g})\qquad \ \qquad\qquad\qquad\qquad~~~(\textbf{h})
\qquad\qquad\qquad\qquad\qquad~(\textbf{i})$\\
\noindent { \small \textbf{Figure 5.} (Color online) Plots of the soliton solution of the equation  with the parameters $q_{-}=1$, $\gamma=0$, $\xi_{1}=2.5i$ and $b_{1}=e^{2+i}$.
$\textbf{(a)}$: the soliton solution with $\alpha(t)=0.01$,
$\textbf{(b)}$: the density plot corresponding to $(a)$,
$\textbf{(c)}$: the contour line of the soliton solution corresponding to $(a)$,
$\textbf{(d)}$: the soliton solution with $\alpha(t)=0.07$,
$\textbf{(e)}$: the density plot corresponding to $(d)$,
$\textbf{(f)}$: the contour line of the soliton solution corresponding to $(d)$,
$\textbf{(g)}$: the soliton solution with $\alpha(t)=0.01t+0.01$,
$\textbf{(h)}$: the density plot corresponding to $(g)$,
$\textbf{(i)}$: the contour line of the soliton solution corresponding to $(g)$.}

The pictures in Figs. 5 illustrates  that  the direction of the propagation of the soliton solution will become farther and farther away from the original direction when the parameter $\alpha(t)$ is a constant and becomes larger. While, it is interesting that when $\alpha(t)$ is a linear function about $t$, the propagation path of the soliton solution will be changed into a path which is similar to a quadratic function curve.

Now, we are more interested in the case that $\alpha(t)\neq0$ and $\gamma\neq0$. Via selecting appropriate parameters, we can get the following images.\\

{\rotatebox{0}{\includegraphics[width=3.6cm,height=3.0cm,angle=0]{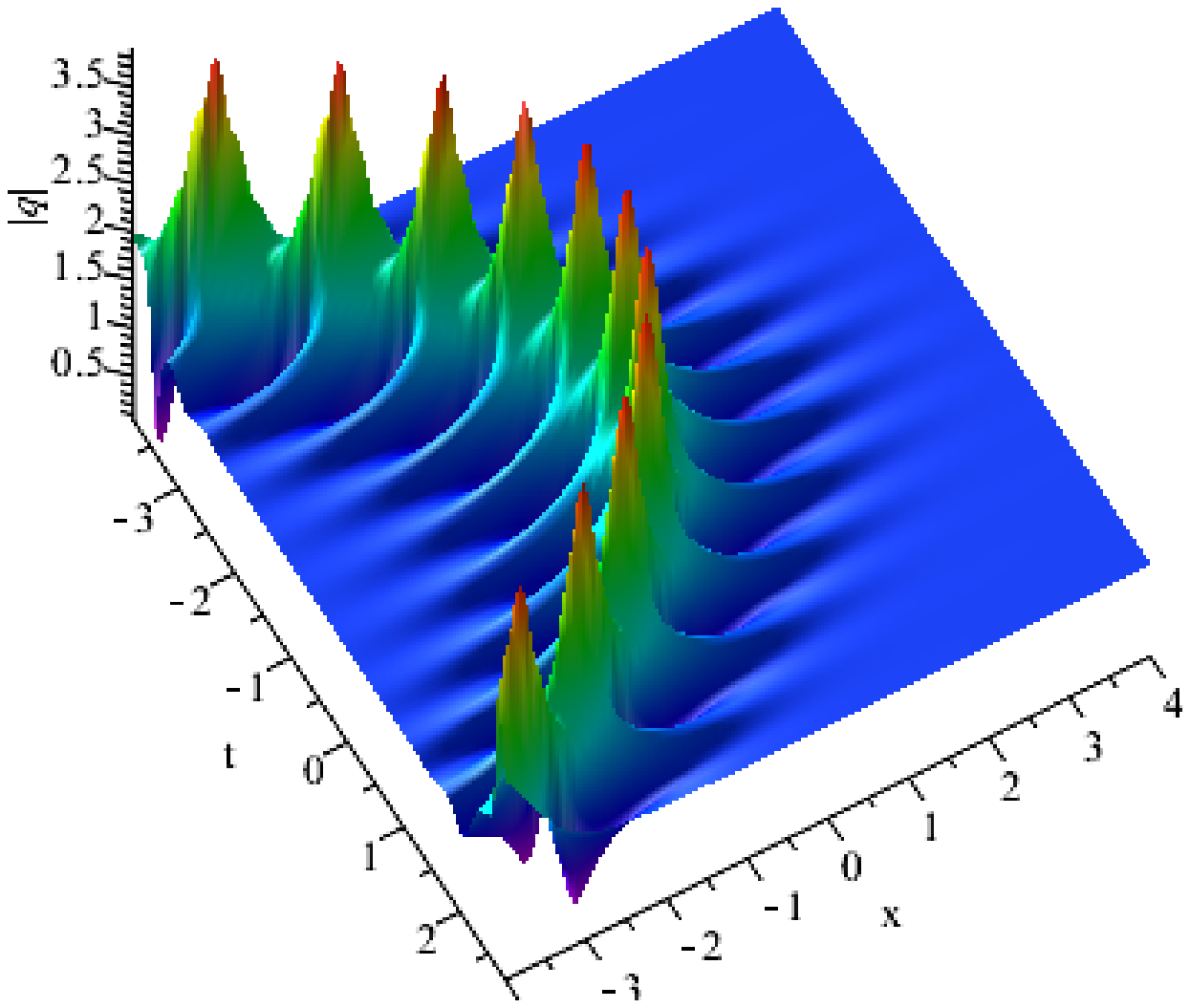}}}
~~~~
{\rotatebox{0}{\includegraphics[width=3.6cm,height=3.0cm,angle=0]{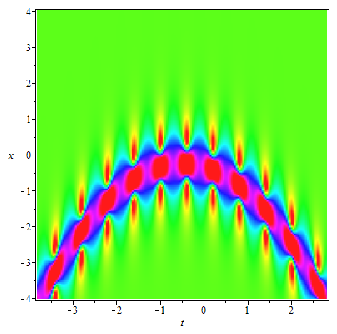}}}
~~~~
{\rotatebox{0}{\includegraphics[width=3.6cm,height=3.0cm,angle=0]{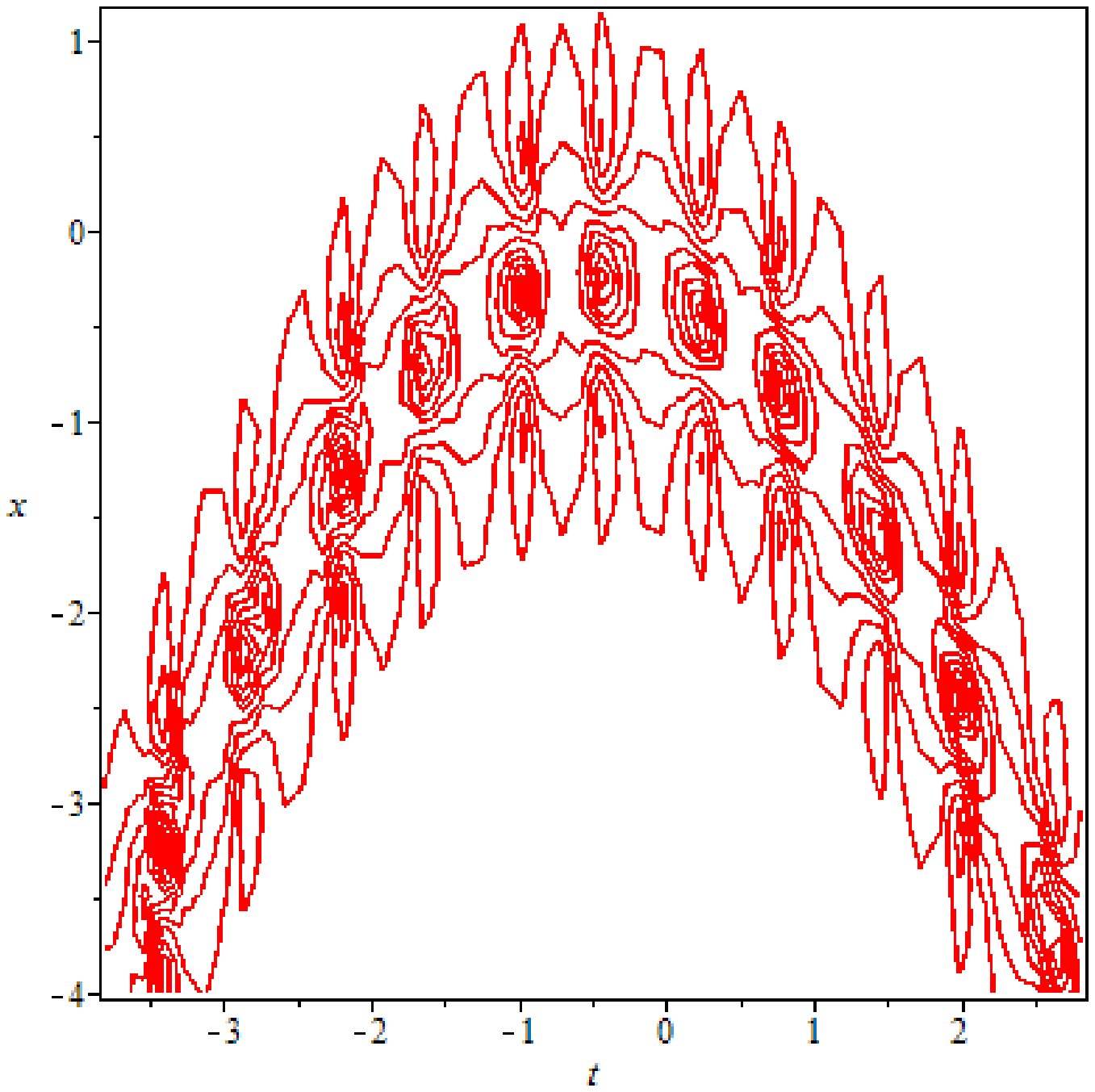}}}

$\qquad~~~~~~~~~(\textbf{a})\qquad \ \qquad\qquad\qquad\qquad~~~(\textbf{b})
\qquad\qquad\qquad\qquad\qquad~(\textbf{c})$\\
\noindent { \small \textbf{Figure 6.} (Color online) Plots of the breather solution of the equation  with the parameters $\gamma=0.01$, $\alpha(t)=0.01t+0.01$, $q_{-}=1$, $\xi_{1}=2.5i$ and $b_{1}=e^{2+i}$.
$\textbf{(a)}$: the soliton solution,
$\textbf{(b)}$: the density plot,
$\textbf{(c)}$: the contour line of the soliton solution.} \\

What we can learn from  the comparison between Fig. 6 and other figures are that the existence of the parameter $\gamma$ only determines the shape of soliton solutions and the existence of the parameter $\alpha(t)$ will change the propagation path of the soliton solution. To summarise, we obtain the dynamic behavior of the soliton solutions and various interesting phenomena via selecting appropriate parameters. It is hoped that these results can contribute to the physical field.

\section{The gvcNLS equation with NZBCs: double poles}
In this section, we will investigate the the double zeros of the analytical scattering coefficients. However, the results of the analysis as usual are not satisfactory. Therefore, we introduce a transformation to make the results more elegant.
\subsection{Discrete spectrum with double poles and residue conditions}
In this section, we will obtain the following main results.

\noindent \textbf {Theorem 8}
\emph{If $z_{n}(n=1,2,...,N)$ are the double zeros of the $s_{11}(z)$ , then }
\begin{subequations}
\begin{gather}
\mu_{+,1}(x,t;z_{n})=e_{n}e^{-2i\theta(x,t;z_{n})}\mu_{-,2}(x,t;z_{n}),\label{DR-1a}\\
\mu'_{+,1}(x,t;z_{n})=e^{-2i\theta(x,t;z_{n})}[(h_{n}-2ie_{n}\theta'(x,t;z_{n}))\mu_{-,2}(x,t;z_{n})
+e_{n}\mu'_{-,2}(x,t;z_{n})];\label{DR-1b}\\
\mu_{+,1}(x,t;\check{z}_{n})=\check{e}_{n}e^{-2i\theta(x,t;\check{z}_{n})}\mu_{-,2}(x,t;\check{z}_{n}),\label{DR-2a}\\
\mu'_{+,1}(x,t;\check{z}_{n})=e^{-2i\theta(x,t;\check{z}_{n})}
[(\check{h}_{n}-2i\check{e}_{n}\theta'(x,t;\check{z}_{n}))\mu_{-,2}(x,t;\check{z}_{n})
+\check{e}_{n}\mu'_{-,2}(x,t;\check{z}_{n})];\label{DR-2b}\\
\mu_{+,2}(x,t;z^{*}_{n})=\tilde{e}_{n}e^{2i\theta(x,t;z^{*}_{n})}\mu_{-,1}(x,t;z^{*}_{n}),\label{DR-3a}\\
\mu'_{+,2}(x,t;z^{*}_{n})=e^{2i\theta(x,t;z^{*}_{n})}
[(\tilde{h}_{n}+2i\tilde{e}_{n}\theta'(x,t;z^{*}_{n}))\mu_{-,1}(x,t;z^{*}_{n})
+\tilde{e}_{n}\mu'_{-,1}(x,t;z^{*}_{n})];\label{DR-3b}\\
\mu_{+,2}(x,t;\hat{z}_{n})=\hat{e}_{n}e^{2i\theta(x,t;\hat{z}_{n})}\mu_{-,1}(x,t;\hat{z}_{n}),\label{DR-4a}\\
\mu'_{+,2}(x,t;\hat{z}_{n})=e^{2i\theta(x,t;\hat{z}_{n})}
[(\hat{h}_{n}+2i\hat{e}_{n}\theta'(x,t;\hat{z}_{n}))\mu_{-,1}(x,t;\hat{z}_{n})
+\hat{e}_{n}\mu'_{-,1}(x,t;\hat{z}_{n})];\label{DR-4b}
\end{gather}
\end{subequations}
\emph{where $e_{n}$, $h_{n}$, $\check{e}_{n}$, $\check{h}_{n}$, $\tilde{e}_{n}$, $\tilde{h}_{n}$, $\hat{e}_{n}$ and $\hat{h}_{n}$ are constants and independent of $x$ and $t$, and the $\hat{z}_{n}$ and $\check{z}_{n}$ are defined in \eqref{DR-6}.}

\noindent \textbf {Theorem 9}
\emph{The constants appeared in Theorem 8 have been fixed relationship by }
\begin{gather}
\tilde{e}_{n}=-e^{*}_{n}, \quad \hat{e}_{n}=-e_{n}\frac{q^{*}_{-}}{q_{+}}, \quad \check{e}_{n}=-\hat{e}^{*}_{n}, \notag\\
\tilde{h}_{n}=-h^{*}_{n},\quad \hat{h}_{n}=-h_{n}\frac{q^{*}_{-}z^{2}_{n}}{q_{+}q^{2}_{0}},\quad
\check{h}_{n}=-\hat{h}^{*}_{n}.\notag
\end{gather}

\noindent \textbf {Theorem 10}
\emph{The residues of $u_{+,1}(x,t;z)/s_{11}(z)$ and $u_{+,2}(x,t;z)/s_{22}(z)$ can be derived by}
\begin{subequations}
\begin{align}
&\mathop{P_{-2}}_{z=z_{n}}\left[\frac{\mu_{+,1}(z)}{s_{11}(z)}\right]=
E_{n}e^{-2i\theta(z_{n})}\mu_{-,2}(z_{n}),\label{DR-5a}\\
&\mathop{Res}_{z=z_{n}}\left[\frac{\mu_{+,1}(z)}{s_{11}(z)}\right]=
E_{n}e^{-2i\theta(z_{n})}\left[\mu'_{-,2}(z_{n})+
\left(H_{n}-2i\theta'(z_{n})\right)\mu_{-,2}(z_{n})\right],\label{DR-5b}\\
&\mathop{P_{-2}}_{z=\check{z}_{n}}\left[\frac{\mu_{+,1}(z)}{s_{11}(z)}\right]=
\check{E}_{n}e^{-2i\theta(\check{z}_{n})}\mu_{-,2}(\check{z}_{n}),\\
&\mathop{Res}_{z=\check{z}_{n}}\left[\frac{\mu_{+,1}(z)}{s_{11}(z)}\right]=
\check{E}_{n}e^{-2i\theta(\check{z}_{n})}\left[\mu'_{-,2}(\check{z}_{n})+
\left(\check{H}_{n}-2i\theta'(\check{z}_{n})\right)\mu_{-,2}(\check{z}_{n})\right],\\
&\mathop{P_{-2}}_{z=z^{*}_{n}}\left[\frac{\mu_{+,2}(z)}{s_{22}(z)}\right]=
\tilde{E}_{n}e^{2i\theta(z^{*}_{n})}\mu_{-,1}(z^{*}_{n}),\\
&\mathop{Res}_{z=z^{*}_{n}}\left[\frac{\mu_{+,2}(z)}{s_{22}(z)}\right]=
\tilde{E}_{n}e^{2i\theta(z^{*}_{n})}\left[\mu'_{-,1}(z^{*}_{n})+
\left(H_{n}+2i\theta'(z^{*}_{n})\right)\mu_{-,1}(z^{*}_{n})\right],\\
&\mathop{P_{-2}}_{z=\hat{z}_{n}}\left[\frac{\mu_{+,2}(z)}{s_{22}(z)}\right]=
\hat{E}_{n}e^{2i\theta(\hat{z}_{n})}\mu_{-,1}(\hat{z}_{n}),\\
&\mathop{Res}_{z=\hat{z}_{n}}\left[\frac{\mu_{+,2}(z)}{s_{22}(z)}\right]=
\hat{E}_{n}e^{2i\theta(\hat{z}_{n})}\left[\mu'_{-,1}(\hat{z}_{n})+
\left(\hat{H}_{n}+2i\theta'(\hat{z}_{n})\right)\mu_{-,1}(\hat{z}_{n})\right],
\end{align}
\end{subequations}
\emph{where $E_{n}$ and $H_{n}$ are defined in \eqref{DR-11}. The $\check{E}_{n}$, $\check{H}_{n}$, $\tilde{E}_{n}$, $\tilde{H}_{n}$, $\hat{E}_{n}$ and $\hat{H}_{n}$ are the similar definitions corresponding to the constants $\check{e}_{n}$, $\check{h}_{n}$, $\tilde{e}_{n}$, $\tilde{h}_{n}$, $\hat{e}_{n}$, and $\hat{h}_{n}$, respectively.}

In the above sections, We know that the discrete spectrum of the scattering problem is a set that is composed of the values $z\in\mathbb{C}\setminus\Sigma$ such that the eigenfunctions exist in $L^{2}(\mathbb{R})$. Here, we suppose that the $z_{n}(\in D_{+}\cap\{z\in\mathbb{C}: Imz>0\}, n=1,2,...,N)$ are the double poles of $s_{11}(z)$ i.e., $s_{11}(z_{n})=s'_{11}(z_{n})=0$, but $s''_{11}(z_{n})\neq0$, $n=1, 2,..., N$. Then, through the symmetries of the elements in scattering matrix that have been shown in \emph{Theorem 4}, it is easy to derive the discrete spectrum as follows
\begin{align}
\bar{Z}=\left\{z_{n}, -\frac{q_{0}^{2}}{z_{n}^{*}},
  z_{n}^{*}, -\frac{q_{0}^{2}}{z_{n}}\right\}, \quad n=1,2,...,N. \notag
\end{align}
For convenience, we define that
\begin{gather}\label{DR-6}
\hat{z}_{n}=-\frac{q^{2}_{0}}{z_{n}}, \quad \check{z}_{n}=-\frac{q^{2}_{0}}{z^{*}_{n}}.
\end{gather}
Now, we are going to prove the theorems that have been given in this section.

The proof of the  Theorem 8 is given as follows.
\begin{proof}
According to the \eqref{2.14}, and $z_{n}$ are the double zeros of $s_{11}(z)$ for $n=1,2,\cdots,N$, we can derive that
\begin{align}\label{DR-7}
\phi_{+,1}(z_{n})=e_{n}\phi_{-,2}(z_{n}),
\end{align}
where $e_{n}$ is constant and independents of $x$ and $t$. By applying   \eqref{2.15}, the \eqref{DR-1a} can be derived obviously. Since the $z_{n}(n=1,2,\cdots,N)$ are the double zeros of $s_{11}(z)$, we have $s'_{11}(z_{n})=0$. Then, according to the \eqref{2.14},  we can derive that
\begin{align}
Wr\left(\phi'_{+,1}(z_{n})-e_{n}\phi'_{-,2}(z_{n}),\phi_{-,2}(z_{n})\right)=0.
\end{align}
Based on the property of the determinant, we have
\begin{align}
\phi'_{+,1}(z_{n})=e_{n}\phi'_{-,2}(z_{n})+h_{n}\phi_{-,2}(z_{n}),
\end{align}
where $h_{n}$ is constant and independents of $x$ and $t$. Then, by applying the \eqref{2.15}, the \eqref{DR-1b} can be derived obviously.
Furthermore, the other formulas can be proved similarly.
\end{proof}

The proof of Theorem 9 is given as follows.
\begin{proof}
Based on the \emph{Theorem 3} and \emph{Theorem 8}, we derive that
\begin{align} \label{DR-8}
-\sigma\mu^{*}_{+,1}(x,t;z_{n})=\tilde{e}_{n}e^{2i\theta(x,t;z^{*}_{n})}\sigma\mu^{*}_{-,2}(x,t;z_{n}).
\end{align}
Note that $\theta(x,t;z^{*}_{n})=\theta^{*}(x,t;z_{n})$. Then, multiplying $\sigma$ to both ends of equation \eqref{DR-8} and taking the complex conjugate, it can be derived that
\begin{align}
\mu_{+,1}(x,t;z_{n})=-\tilde{e}^{*}_{n}e^{-2i\theta(x,t;z_{n})}\mu_{-,2}(x,t;z_{n}).
\end{align}
Therefore, the relationship that $\tilde{e}_{n}=-e^{*}_{n}$ is proved.

Furthermore, the other relationships among the constants can be proved similarly.
\end{proof}
Then, we pay attention to the residue condition, i.e., \emph{Theorem 10}. We  first  derive a fact.

\noindent \textbf {Proposition 1.}
\emph{If the functions $f$ and $g$ are analytic in a complex region $\Omega\in\mathbb{C}$ , $g$ has a double poles at $z_{0}\in\Omega$, i.e., $g(z_{0})=g'(z_{0})=0$, $g''(z_{0})\neq0$, and $f(z_{0})\neq0$. Thus the residue of $f/g$ can be calculated by the Laurent expansion at $z=z_{0}$, namely,}
\begin{align}\label{DR-9}
\mathop{Res}_{z=z_{0}}\left[\frac{f}{g}\right]=\frac{2f'(z_{0})}{g''(z_{0})}-
\frac{2f(z_{0})g'''(z_{0})}{3(g''(z_{0}))^{2}},\quad
\mathop{P_{-2}}_{z=z_{0}}\left[\frac{f}{g}\right]=\frac{2f(z_{0})}{g''(z_{0})}.
\end{align}

Next, we prove the \emph{Theorem 10}.
\begin{proof}
Based on the \emph{Theorem 8} and the \emph{Proposition 1}, the residue of the $\frac{\mu_{+,1}(x,t;z)}{s_{11}(z)}$ can be derived as
\begin{subequations}
\begin{align}
&\mathop{P_{-2}}_{z=z_{n}}\left[\frac{\mu_{+,1}(x,t;z)}{s_{11}(z)}\right]=
\frac{2\mu_{+,1}(x,t;z_{n})}{s''_{11}(z_{n})}=
\frac{2e_{n}}{s''_{11}(z_{n})}e^{-2i\theta(x,t;z_{n})}\mu_{-,2}(x,t;z_{n}),\label{DR-10a}\\
&\mathop{Res}_{z=z_{n}}\left[\frac{\mu_{+,1}(x,t;z)}{s_{11}(z)}\right]=
\frac{2\mu'_{+,1}(x,t;z_{n})}{s''_{11}(z_{n})}
-\frac{2\mu_{+,1}(x,t;z_{n})s'''_{11}(z_{n})}{3(s''_{11}(z_{n}))^{2}} \notag\\
&=\frac{2e_{n}}{s''_{11}(z_{n})}e^{-2i\theta(x,t;z_{n})}\left[\mu'_{-,2}(x,t;z_{n})+
\left(\frac{h_{n}}{e_{n}}-2i\theta'(x,t;z_{n})-\frac{s'''_{11}(z_{n})}{3s''_{11}(z_{n})}
\right)\mu_{-,2}(x,t;z_{n})\right].\label{DR-10b}
\end{align}
\end{subequations}
Then, we introduce the representation
\begin{align}
E_{n}=\frac{2e_{n}}{s''_{11}(z_{n})}, \quad H_{n}=\frac{h_{n}}{e_{n}}-\frac{s'''_{11}(z_{n})}{3s''_{11}(z_{n})}.\label{DR-11}
\end{align}
Obviously,   \eqref{DR-5a} and the \eqref{DR-5b} can be derived.
Furthermore, the other formulas can be proved similarly.
\end{proof}

\subsection{Reconstruction formula for the potential}
In this section, we will obtain the following main results.
\begin{itemize}
  \item \emph{The reconstruction formula for the potential are that}
  \begin{align}\label{RC-1}
  \begin{split}
q(x,t)=&iw(x,t)=i\left(w_{-}+\frac{1}{2\pi}\int_{\Sigma}(M^{+}(x,t;s)G(x,t;s))_{21}\,ds\right. \\
&\left.+i\sum_{n=1}^{2N}E_{n}e^{-2i\theta(x,t;\xi_{n})}
[\mu'_{-,22}(x,t;\xi_{n})+\mu_{-,22}(x,t;\xi_{n})D_{n}(x,t)]\right),
\end{split}
\end{align}
\emph{where the $w(x,t)$ and $D_{n}(x,t)$ are defined in \eqref{RC-4}.}
\end{itemize}

For convenience, firstly, we introduce a substitution
\begin{align}
\xi_{n}:=z_{n},\quad \xi_{N+n}:=\check{z}_{n},\quad \hat{\xi}_{n}:=\hat{z}_{n}, \quad \hat{\xi}_{N+n}:=z_{n}^{*}.
\end{align}
Therefore, the $\xi_{n}, n=1,2,\cdots, 2N$ are the double zeros of $s_{11}$ in $D_{+}$ and $\hat{\xi}_{n}, n=1,2,\cdots, 2N$ are the double zeros of $s_{22}$ in $D_{-}$. Similar to the case of simple pole, through subtracting out the asymptotic behavior and the double poles contributions, we obtain a regular RHP. Then, we have
\begin{align}\label{RC-2}
\begin{split}
&M^{-}-\mathbb{I}-\frac{i}{z}\sigma_{3}Q_{-}-\sum_{n=1}^{2N}\left\{
\frac{\mathop{Res}\limits_{z=\hat{\xi}_{n}}M^{-}}{z-\hat{\xi}_{n}}
+\frac{\mathop{P_{-2}}\limits_{z=\hat{\xi}_{n}}M^{-}}{(z-\hat{\xi}_{n})^{2}}
+\frac{\mathop{Res}\limits_{z=\xi_{n}}M^{+}}{z-\xi_{n}}
+\frac{\mathop{P_{-2}}\limits_{z=\xi_{n}}M^{+}}{(z-\xi_{n})^{2}}\right\}\\&=
M^{+}-\mathbb{I}-\frac{i}{z}\sigma_{3}Q_{-}-\sum_{n=1}^{2N}\left\{
\frac{\mathop{Res}\limits_{z=\hat{\xi}_{n}}M^{-}}{z-\hat{\xi}_{n}}
+\frac{\mathop{P_{-2}}\limits_{z=\hat{\xi}_{n}}M^{-}}{(z-\hat{\xi}_{n})^{2}}
+\frac{\mathop{Res}\limits_{z=\xi_{n}}M^{+}}{z-\xi_{n}}
+\frac{\mathop{P_{-2}}\limits_{z=\xi_{n}}M^{+}}{(z-\xi_{n})^{2}}\right\}
-M^{+}G.
\end{split}
\end{align}
Based on   \eqref{Matrix}, we can acquire that
\begin{align}
\mathop{Res}_{z=\xi_{n}}[M^{+}]=(\mathop{Res}_{z=\xi_{n}}\left[\frac{\mu_{+,1}(x,t;z)}{s_{11}(z)}\right],0), \quad \mathop{P_{-2}}_{z=\xi_{n}}M^{+}=(\mathop{P_{-2}}_{z=\xi_{n}}\left[\frac{\mu_{+,1}(x,t;z)}{s_{11}(z)}\right],0), \notag \\
\mathop{Res}_{z=\hat{\xi}_{n}}[M^{-}]=(0,\mathop{Res}_{z=\hat{\xi}_{n}}\left[\frac{\mu_{+,2}(x,t;z)}
{s_{22}(z)}\right]), \quad \mathop{P_{-2}}_{z=\hat{\xi}_{n}}M^{-}=(0,\mathop{P_{-2}}_{z=\hat{\xi}_{n}}\left[\frac{\mu_{+,2}(x,t;z)}
{s_{22}(z)}\right]).
\end{align}
It is obvious that the left side of \eqref{RC-2} is analytic in $D_{-}$ and the right side of \eqref{R-4}, except the item $M^{+}(z)G(z)$, is analytic in $D_{-}$. At the same time, combining \emph{Theorem 6} and \emph{Theorem 7}, it is apparent that the asymptotic behavior of both sides of the equation \eqref{R-4} are $O(1/z)(z\rightarrow\infty)$ and $O(1)(z\rightarrow0)$. Then, by using the projection operators \eqref{Cauchy}, we can obtain the solution of the RHP as follows
\begin{align}\label{RC-3}
M(x,t;z)=&\mathbb{I}+\frac{i}{z}\sigma_{3}Q_{-}+\sum_{n=1}^{2N}\left\{
\frac{\mathop{Res}\limits_{z=\hat{\xi}_{n}}M^{-}}{z-\hat{\xi}_{n}}
+\frac{\mathop{P_{-2}}\limits_{z=\hat{\xi}_{n}}M^{-}}{(z-\hat{\xi}_{n})^{2}}
+\frac{\mathop{Res}\limits_{z=\xi_{n}}M^{+}}{z-\xi_{n}}
+\frac{\mathop{P_{-2}}\limits_{z=\xi_{n}}M^{+}}{(z-\xi_{n})^{2}}\right\} \notag\\
&+\frac{1}{2\pi i}\int_{\Sigma}\frac{M^{+}(x,t;s)G(x,t;s)}{s-z}\,ds,\quad z\in\mathbb{C}\setminus\Sigma.
\end{align}
However, to make the results more elegant, we introduce the notation $E_{N+n}:=\check{E}_{n}$ and $H_{N+n}:=\check{H}_{n}$, and define that
\begin{gather}\label{RC-4}
iq\triangleq-w;\notag\\
C_{n}(x,t;z)=\frac{\hat{E}_{n}}{z-\hat{\xi}_{n}}e^{-2i\theta(x,t;\hat{\xi}_{n})},\quad
D_{n}(x,t)=\hat{H}_{n}-2i\theta'(x,t;\hat{\xi}_{n}).
\end{gather}
Correspondingly, we have $q_{\pm}=iw_{\pm}$ as $x\rightarrow\pm\infty$.
To get a closed linear algebraic integral system for the solution of the RHP, we evaluate the first column of the \eqref{RC-3} at $z=\hat{\xi}_{n}$ in $D_{-}$ through the concrete expression of $\mathop{Res}_{z=\xi_{n}}M^{+}$ and $\mathop{Res}_{z=\hat{\xi}_{n}}M^{-}$ i.e., \emph{Theorem 10}. Then, it is easy to get that
\begin{align}\label{RC-5}
\mu_{-,1}(x,t;\hat{\xi}_{k})=&\left(\begin{array}{cc}
                        1\\
                        \frac{iw_{-}}{\hat{\xi}_{k}}
                     \end{array}\right)
+\frac{1}{2\pi i}\int_{\Sigma}\frac{(M^{+}G)_{1}(s)}{s-\hat{\xi}_{k}}\,ds \notag \\
&+\sum_{n=1}^{2N}C_{n}(\hat{\xi}_{k})
\left[\mu'_{-,2}(x,t;\xi_{n})+\left(D_{n}(x,t)+\frac{1}{\hat{\xi}_{k}-\xi_{n}}\right)
\mu_{-,2}(x,t;\xi_{n})\right].
\end{align}
Via applying the \emph{Theorem 3}, we can derive that
\begin{align}\label{RC-6}
&\sum_{n=1}^{2N}\left(C_{n}(\hat{\xi}_{k})\mu'_{-,2}(x,t;\xi_{n})+
\left[C_{n}(\hat{\xi}_{k})\left(D_{n}(x,t)+\frac{1}{\hat{\xi}_{k}-\xi_{n}}\right)-
\frac{\xi_{k}}{iw^{*}_{-}}\delta_{kn}\right]\mu_{-,2}(x,t;\xi_{n})\right) \notag\\
&+\left(\begin{array}{cc}
         1 \\
         \frac{iw_{-}}{\hat{\xi}_{k}}
       \end{array}\right)
+\frac{1}{2\pi i}\int_{\Sigma}\frac{(M^{+}G)_{1}(s)}{s-\hat{\xi}_{k}}\,ds=0,
\end{align}
where the $\delta_{ij}$ is Kronecker delta. Then, by taking the first-order derivative of $\mu_{-,1}(x,t;z)$ and \eqref{S-4} in \emph{Theorem 3} with respect to $z$, and evaluating at $z=\hat{\xi}_{k}$, we can obtain that
\begin{align}\label{RC-7}
&\sum_{n=1}^{2N}\left(\left[\frac{C_{n}(\hat{\xi}_{k})}{\hat{\xi}_{k}-\xi_{n}}+
\frac{\xi_{k}^{3}}{iw^{*}_{-}q_{0}^{2}}\delta_{kn}\right]\mu'_{-,2}(\xi_{n})+
\left[\frac{C_{n}(\hat{\xi}_{k})}{\hat{\xi}_{k}-\xi_{n}}\left(D_{n}(x,t)+\frac{2}{\hat{\xi}_{k}-\xi_{n}}\right)
+\frac{\xi_{k}^{2}}{iw^{*}_{-}q_{0}^{2}}\delta_{kn}\right]\mu_{-,2}(\xi_{n})\right) \notag\\
&+\left(\begin{array}{cc}
         0\\
         \frac{iw_{-}}{\hat{\xi}_{k}^{2}}
       \end{array}\right)
-\frac{1}{2\pi i}\int_{\Sigma}\frac{(M^{+}G)_{1}(s)}{(s-\hat{\xi}_{k})^{2}}\,ds=0.
\end{align}
Furthermore, through considering the asymptotic behavior of   \eqref{RC-3}, we can obtain that
\begin{align}\label{RC-8}
\begin{split}
M(x,t;z)=\mathbb{I}&+\frac{1}{z}\left\{i\sigma_{3}Q_{-}+\sum_{n=1}^{2N}\left[
\mathop{Res}_{z=\hat{\xi}_{n}}M^{-}
+\mathop{Res}_{z=\xi_{n}}M^{+}\right]\right. \\
&\left.-\frac{1}{2\pi i}\int_{\Sigma}M^{+}(x,t;s)G(x,t;s)\,ds\right\}+O(z^{-2}),\quad z\rightarrow\infty.
\end{split}
\end{align}
 By taking $M=M^{-}$ and combining the $2,1$ element of \eqref{RC-8} and the \emph{Theorem 5}, we can get that
\begin{align}\label{RC-9}
w(x,t)=&w_{-}+\frac{1}{2\pi}\int_{\Sigma}(M^{+}(x,t;s)G(x,t;s))_{21}\,ds \notag \\
&+i\sum_{n=1}^{2N}E_{n}e^{-2i\theta(x,t;\xi_{n})}
[\mu'_{-,22}(x,t;\xi_{n})+\mu_{-,22}(x,t;\xi_{n})D_{n}(x,t)].
\end{align}
Therefore, it is obvious to acquire the reconstruction formula for the potential as shown at the begining of this subsection.

\subsection{Trace formulate and theta condition}
In this subsection, we will get the following main results.
\begin{itemize}
  \item \emph{The trace formulate are that}
  \begin{align}
s_{11}(z)&=exp\left(-\frac{1}{2\pi i }\int_{\Sigma}\frac{\log[1-\rho(s)\tilde{\rho}(s)]}{s-z}
\,d\zeta\right)\prod_{n=1}^{N}\frac{(z-z_{n})^{2}(z+q_{0}^{2}/z_{n}^{*})^{2}}
{(z-z_{n}^{*})^{2}(z+q_{0}^{2}/z_{n})^{2}},\label{dTT-1}\\
s_{22}(z)&=exp\left(-\frac{1}{2\pi i }\int_{\Sigma}\frac{\log[1-\rho(s)\tilde{\rho}(s)]}{s-z}
\,ds\right)\prod_{n=1}^{N}\frac{(z-z_{n}^{*})^{2}(z+q_{0}^{2}/z_{n})^{2}}
{(z-z_{n})^{2}(z+q_{0}^{2}/z_{n}^{*})^{2}}.\label{dTT-2}
\end{align}
  \item \emph{The theta condition are that}
  \begin{align}
\arg\frac{q_{+}}{q_{-}}=\arg q_{+}-\arg q_{-}=8\sum_{n=1}^{N}\arg z_{n}+\frac{1}{2\pi}\int_{\Sigma}
\frac{\log[1-\rho(s)\tilde{\rho}(s)]}{s}\,ds.
\end{align}
\end{itemize}

According to the analytic properties of $s_{11}$ and $s_{22}$ that have been shown in \emph{Theorem 2}, and the discrete spectrum, i.e.,
\begin{align}
\bar{Z}=\left\{z_{n}, -\frac{q_{0}^{2}}{z_{n}^{*}},
  z_{n}^{*}, -\frac{q_{0}^{2}}{z_{n}}\right\}, \quad n=1,2,...,N,
\end{align}
which have been analyzed in the \emph{subsection 8.1}, we can construct the following function
\begin{align}\label{dTT-3}
\zeta^{+}_{2}(z)=s_{11}(z)\prod_{n=1}^{N}\frac{(z-z_{n}^{*})^{2}(z+q_{0}^{2}/z_{n})^{2}}
{(z-z_{n})^{2}(z+q_{0}^{2}/z_{n}^{*})^{2}},\notag\\
\zeta^{-}_{2}(z)=s_{22}(z)\prod_{n=1}^{N}\frac{(z-z_{n})^{2}(z+q_{0}^{2}/z_{n}^{*})^{2}}
{(z-z_{n}^{*})^{2}(z+q_{0}^{2}/z_{n})^{2}}.
\end{align}
Therefore, the $\zeta^{+}_{2}$ and $\zeta^{-}_{2}$ are analytic in $D_{\pm}$, respectively, and have no zeros. Meanwhile, based on the asymptotic behavior of $S(z)$ in \emph{Theorem 6}, it is obvious that $\zeta^{\pm}_{2}(z)\rightarrow1$ as $z\rightarrow\infty$. Then, considering that $\det S(z)=1$ and the expression of the reflection coefficients, we obtain that
\begin{align}\label{dTT-4}
\zeta^{+}_{2}(z)\zeta_{2}^{-}(z)=\frac{1}{1-\rho(z)\tilde{\rho}(z)},\quad z\in\Sigma.
\end{align}
Furthermore, by taking the logarithm of   \eqref{dTT-4} and applying the Plemelj's formulae and projection operators, we obtain that
\begin{align}\label{dTT-5}
\log\zeta^{\pm}_{2}(z)=-\frac{1}{2\pi i}\int_{\Sigma}
\frac{\log[1-\rho(s)\tilde{\rho}(s)]}{s-(z\pm i0)}\,ds,\quad z\in D_{\pm}.
\end{align}
Then, through substituting the \eqref{dTT-5} into \eqref{dTT-3},we can obtain the trace formula, i.e., the first result in this section.

Then we consider the theta condition. According to the \emph{Theorem 6}, we know that $s_{11}(z)\rightarrow q_{+}/q_{-}$ as $z\rightarrow0$. Meanwhile, since
\begin{align}
\prod_{n=1}^{N}\frac{(z-z_{n}^{*})^{2}(z+q_{0}^{2}/z_{n})^{2}}
{(z-z_{n})^{2}(z+q_{0}^{2}/z_{n}^{*})^{2}}\rightarrow 1, \quad as \quad z\rightarrow0, \notag
\end{align}
we can obtain the theta condition, i.e., the second result that has been shown at the beginning of this section.

\subsection{Solving the Generalized Riemann-Hilbert problem}
\subsubsection{Soliton solutions}
In this subsection, we will obtain a type of $N$-soliton solutions.
\begin{itemize}
  \item \emph{The specific formula of the $N$-soliton solutions is}
  \begin{align}\label{dSS-1}
&q(x,t)=iw(x,t)\notag \\
&=i\left(w_{-}+i\sum_{n=1}^{2N}E_{n}e^{-2i\theta(x,t;\xi_{n})}
[\mu'_{-,22}(x,t;\xi_{n})+\mu_{-,22}(x,t;\xi_{n})D_{n}(x,t)]\right).
\end{align}
\emph{where the $\mu'_{-,22}(x,t;\xi_{n})$ and $\mu_{-,22}(x,t;\xi_{n})$ can be solved by  equation \eqref{dSS-4}.}
\end{itemize}

In this subsection, we are going to consider a case that the reflection coefficients $\rho(z)$ and $\tilde{\rho}(z)$ disappear. Thus, the jump from $M^{+}$ to $M^{-}$ vanishes, i.e. $G(x,t;z)=0$. Under this conditions and from the \eqref{RC-1}, we can get the $N$-soliton solutions which have been shown in \eqref{dSS-1}. Now, we confirm the $\mu'_{-,22}$ and $\mu_{-,22}$. According to the assumption and from the \eqref{RC-6} and \eqref{RC-7}, we can derive that
\begin{gather}
\sum_{n=1}^{2N}\left(C_{n}(\hat{\xi}_{k})\mu'_{-,22}(\xi_{n})+
\left[C_{n}(\hat{\xi}_{k})\left(D_{n}(x,t)+\frac{1}{\hat{\xi}_{k}-\xi_{n}}\right)-
\frac{\xi_{k}}{iw^{*}_{-}}\delta_{kn}\right]\mu_{-,22}(\xi_{n})\right)+\frac{iw_{-}}{\hat{\xi}_{k}}=0, \label{dSS-2} \\
\sum_{n=1}^{2N}\left[\frac{C_{n}(\hat{\xi}_{k})}{\hat{\xi}_{k}-\xi_{n}}\left(D_{n}(x,t)
+\frac{2}{\hat{\xi}_{k}-\xi_{n}}\right)
+\frac{\xi_{k}^{2}}{iw^{*}_{-}q_{0}^{2}}\delta_{kn}\right]\mu_{-,22}(\xi_{n}) \notag \\
+\sum_{n=1}^{2N}\left(\frac{C_{n}(\hat{\xi}_{k})}{\hat{\xi}_{k}-\xi_{n}}+
\frac{\xi_{k}^{3}}{iw^{*}_{-}q_{0}^{2}}\delta_{kn}\right)\mu'_{-,22}(\xi_{n})
+\frac{iw_{-}}{\hat{\xi}_{k}^{2}}=0. \label{dSS-3}
\end{gather}
Then, we introduce that
\begin{align}
X_{n}=\mu_{-,11}(x,t;\hat{\xi}_{n}), \quad X_{2N+n}=\mu'_{-,11}(x,t;\hat{\xi}_{n}),\\
V_{n}=\frac{iq_{-}}{\xi_{n}}, \quad V_{2N+n}=\frac{iq_{-}}{\xi^{2}_{n}}, \quad n=1,2,\cdots,2N.
\end{align}
Consequently, Eqs. \eqref{dSS-2} and \eqref{dSS-3} can be expressed as
\begin{align}\label{dSS-4}
MX=V,
\end{align}
where $M$ is a $4N\times4N$ matrix and the elements of $M$ can be found easily from Eqs. \eqref{dSS-2} and \eqref{dSS-3}.

\subsubsection{The phenomenon of the soliton solutions}
In this subsection, we will analyze the dynamic behavior of the soliton solutions for the double case. Firstly, as usual, we investigate the case that the coefficient $\alpha(t)$ and $\gamma$ are zeros.
When $N=1$, by selecting the appropriate parameters, the following images can be obtained.
\\

{\rotatebox{0}{\includegraphics[width=3.6cm,height=3.0cm,angle=0]{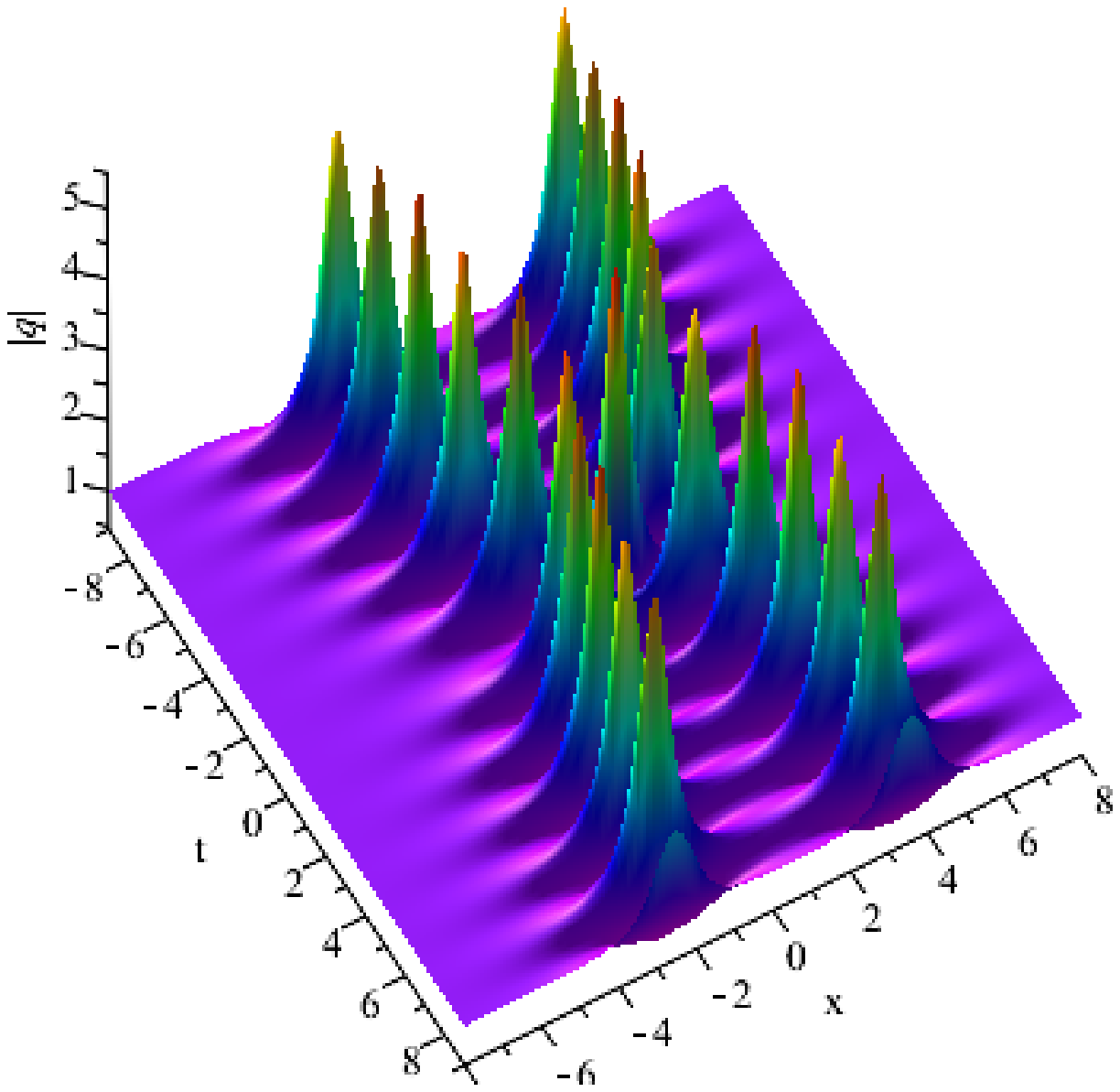}}}
~~~~
{\rotatebox{0}{\includegraphics[width=3.6cm,height=3.0cm,angle=0]{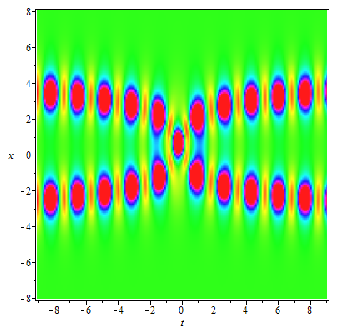}}}
~~~~
{\rotatebox{0}{\includegraphics[width=3.6cm,height=3.0cm,angle=0]{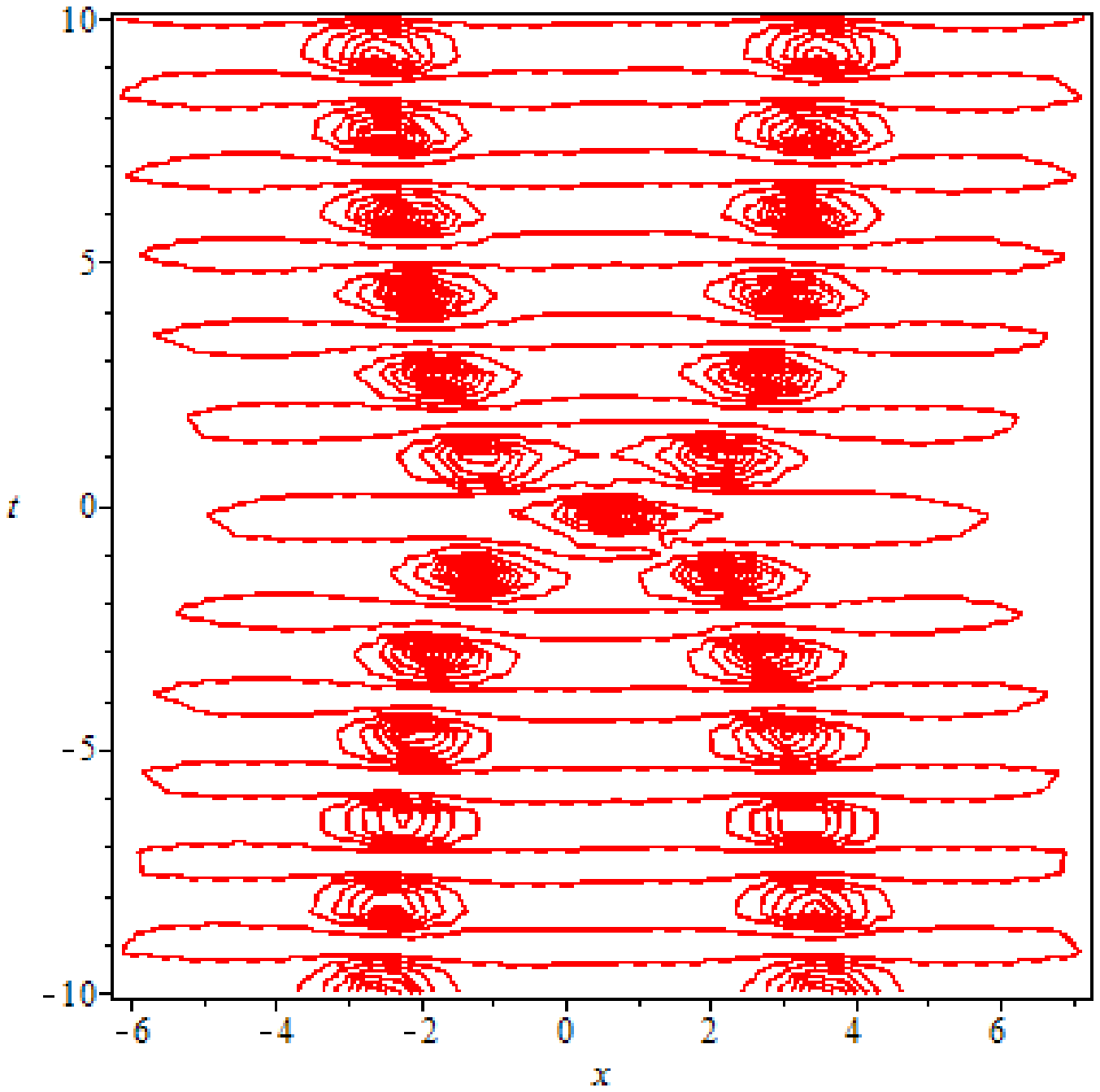}}}

$\qquad~~~~~~~~~(\textbf{a})\qquad \ \qquad\qquad\qquad\qquad~~~(\textbf{b})
\qquad\qquad\qquad\qquad\qquad~(\textbf{c})$\\
\noindent { \small \textbf{Figure 7.} (Color online) Plots of the soliton solution of the equation  with the parameters $\alpha(t)=\gamma=0$, $q_{-}=1$, $\xi_{1}=2i$ and $e_{1}=h_{1}=e^{1+i}$.
$\textbf{(a)}$: the soliton solution,
$\textbf{(b)}$: the density plot ,
$\textbf{(c)}$: the contour line of the soliton solution.} \\

The Fig. 7 illustrates an interesting phenomenon that there are two columns of breather solutions interact in the process of propagation. Next, by changing the boundary value $q_{-}$, we can get the following graphs.\\

{\rotatebox{0}{\includegraphics[width=3.6cm,height=3.0cm,angle=0]{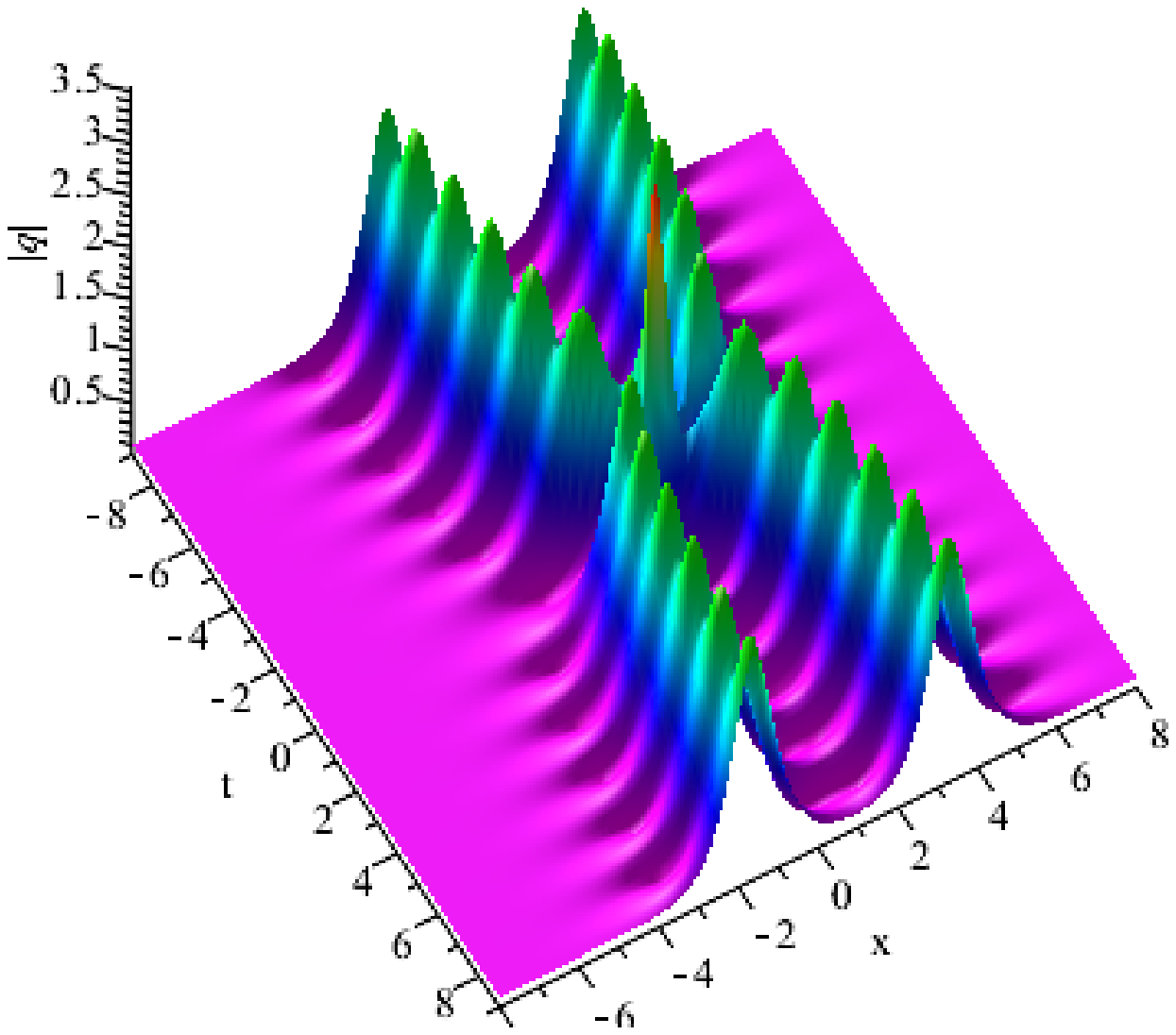}}}
~~~~
{\rotatebox{0}{\includegraphics[width=3.6cm,height=3.0cm,angle=0]{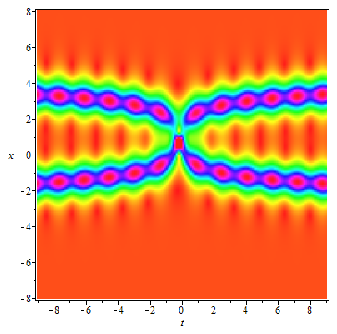}}}
~~~~
{\rotatebox{0}{\includegraphics[width=3.6cm,height=3.0cm,angle=0]{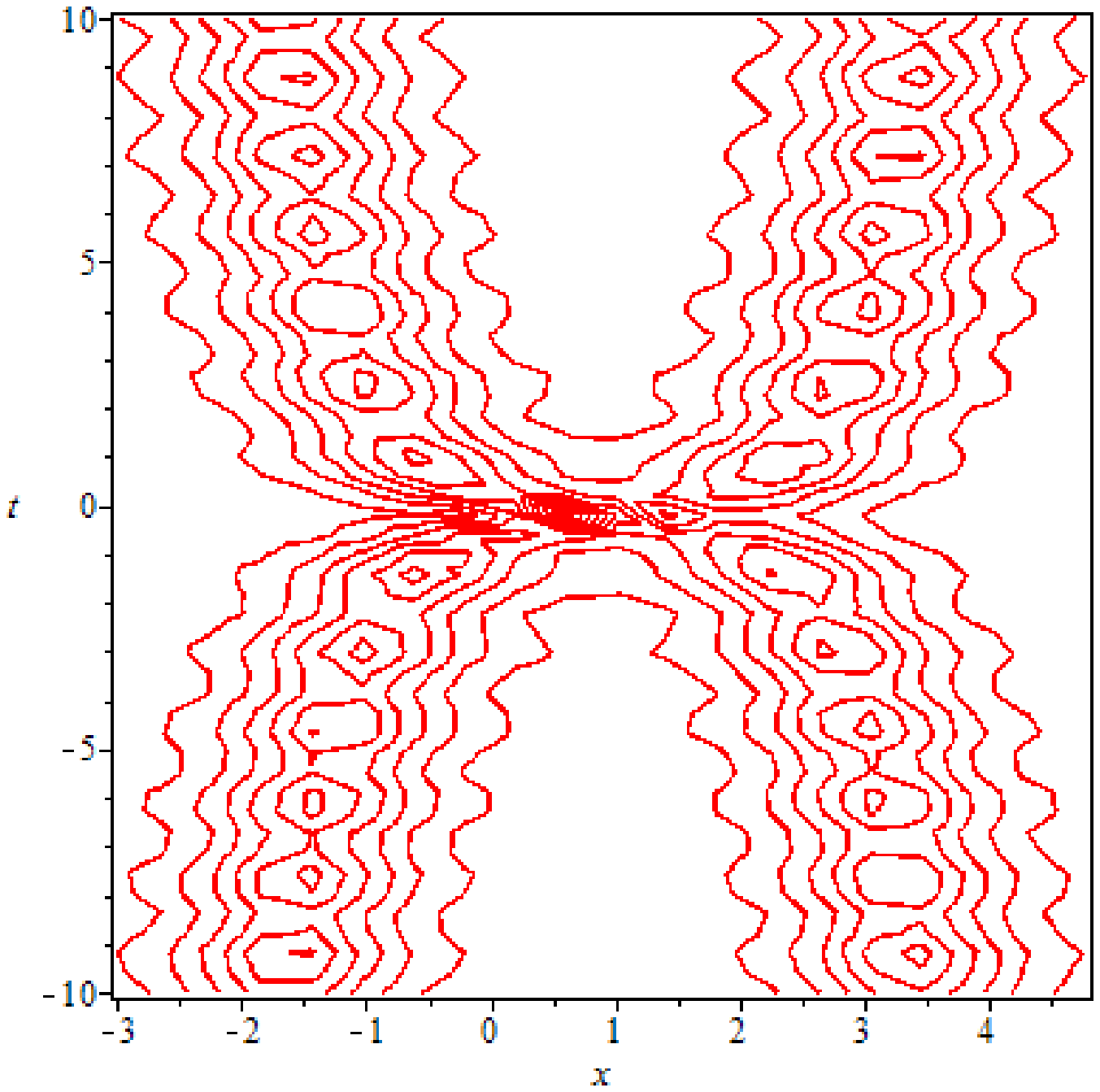}}}

$\qquad~~~~~~~~~(\textbf{a})\qquad \ \qquad\qquad\qquad\qquad~~~(\textbf{b})
\qquad\qquad\qquad\qquad\qquad~(\textbf{c})$\\

{\rotatebox{0}{\includegraphics[width=3.6cm,height=3.0cm,angle=0]{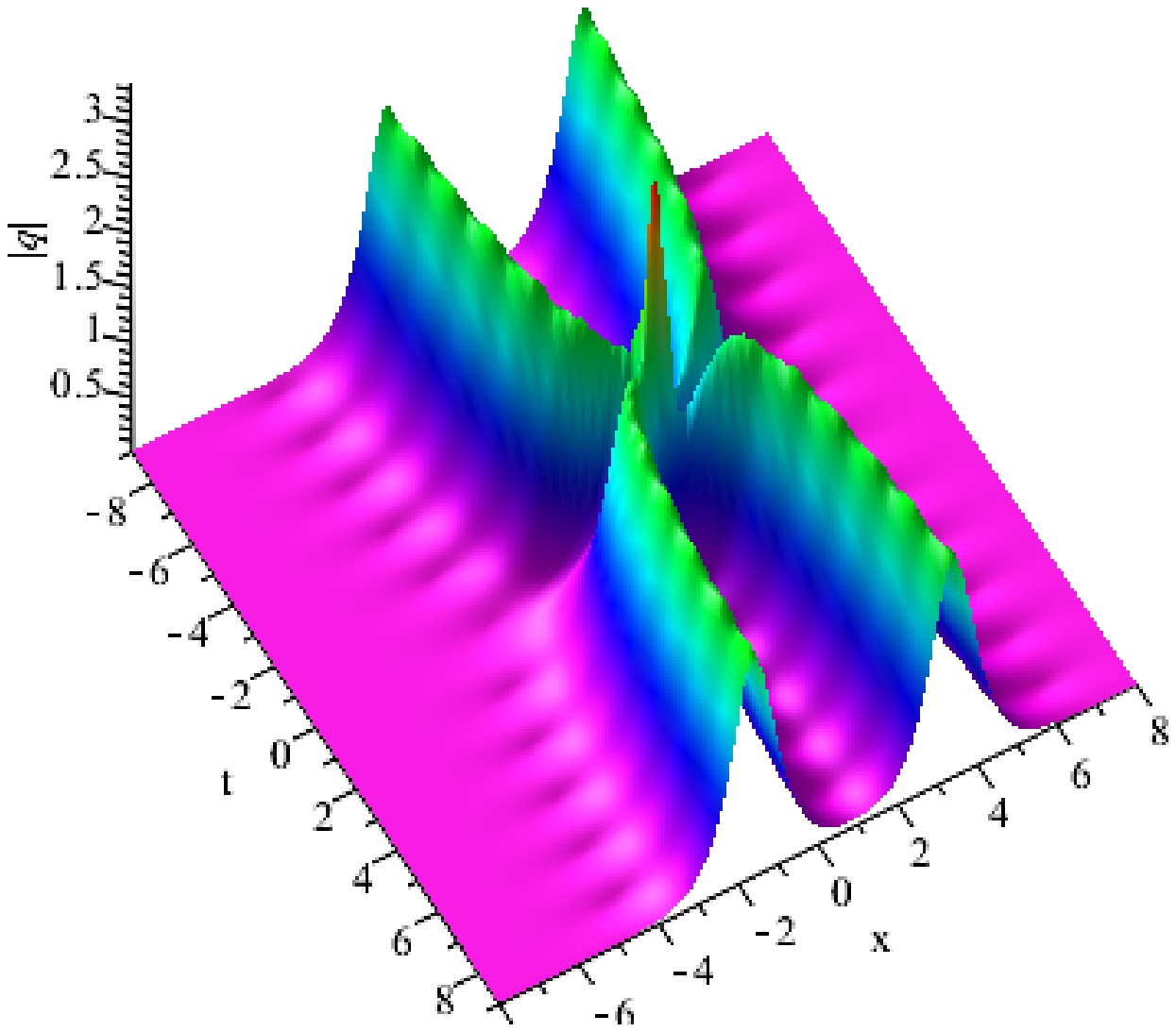}}}
~~~~
{\rotatebox{0}{\includegraphics[width=3.6cm,height=3.0cm,angle=0]{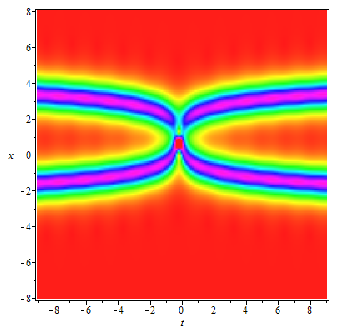}}}
~~~~
{\rotatebox{0}{\includegraphics[width=3.6cm,height=3.0cm,angle=0]{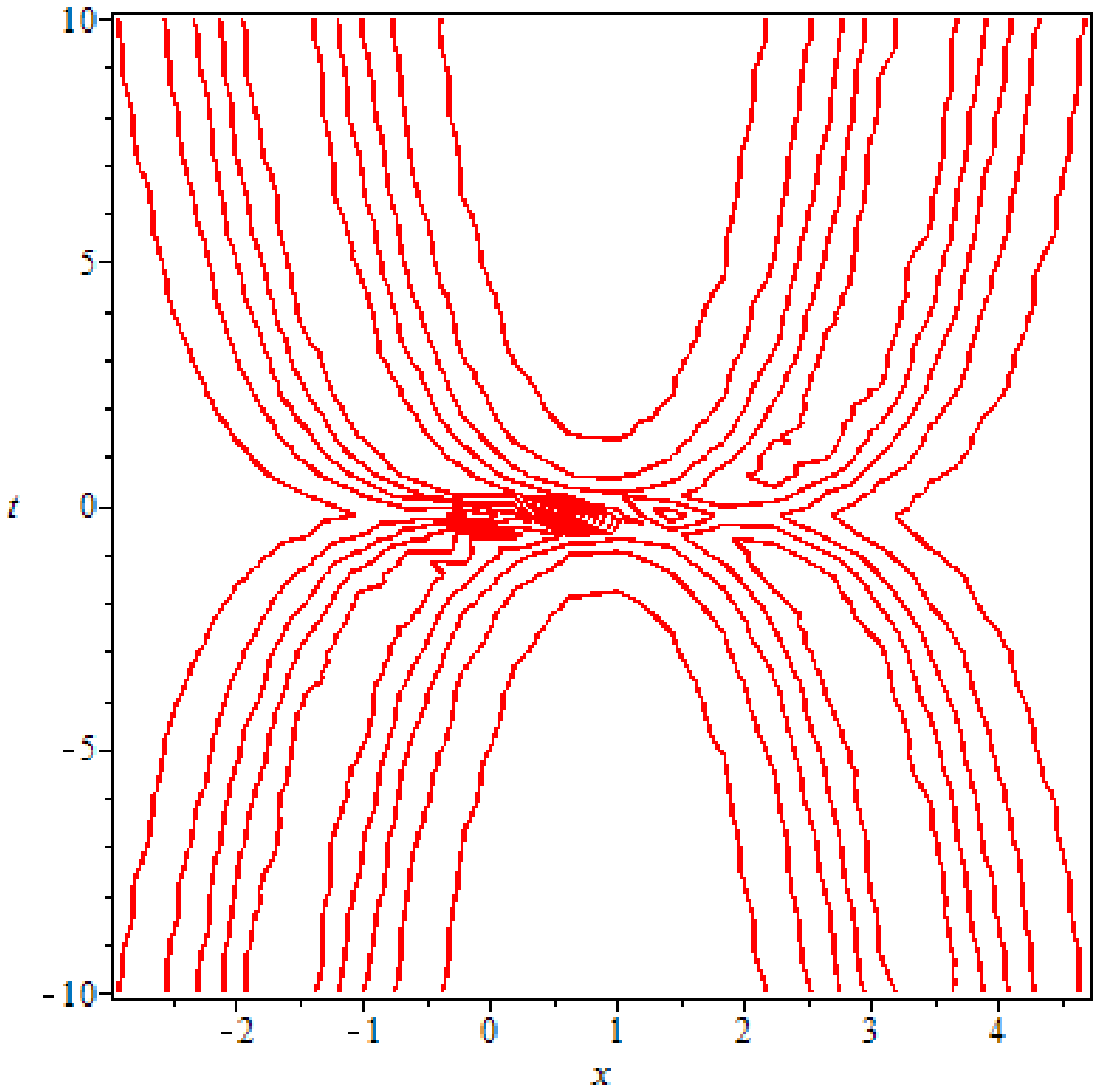}}}

$\qquad~~~~~~~~~(\textbf{d})\qquad \ \qquad\qquad\qquad\qquad~~~(\textbf{e})
\qquad\qquad\qquad\qquad\qquad~(\textbf{f})$\\
\noindent { \small \textbf{Figure 8.} (Color online) Plots of the soliton solutions of the equation  with the parameters $\alpha(t)=\gamma=0$, $\xi_{1}=2i$ and $e_{1}=h_{1}=e^{1+i}$.
$\textbf{(a)}$: the soliton solution with $q_{-}=0.1$,
$\textbf{(b)}$: the density plot corresponding to $(a)$,
$\textbf{(c)}$: the contour line of the soliton solution corresponding to $(a)$,
$\textbf{(d)}$: the soliton solution with $q_{-}=0.01$,
$\textbf{(e)}$: the density plot corresponding to $(d)$,
$\textbf{(f)}$: the contour line of the soiton solution corresponding to $(d)$.} \\

The Fig. 8 shows that the breathing phenomenon gradually disappears and only a sharp soliton exists when the boundary value $q_{-}$ becomes smaller gradually. This is an interesting phenomenon. Furthermore, we study the case that the parameter $\alpha(t)$ and $\gamma$ are changing.\\

{\rotatebox{0}{\includegraphics[width=3.6cm,height=3.0cm,angle=0]{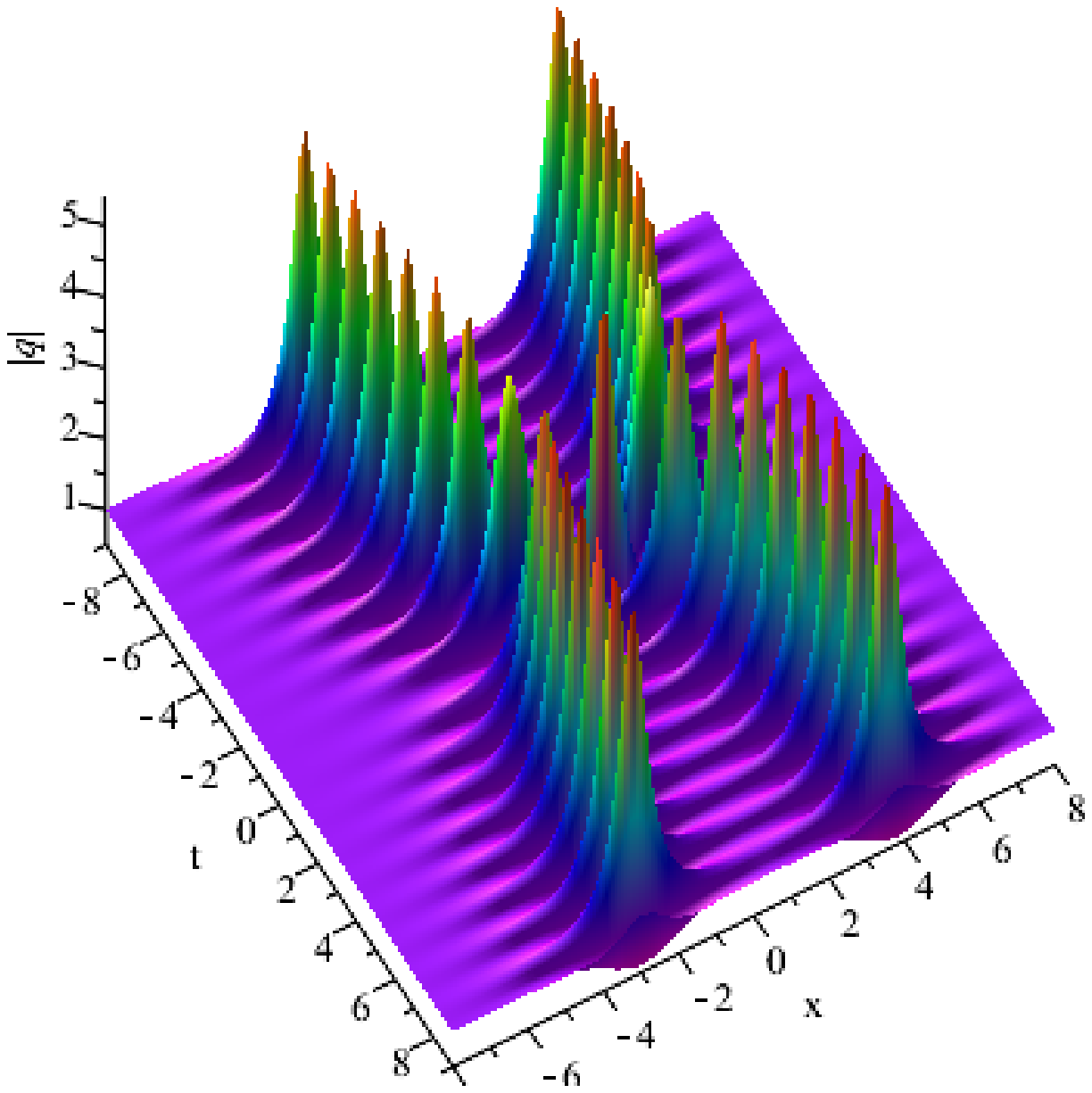}}}
~~~~
{\rotatebox{0}{\includegraphics[width=3.6cm,height=3.0cm,angle=0]{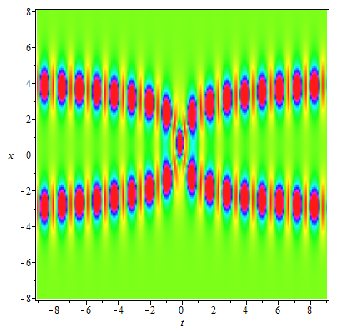}}}
~~~~
{\rotatebox{0}{\includegraphics[width=3.6cm,height=3.0cm,angle=0]{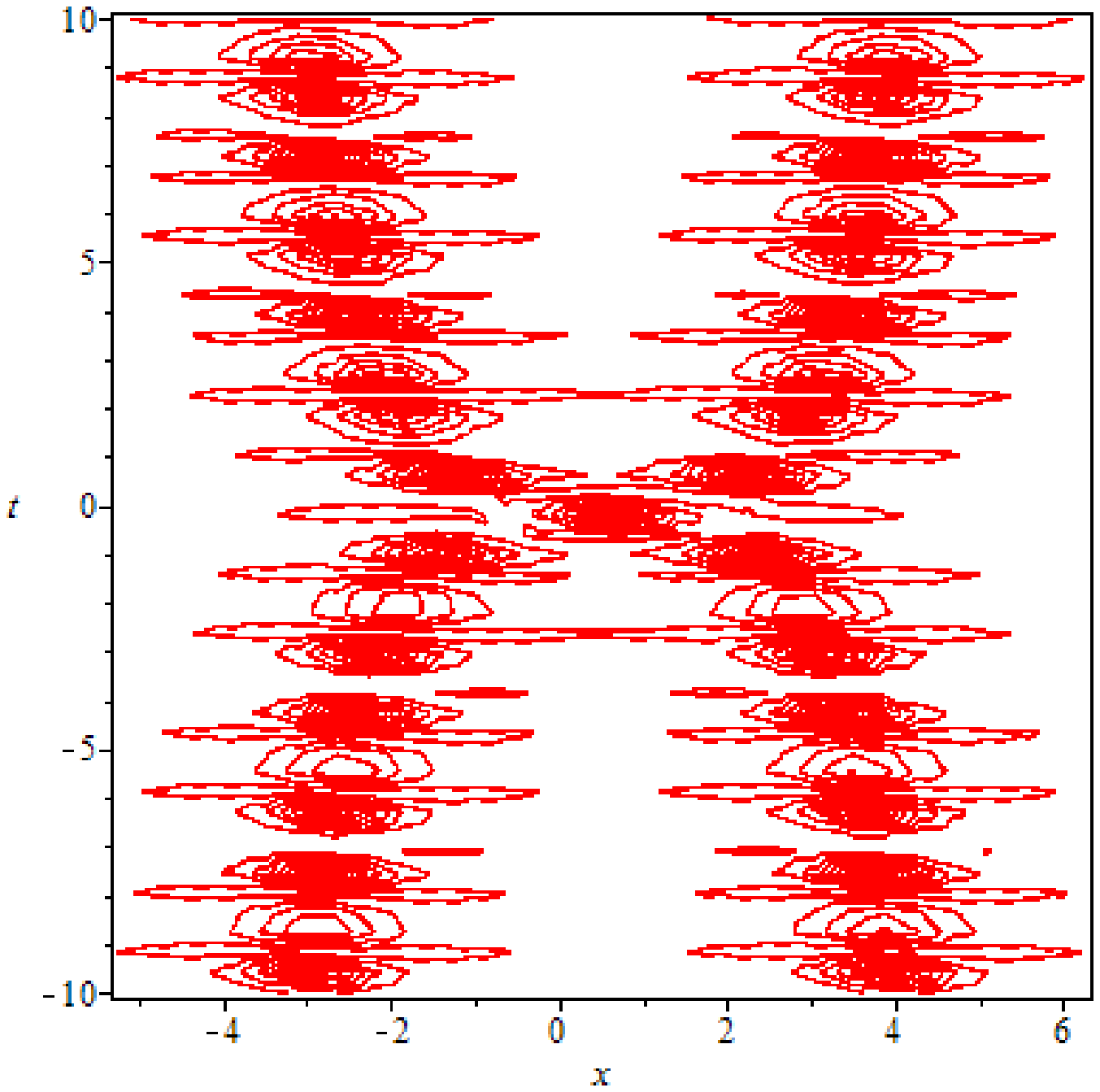}}}

$\qquad~~~~~~~~~(\textbf{a})\qquad \ \qquad\qquad\qquad\qquad~~~(\textbf{b})
\qquad\qquad\qquad\qquad\qquad~(\textbf{c})$ \\

{\rotatebox{0}{\includegraphics[width=3.6cm,height=3.0cm,angle=0]{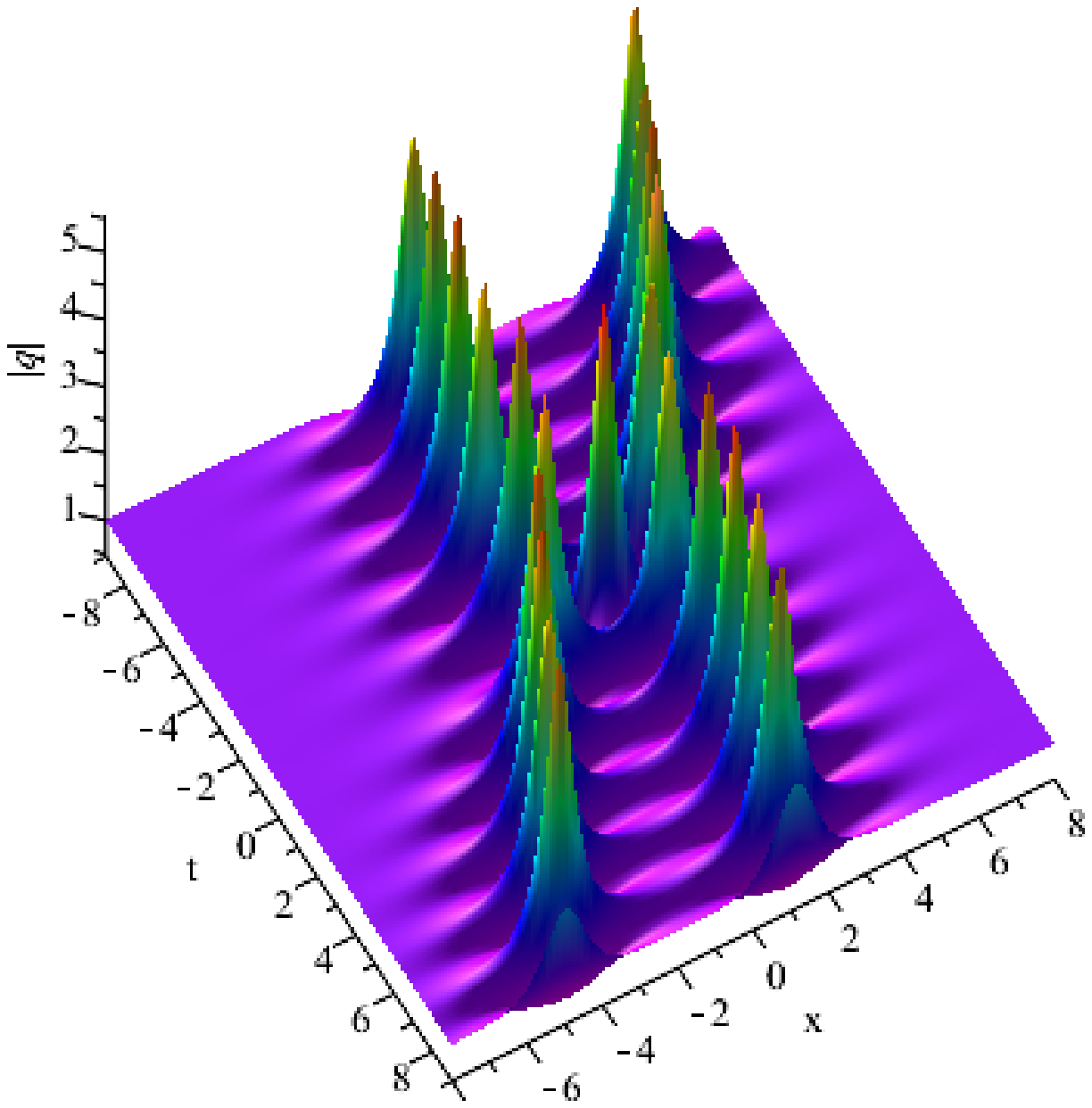}}}
~~~~
{\rotatebox{0}{\includegraphics[width=3.6cm,height=3.0cm,angle=0]{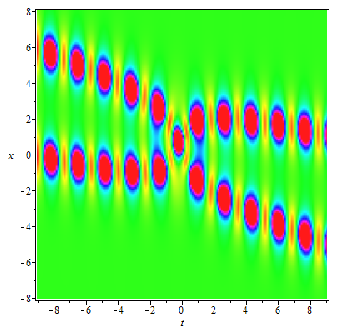}}}
~~~~
{\rotatebox{0}{\includegraphics[width=3.6cm,height=3.0cm,angle=0]{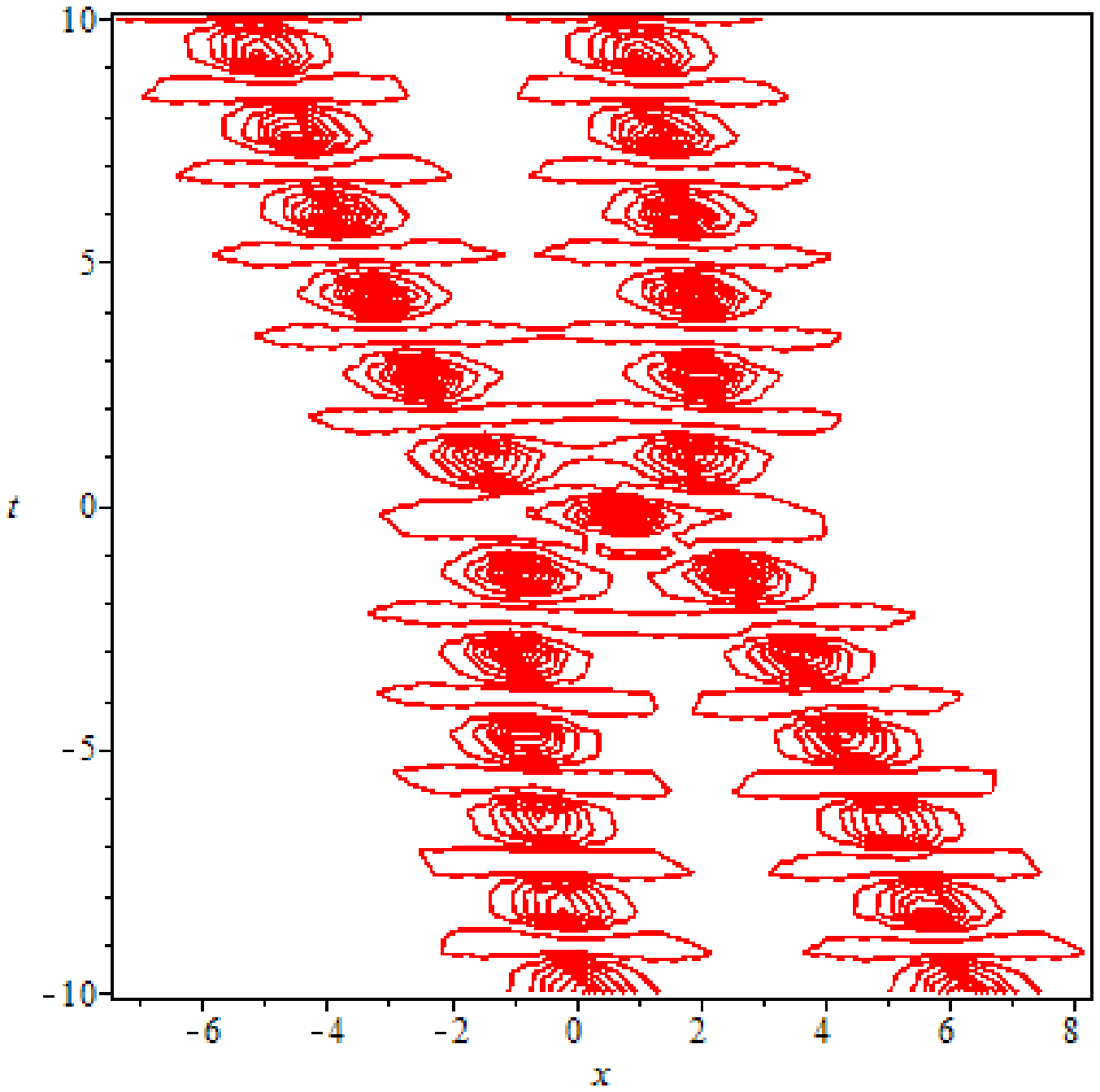}}}

$\qquad~~~~~~~~~(\textbf{d})\qquad \ \qquad\qquad\qquad\qquad~~~(\textbf{e})
\qquad\qquad\qquad\qquad\qquad~(\textbf{f})$\\

{\rotatebox{0}{\includegraphics[width=3.6cm,height=3.0cm,angle=0]{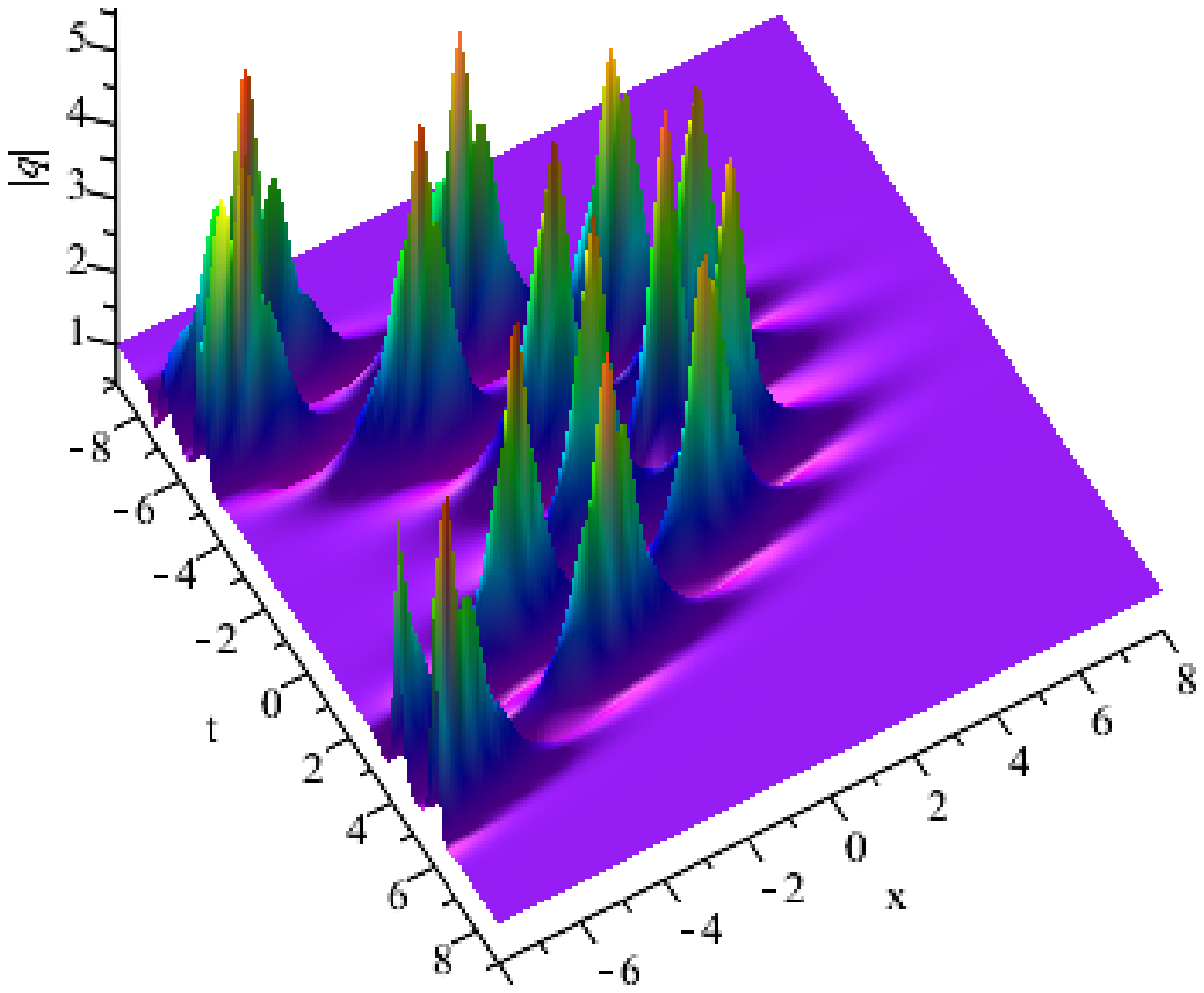}}}
~~~~
{\rotatebox{0}{\includegraphics[width=3.6cm,height=3.0cm,angle=0]{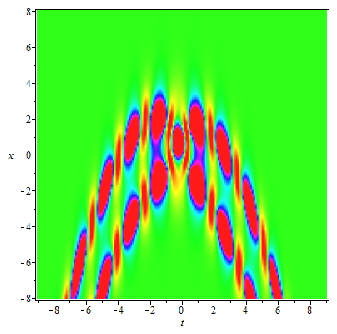}}}
~~~~
{\rotatebox{0}{\includegraphics[width=3.6cm,height=3.0cm,angle=0]{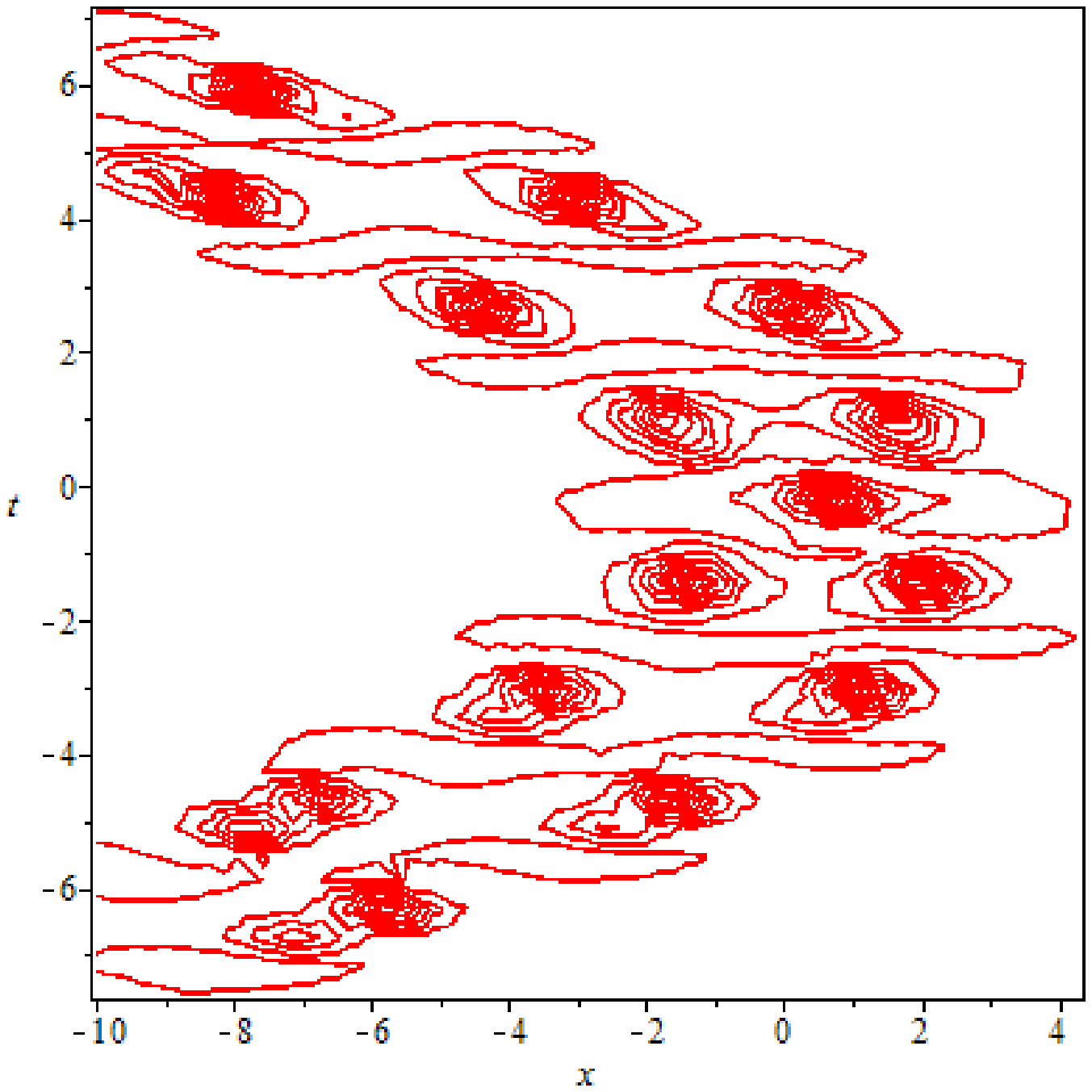}}}

$\qquad~~~~~~~~~(\textbf{g})\qquad \ \qquad\qquad\qquad\qquad~~~(\textbf{h})
\qquad\qquad\qquad\qquad\qquad~(\textbf{i})$\\
\noindent { \small \textbf{Figure 9.} (Color online) Plots of the soliton solutions of the equation  with the parameters $q_{-}=1$, $\xi_{1}=2i$ and $e_{1}=h_{1}=e^{1+i}$.
$\textbf{(a)}$: the soliton solution with $\alpha(t)=0$ and $\gamma=0.01$,
$\textbf{(b)}$: the density plot corresponding to $(a)$,
$\textbf{(c)}$: the contour line of the soliton solution corresponding to $(a)$,
$\textbf{(d)}$: the soliton solution with $\alpha(t)=0.01$ and $\gamma=0$,
$\textbf{(e)}$: the density plot corresponding to $(d)$,
$\textbf{(f)}$: the contour line of the soliton solution corresponding to $(d)$,
$\textbf{(g)}$: the soliton solution with $\alpha(t)=0.01t+0.01$ and $\gamma=0$,
$\textbf{(h)}$: the density plot corresponding to $(g)$,
$\textbf{(i)}$: the contour line of the soliton solution corresponding to $(g)$.} \\

In Fig. 9, $(a)$ reveals that  both of the two column breather soliton solutions are more closely arranged when the parameter $\gamma$ exists; $(d)$ reveals that the direction of the propagation of the soliton solution will deviate from the original direction when the parameter $\alpha(t)$ exists and is a real constant; and $(g)$ reveals that  the propagation path of the soliton solution will be changed into a path which is similar to a quadratic function curve when $\alpha(t)$ is a linear function about $t$. Furthermore, we evaluate the case that $\alpha(t)$ and $\gamma$ are not zeros.\\

{\rotatebox{0}{\includegraphics[width=3.6cm,height=3.0cm,angle=0]{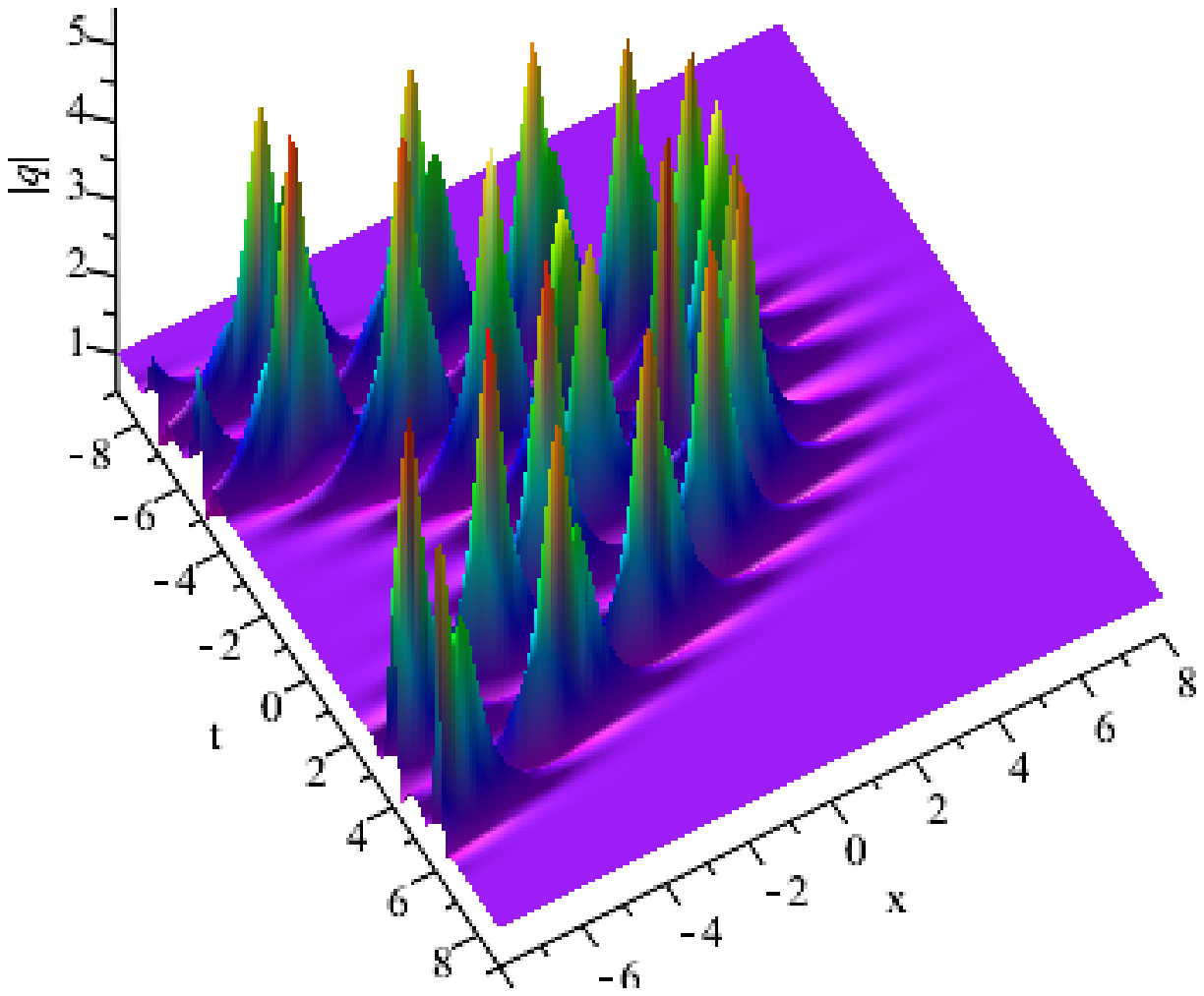}}}
~~~~
{\rotatebox{0}{\includegraphics[width=3.6cm,height=3.0cm,angle=0]{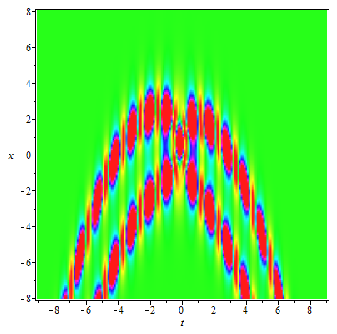}}}
~~~~
{\rotatebox{0}{\includegraphics[width=3.6cm,height=3.0cm,angle=0]{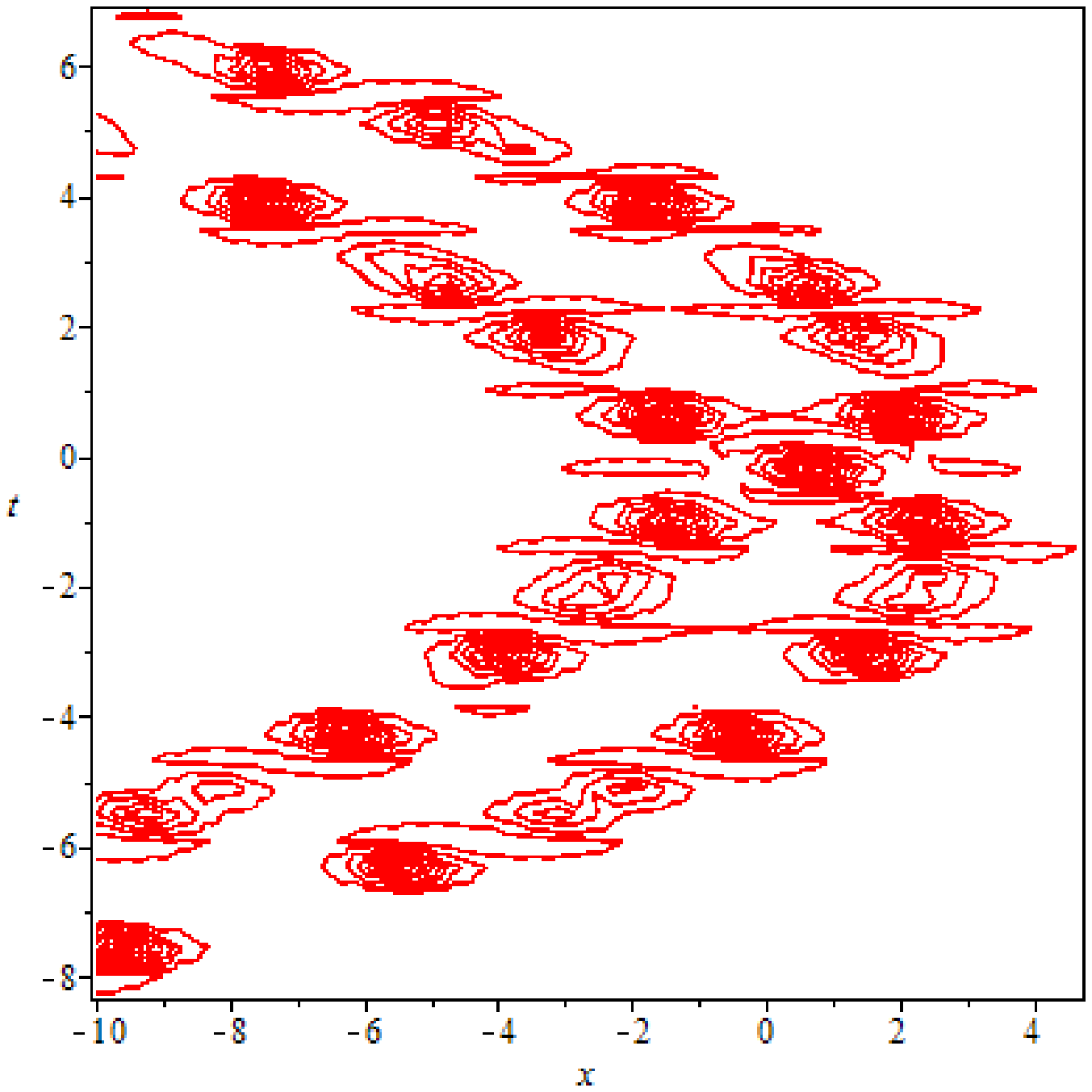}}}

$\qquad~~~~~~~~~(\textbf{a})\qquad \ \qquad\qquad\qquad\qquad~~~(\textbf{b})
\qquad\qquad\qquad\qquad\qquad~(\textbf{c})$\\
\noindent { \small \textbf{Figure 10.} (Color online) Plots of the soliton solution of the equation  with the parameters $q_{-}=1$, $\xi_{1}=2i$ and $e_{1}=h_{1}=e^{1+i}$.
$\textbf{(a)}$: the soliton solution with $\alpha(t)=0.01t+0.01$ and $\gamma=0.01$,
$\textbf{(b)}$: the density plot ,
$\textbf{(c)}$: the contour line of the soliton solution.} \\

The Fig. 10 shows an interesting phenomenon that the propagation path of the soliton solution is not only changed into a path which is similar to a quadratic function curve, but also both of the two column breather soliton solutions are more closely arranged due to the influence of the existence of the parameter $\gamma$ and $\alpha(t)$.

From the above figures, in the double poles case, we can get a results which are similar to the simple poles case, i.e., the existence of the parameter $\gamma$ only determines the shape of soliton solutions and the existence of the parameter $\alpha(t)$ will change the propagation path of the soliton solution. To summarise, we obtain the dynamic behavior of the soliton solutions and various interesting phenomena via selecting appropriate parameters. It is hoped that these results can contribute to the physical field.

\section{Conclusions and discussions}
For the gvcNLS equation \eqref{1.1}, Zuo, et al. have obtained the Lax pair, soliton and rogue-wave solutions via the Darboux transformation and generalized Darboux transformation. However, what we have studied here is the gvcNLS equation with non-zero boundary conditions via using the inverse scattering transform and RH approach. Meanwhile, we not only get the specific expression of the analytical solution, but also fully evaluate the impact of the parameter $\alpha(t)$ and $\gamma$ on the soliotn solutions. Furthermore, some interesting phenomena are presented to study the dynamic behavior of the soliton solutions

In this work, the IST for the gvcNLS equation with NBCs at infinity is presented. Firstly, by introducing a appropriate Riemann surface and uniformization variable, we deal with the difficulty that the double-valued functions occur in the process of direct scattering. Then, for the Jost function and scattering matrix, their analytical properties, symmetries and asymptotic behavior are well provided. Based on these results, we  establish a generalized Riemann-Hilbert problem. Furthermore, under the condition of reflection-less potentials, a kind of special soliton solutions is given for both of the two case, i.e., simple and double poles. Additionally, based on the concrete expression of the solution, some graphic analysis are presented to reveal the interesting phenomena by selecting some appropriate parameters, and changing the parameter $\alpha(t)$ and $\gamma$.

\section*{Acknowledgements}
%\textcolor[rgb]{1.00,0.00,0.00}{We express our sincere thanks to the Editor and Reviewers for their valuable comments.}
This work was supported by the Postgraduate Research and Practice of Educational Reform for Graduate students in CUMT under Grant No. 2019YJSJG046, the Natural Science Foundation of Jiangsu Province under Grant No. BK20181351, the Six Talent Peaks Project in Jiangsu Province under Grant No. JY-059, the Qinglan Project of Jiangsu Province of China, the National Natural Science Foundation of China under Grant No. 11975306, the Fundamental Research Fund for the Central Universities under the Grant Nos. 2019ZDPY07 and 2019QNA35, and the General Financial Grant from the China Postdoctoral Science Foundation under Grant Nos. 2015M570498 and 2017T100413.

\renewcommand{\baselinestretch}{1.2}

\end{document}